\documentclass[a4paper]{issibook} 		 
\usepackage[cp1250]{inputenc}
\usepackage{amsmath, amssymb, graphicx, natbib}
\bibpunct{(}{)}{;}{a}{}{,}
\usepackage{txfonts}
\usepackage[bookmarks=false]{hyperref}

\begin{document}

\setcounter{chapter}{1}
\chapter[Solar parameters]{Solar parameters for modeling interplanetary background}

\chapauthor{M. Bzowski, J.M. Sok\'{o}{\l} \\
Space Research Center Polish Academy of Sciences, Warsaw, Poland}
\chapauthor{M. Tokumaru, K. Fujiki\\
Solar-Terrestrial Environment Laboratory, Nagoya University, Nagoya, Japan}
\chapauthor{E. Quemerais, R. Lallement\\
LATMOS-IPSL, Universite Versailles Saint-Quentin, Guyancourt, France}
\chapauthor{S. Ferron\\
ACRI-ST, Sophia Antipolis, France}
\chapauthor{P. Bochsler \\ Space Science Center \& Department of Physics, University of New Hampshire, Durham NH \\ Physikalisches Institut, University of Bern, Bern, Switzerland}
\chapauthor{D.J. McComas \\ Southwest Research Institute, San Antonio TX \\ University of Texas at San Antonio, San Antonio, TX 78249, USA}

\section*{Abstract}
The goal of the Fully Online Datacenter of Ultraviolet Emissions (FONDUE) Working Team of the International Space Science Institute (ISSI) in Bern, Switzerland, was to establish a common calibration of various UV and EUV heliospheric observations, both spectroscopic and photometric. Realization of this goal required a credible and up-to-date model of spatial distribution of neutral interstellar hydrogen in the heliosphere, and to that end, a credible model of the radiation pressure and ionization processes was needed. This chapter describes the latter part of the project: the solar factors responsible for shaping the distribution of neutral interstellar H in the heliosphere. Presented are the solar Lyman-alpha flux and the question of solar Lyman-alpha resonant radiation pressure force acting on neutral H atoms in the heliosphere, solar EUV radiation and the process of photoionization of heliospheric hydrogen, and their evolution in time and the still hypothetical variation with heliolatitude. Further, solar wind and its evolution with solar activity is presented, mostly in the context of the charge exchange ionization of heliospheric neutral hydrogen, and in the context of dynamic pressure variations. Also the question of electron ionization and its variation with time, heliolatitude, and solar distance is presented. After a review of the state of the art in all of those topics, we proceed to present an interim model of solar wind and the other solar factors based on up-to-date in situ and remote sensing observations of solar wind. This model was used by Izmodenov et al. to calculate the distribution of heliospheric hydrogen, which in turn was basis for intercalibration of the heliopsheric UV and EUV measurements discussed in the other chapters of the book. Results of this joint effort will further be utilised to improve on the model of solar wind evolution, which will be an invaluable asset in all heliospheric measurements, including, among others, the observations of Energetic Neutral Atoms by the Interstellar Boundary Explorer (IBEX).

\section{Brief description of the physics of neutral interstellar gas in the inner heliosphere}

Distribution of neutral interstellar hydrogen and ultraviolet radiation in the inner heliosphere are closely interrelated. Absolute calibration of observations of the heliospheric backscatter Lyman-alpha glow requires both the knowledge of the well-calibrated solar EUV output and of other solar forcing factors, mainly the solar wind. The role of those factors and their variabilities in shaping the distribution of neutral interstellar hydrogen can be apprehended from modeling papers cited in the remaining portion of this section.

If one takes an inflow with a finite velocity $\vec{v}_{\infty}$ of a fully neutral gas at a temperature $T_{\infty}$ far away from the Sun, then when the ionization rate $\beta$ at $\vec{r}$ and the effective force $\vec{F}\left(\vec{r}\right)$ acting on the atoms are spherically symmetric and time-independent, the distribution function of the gas at a distance $r$ from the Sun will feature axial symmetry around the inflow direction and be given by the equation:
\begin{equation}
\label{eqBoltz}
\vec{v} \cdot \nabla_{\vec{r}} f\left( \vec{v}, \vec{r} \right)+\frac{\vec{F}\left( \vec{r} \right)}{m_H}\cdot \nabla_{\vec{v}}{f\left(\vec{v}, \vec{r}\right)}=-\beta\, f\left(\vec{v},\vec{r}\right)
\end{equation}
where $\nabla_x$ is the gradient operation in the $x$-space, $\vec{r}$ and $\vec{v}$ are, respectively, position and velocity vectors of the gas cell element, and $m_H$ is hydrogen atom mass. Together with the assumption that the gas ``at infinity'' (in practice: a few hundreds of AU from the Sun) is homogeneous and Maxwellian and that the gas is collisionless at the distances relevant for the model \citep{izmodenov_etal:00a} this is the basis for the classical hot model of the distribution of neutral interstellar gas in the heliosphere \citep{thomas:78, fahr:78, fahr:79, wu_judge:79a, lallement_etal:85b}. Based on Liouville theorem, solution of this equation \citep{danby_camm:57} for the distribution function $f\left(\vec{v}, \vec{r}, t \right)$ for time $t$, location $\vec{r}$ and velocity $\vec{v}$ can be expressed as: 
		\begin{equation}
		\label{eqDiFu}
		f\left(\vec{v},\vec{r},t\right)= f_{\infty }\left( \vec{v}_{\infty}\left(\vec{v},r\right),\vec{r}_{\infty }\left(\vec{v},\vec{r}\right)\right) W\left(\vec{v},\vec{r},t\right),
		\end{equation}
where $W\left(\vec{v}, \vec{r}, t\right)$ is the survival probability of an atom that at the time $t$ and location $\vec{r}$ has velocity $\vec{v}$ and which at a distant location $\vec{r_{\infty}}$ had velocity $\vec{v_{\infty}}$. For now, $t$ is only a formal parameter here. The probability of existence of such an atom in the distant region of the heliosphere is given by the distribution function $f_{\infty}\left(\vec{v}_{\infty},\vec{r}_{\infty}\right)$, and the link between the local velocity and position vectors $\vec{v}\left(t\right)$, $\vec{r}\left(t\right)$ and the corresponding vectors in the so-called source region of the atoms can be obtained from solution of the equation of motion of hydrogen atoms in the heliosphere:
		\begin{equation}
		\label{eqEqMo}
		\vec{F}\left(\vec{r},t,v_r\right)=-\frac{G m M \left(1-\mu \left(v_r,t\right)\right)}{r^2}\,\frac{\vec{r}}{r}
		\end{equation}
where $\vec{F}$ is the total force acting on the atom with a mass $m$, $G$ is gravitational constant, $M$ solar mass, $v_r$ radial velocity of the atom at time $t$, and $\mu$ is the ratio of solar resonant radiation pressure force to the solar gravity, which will be discussed further on in this chapter.

As can be seen from this description, the distribution of neutral interstellar hydrogen in the inner heliosphere is determined on one hand by the dynamical influence of the Sun by a balance between its gravity and radiation pressure forces and on the other hand by the ionization losses. Both will be extensively discussed in the following parts of this chapter. Here we only say that the ionization processes include charge exchange between the incoming neutral atoms and solar wind protons, ionization by impact of solar wind electrons, and ionization by the solar EUV radiation.

The hot model, however handy it is being almost analytical (in fact, numerical calculations are needed only when integrating the local distribution function to yield its moments, like density and mean velocity), is not perfect because none of its assumptions is fulfilled. First, the interstellar gas in the Local Cloud is not fully neutral and the interaction of its ionized component with the plasma of the solar wind creates a boundary region of the heliosphere (the heliospheric interface), which begins at the termination shock of the solar wind, where the solar wind becomes subsonic and eventually turns back at the heliopause, which can be approximated as a thin layer separating the solar wind plasma from the interstellar plasma. Beyond the heliopause there is the outer heliosheath, where the pristine neutral interstellar gas becomes altered due to charge exchange interactions with the protons from the piled and heated up interstellar plasma. History of the development of modeling of this region of the heliosphere can be found in \citet{baranov:06b} and modern views on this topic have been recently reviewed by \citet{fahr:04a, baranov:06a, izmodenov_baranov:06a} and by Izmodenov et al. in this book. 

Quantitative interpretation of the helioglow measurements required improvements in the classical hot model, which were realized quite early in the history of heliospheric research. \citet{lallement_etal:85a} allowed for latitudinal modulation of the charge exchange rate, approximating it with a one-parameter formula $1 - A \sin^2 \phi$, which enabled them to vary the equator-to-pole ratio of the ionization rates, but required keeping the width and range of the equatorial region of enhanced ionization fixed. 

A different extension of the hot model was proposed by \citet{rucinski_fahr:89, rucinski_fahr:91} who pointed out that the rate of electron-impact ionization does not change with $r^{-2}$, even though its effects on the distribution of neutral interstellar hydrogen in the heliosphere are noticeable only within a few AU from the Sun, where its density is already strongly reduced. Therefore this aspect of the heliospheric physics has been left out afterwards, until \citet{bzowski:08a, bzowski_etal:08a} reintroduced it in a refined, latitude-dependent manner (see also further down in this chapter). 

The next generation of heliospheric models abandoned the assumption of invariability of the solar radiation pressure force and of the ionization rate. The first, though very much simplified model was proposed by \citet{kyrola_etal:94}, followed by \citet{rucinski_bzowski:95a, bzowski_rucinski:95a, bzowski_rucinski:95b, bzowski_etal:97}, who studied variations in  density, bulk velocity, and temperature of neutral interstellar hydrogen near the Sun, as well as variations in the helioglow intensity assuming (lacking sufficient data at that time) an analytic model of evolution of the radiation pressure and ionization rate during the solar cycle. 

Another modification was introduced by \citet{scherer_etal:99}, who following \citet{osterbart_fahr:92, baranov_etal:91, baranov_malama:93} \citep[see also][]{malama_etal:06} realized that the charge exchange processes at the boundary layer of the heliosphere create a new, so-called secondary collisionless population of neutral H atoms, which can be taken into account by approximating the distribution function $f_{\infty}$ in Eq.~\ref{eqDiFu} by a sum of two two-dimensional Maxwellian functions with anisotropic temperatures. The values of temperatures, densities and bulk velocities in these Maxwellian functions, however, had to be adopted as some functions of the angular separation of the point $\vec{r}_{\infty}$ in Eq.~\ref{eqDiFu} from the upwind direction. Such an approach was later expanded and improved by \citet{katushkina_izmodenov:10}. In this approach, however, the parameters of the distribution function $f_{\infty}$ must be obtained from an external model, such as the Moscow Monte Carlo model of the heliosphere \citep[see, e.g.,][]{izmodenov_etal:09a}. Along with the two-populations non-Gaussian model, an approximation of the radiation pressure and ionization rate by spherically symmetric series of sines and cosines fitted to actual measurements of the ionization rate and radiation pressure in the ecliptic was added. These approximations were described by \citet{scherer_etal:99} and \citet{bzowski:01a, bzowski:01b}.

The most recent development in the modeling was addition by \citet{tarnopolski:07, tarnopolski_bzowski:08a} of the radiation pressure force being function of radial velocity of an atom relative to the Sun; this effect will be discussed further in this chapter.

The list of modifications to the classical hot model presented above is also a (perhaps not full) list of effects that need to be taken into account at the solar side to facilitate inter-calibration of the measurements of the helioglow with other UV observations in space. Apart from the heliospheric side, there is also physics of the heliospheric interface and the conditions in the Local Interstellar Cloud (\citep[see, e.g., ][ for review]{frisch_etal:09a, frisch_etal:11a} that must be taken into account, which, however, are beyond the scope of this chapter.

From the above description it is clear that accurate modeling of neutral interstellar hydrogen in the inner heliosphere requires accurate knowledge of the factors contributing to the ionization and radiation pressure. For the radiation pressure force one needs to know the evolution of the shape of the solar Lyman-alpha spectral line in time. The ionization processes include charge exchange with solar wind protons (and perhaps also alpha particles), ionization by electron impact, and ionization by solar EUV radiation. Recombination is not important here because of two reasons: (1) its rate is small as compared with the ionization rate \citep{wachowicz:06}, and (2) the recombination products maintain their velocity, which is equal to the solar wind velocity, so they do not contribute to the populations of heliospheric atoms that are capable of scattering the solar EUV radiation responsible for the helioglow.

\section{Radiation pressure}
\subsection{Evolution of the total solar flux in Lyman-alpha in the ecliptic}

\begin{figure}
		\centering
		\includegraphics[scale=0.7]{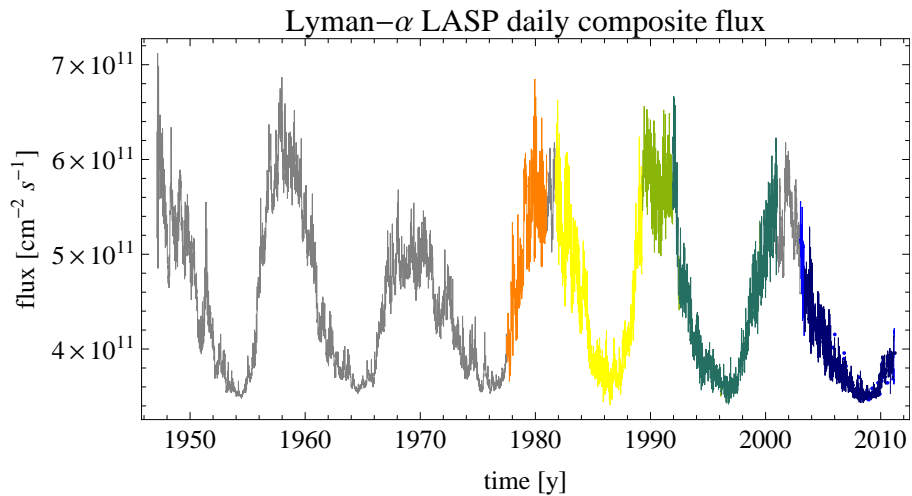}		
  	\caption{Wavelength- and disk-integrated solar Lyman-alpha flux from the Laboratory for Atmospheric and Space Physics (LASP), referred to as the total Lyman-alpha flux $I_{\mathrm{tot}}$. The daily time series is a composite of actual measurements from various experiments re-scaled to the common calibration of UARS/SOLSTICE, with the gaps filled by the proxies. Color codes: gray: F$_{10.7}$, orange: AE-E, yellow: SME, green: MgII$_{\mathrm{c/w}}$ proxy, aquamarine: UARS/SOLSTICE version 18, light blue: TIMED/SEE, dark blue: SORCE/SOLSTICE. Based on \citet{woods_etal:00}.}
 		\label{figCompoLyaDaily}
		\end{figure}

The radiation pressure force acting on neutral interstellar H atoms in the heliosphere is proportional to the total flux in the solar Lyman-alpha spectral line, which is defined as the spectral flux integrated over a 1 nm interval from 120 to 121~nm. It has been measured since the middle of 1970s \citep[see][for the history of measurements]{woods_etal:00}. Despite all efforts, while precision of the measurements was good, the problem of absolute calibration, prone to changes with time, has affected the accuracy from the very beginning until present and a measure of progress in this field is the reduction of discrepancies from a factor of 4 in the 1970s to $\sim 15$\% nowadays. The total composite Lyman-alpha time series, available from the Laboratory for Atmospheric and Space Physics (LASP) at the University of Colorado in Boulder, Co, is scaled to the absolute calibration of UARS/SOLSTICE \citep{woods_etal:96a, woods_etal:00}. The cadence of measurements is presently 1 day and the inevitable gaps are usually filled using a hierarchy of proxies, as illustrated in Fig.~\ref{figCompoLyaDaily}. 

The most widely used is the proxy based on the solar radio flux at the 10.7 cm wavelength, the so-called $F_{10.7}$ flux \citep{covington:47, tapping:87}. Another frequently used is the Mg~II core-to-wing (MgII$_{\mathrm{c/w}}$) index \citep{heath_schlesinger:86, viereck_puga:99}. 

\begin{figure}[t]
		\centering
		\includegraphics[scale=0.5]{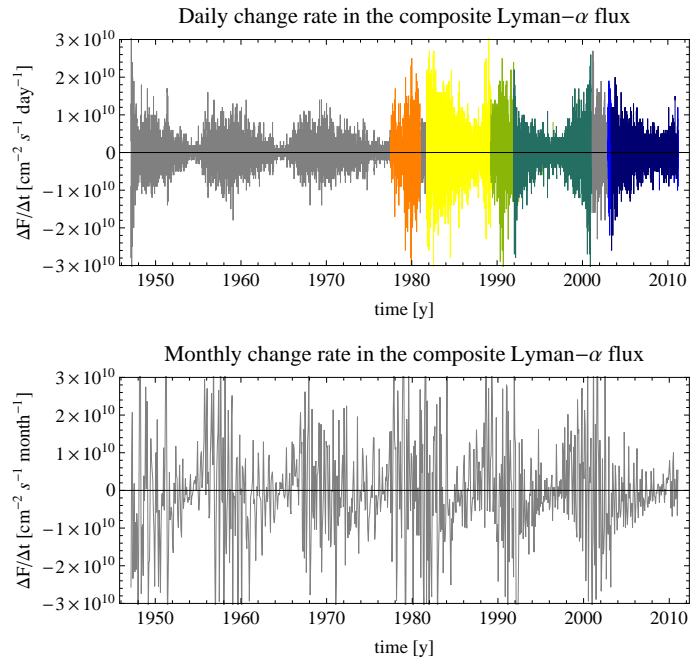}		
  	\caption{Rates of change of the total Lyman-alpha flux $I_{\mathrm{tot}}$. Shown are the rates of change per unit time $\frac{\Delta I_{\mathrm{tot}}}{\Delta t}=\frac{I_{\mathrm{tot}}\left(t_{i+1}\right)-I_{\mathrm{tot}}\left(t_i\right)}{t_{t+1}-t_i}$ of the daily (upper panel) and monthly (lower panel) composite Lyman-alpha flux shown in Fig. \ref{figCompoLyaDaily}, with identical color coding.}
 		\label{figLyaChangeRate}
		\end{figure}

The use of proxies raises the question of credibility of the results \citep{floyd_etal:02a, floyd_etal:05a}. The solar EUV radiation is variable on many time scales, from hours to more than solar cycle length. Proxies generally follow the variability of the quantity being approximated, but not precisely and not at all time scales. In particular, even though the correlation coefficients for daily values may exceed 0.9, at the shortest time scales the agreement can be problematic because -- depending on the location of the source region of a given proxy quantity -- they can show the effect of limb brightening or limb darkening different from the quantity they are used to approximate \citep{floyd_etal:05a}; an illustration can be found in the upper panel of Fig.~\ref{figLyaChangeRate}. In consequence, the quality of the approximation at these short scales is reduced.

On the other hand, the rates averaged by Carrington rotation (alternatively referred to as ``monthly'' in this text) seem to be reproduced quite well, as judged from the behavior of the monthly change rates shown in the lower panel of this figure and as suggested by \citet{dudokdewit_etal:09a}. For modeling of the distribution of neutral interstellar gas in the heliosphere the Carrington period is anyway the finest time scale presently in use, so the proxy imperfections are not a big problem in this respect. 

\subsection{Variation of the Lyman-alpha flux with heliolatitude}

The solar disk-averaged Lyman-alpha flux is made of at least three components \citep{amblard_etal:08a}: from the quiet Sun, from the coolest part of the chromosphere, and from the hot lower corona. The inhomogeneous distribution in heliolatitude of active regions was pointed out by \citet{cook_etal:80a}, who constructed a two-component latitude-dependent model of disk-averaged solar UV irradiance. \citet{cook_etal:81a} considered the solar Lyman-alpha emission and suggested that the ratio of the disk-integrated solar flux at the pole to the flux at the equator should be about $a_{Lya} = 0.8$ during solar minimum, when the active regions are distributed in latitudinal bands. These suggestions were supported by direct solar minimum observations of the solar corona by \citet{auchere:05}. Such a ratio was suggested as valid also for solar maximum by \citet{pryor_etal:92}, based on indirect evidence from observations of the Lyman-alpha helioglow. Thus it seems that the latitudinal anisotropy of the Lyman-alpha flux does not change substantially during the solar cycle, although this conclusion certainly needs further verification. \citet{bzowski:08a} suggested that the latitude dependence of the disk-integrated solar Lyman-alpha flux may be approximated by the formula
		\begin{equation}
		\label{eqLyaFlat}
		I_{\mathrm{tot}}\left(\phi \right)=I_{\mathrm{tot}}\left( 0 \right) \sqrt{a_{\mathrm{Lya}}\sin^2\left( \phi \right) + \cos^2 \left( \phi \right)}	
		\end{equation}
where $\phi$ is heliolatitude, $a_{\mathrm{Lya}}$ the ``flattening'' factor and $I_{\mathrm{tot}}\left( 0 \right)$ is a equatorial Lyman-alpha flux.

Heliolatitude variation of the disk-integrated Lyman-alpha line profile is, to our knowledge, unexplored. On one hand, one should expect some form of variation because, as shown by \citet{tian_etal:09e, tian_etal:09c, tian_etal:09d}, the line profile depends on the features on the solar disk that are being observed and the latitude distribution of these features is inhomogeneous and varies during solar cycle. On the other hand, the apparent lack of strong variability of the disk-integrated flux might suggest that the spectral variation with heliolatitude is mild. Recent investigation of the variation of solar spectrum with heliolatitude by \citet{kiselman_etal:11a} seems to bring a negative result (i.e., no variation). Should the disk-integrated spectral flux indeed vary with heliolatitude, this would potentially have consequences both for the photoionization rate and the radiation pressure force.

\subsection{Mechanism of radiation pressure}

Mechanism of resonant interaction of a H atom with solar radiation, which leads to the force of resonant radiation pressure acting on neutral H atoms in the heliosphere, was extensively discussed by \citet{brasken_kyrola:98}. In brief, the probability $f_{\mathrm{abs}}\left(\lambda\right)$ that a hydrogen atom in the ground state, whose base wavelength is $\lambda_0$, absorbs an incoming photon at wavelength $\lambda$ is equal to: 
		\begin{equation}
		\label{eqAtomProfile}
		f_{\mathrm{abs}}\left(\lambda\right)=\frac{\Gamma_{R}}{2\pi c\left(\frac{1}{\lambda}-\frac{1}{\lambda_0}\right)^2+\frac{\Gamma_R^2}{4}}
		\end{equation}
where $\Gamma_R$ is the energetic width of the second orbital of the atom, corresponding to about $\pm 25$~m~s$^{-1}$ around the base wavelength of the Lyman-alpha transition. The atom whose radial velocity relative to the Sun is 0 will absorb photons from the very center of the solar line, but if its radial velocity $v_r$ is non-zero, then due to Doppler effect it will be tuned to a different portion of the solar line profile, namely to the wavelength $\lambda = \lambda_0\left(1 - v_r/c\right)$. Within about $10^{-15}$~s after absorption, the atom will re-emit the Lyman-alpha photon at an angle $\omega$ to the impact direction with the scatter-angle probability $p\left(\omega\right)$ described by:		
		\begin{equation}
		\label{eqScattFun}
		p\left(\omega \right)=\frac{\cos \left(\omega\right)}{4}+\frac{11}{12}
		\end{equation}
Hence a resonant interaction of the atom with a suitable photon results in a change of atomic momentum at the moment of absorption by $\Delta p = h v = c\, h/\lambda$ in the antisolar direction, followed after a time of $10^{-15}$~s by another momentum change in the direction described by Eq.~\ref{eqScattFun}. However, typical interactions frequencies at 1 AU, which are proportional to the solar spectral flux, are on the order of 1/500~Hz \citep{quemerais:06a}, so statistically at time scales which are shorter than the time scales of a change in atomic velocity relative to the Sun the only net effect of the interaction of the atom with solar radiation is the antisolar momentum change. Since the interplanetary medium is almost optically thin within a few AU from the Sun \citep{quemerais:06a}, the solar spectral flux scales with inverse square of solar distance and consequently also the solar radiation pressure force scales with solar distance as $1/r^2$, which gives effect of compensation of solar gravity by a factor $\mu$. Hence the effective solar force acting on the atom is conveniently expressed as the factor $\mu\left(I_{\mathrm{tot}}\left( t\right), v_r \right)$ of solar gravity. It is proportional to the spectral flux $F_{\lambda}$ corresponding to the Doppler-shifted wavelength $\lambda = \lambda_0\left(1+v_r/c\right)$, resulting from the instantaneous radial speed $v_r$ of the atom relative to the Sun. Since the spectral flux varies with time, effectively the $\mu$ factor is a function of radial velocity and time, as expressed in Eq.~\ref{eqEqMo}.

\subsection{Profile of solar Lyman-alpha line and the solar radiation pressure factor $\mu$}

Measurements of the solar Lyman-alpha line profile, although scarce, date back to the 1970s \citep{vidal-madjar:75, artzner_etal:78a, bonnet_etal:78a, lemaire_etal:78a, woods_etal:95a}. They were performed from within the Earth exosphere and hence suffered from the absorption by geocoronal hydrogen in the spectral region most relevant for the helioglow. Only after launch of SOHO, which orbits at the L1 Lagrange point, an unobscured view on the full spectral range of the disk-integrated solar line was obtained by \citet{warren_mariska:98a, lemaire_etal:98, lemaire_etal:02, lemaire_etal:05}, who used a novel, ingenuous observations technique proposed by J.-L. Bertaux. Owing to the latter measurements it was possible to study the evolution of the profile at all solar activity phases from minimum to maximum.

The solar Lyman-alpha line features a self-reversed shape that used to be approximated by two Gaussian functions \citep{fahr:79, chabrillat_kockarts:97, scherer_etal:00a}. Recently, \citet{tarnopolski:07, tarnopolski_bzowski:09} showed that the measurements by \citet{lemaire_etal:02} can all be fit by a three-Gaussian model parameterized by the disk-integrated flux:
		\begin{equation}
		\label{eqSolProfileFit}
		\mu \left( v_r, I_{\mathrm{tot}}\left( t\right) \right) = A \left[ 1+B I_{\mathrm{tot}}\left( t \right) \right] \mathrm{exp}\left( -Cv_r^2\right)\left[ 1+D\mathrm{exp}\left( Fv_r-Gv_r^2\right) +H\mathrm{exp}\left( -Pv_r-Qv_r^2\right) \right] 
		\end{equation}
with the following parameters:
		\begin{displaymath}
		\begin{array}{lll}
		A = 2.4543 \times 10^{-9}, & B = 4.5694 \times 10^{-4}, & C = 3.8312 \times 10^{-5}\\
		D = 0.73879, & F = 4.0396 \times 10^{-2}, & G = 3.5135 \times 10^{-4}\\
		H = 0.47817, & P = 4.6841 \times 10^{-2}, & Q = 3.3373 \times 10^{-4}\\
		\end{array}
		\end{displaymath}
The accuracy of the fit is similar to the accuracy of the measurements, assessed to be $\sim 10$\%.

\begin{figure}[t]
		\centering
		\includegraphics[scale=0.7]{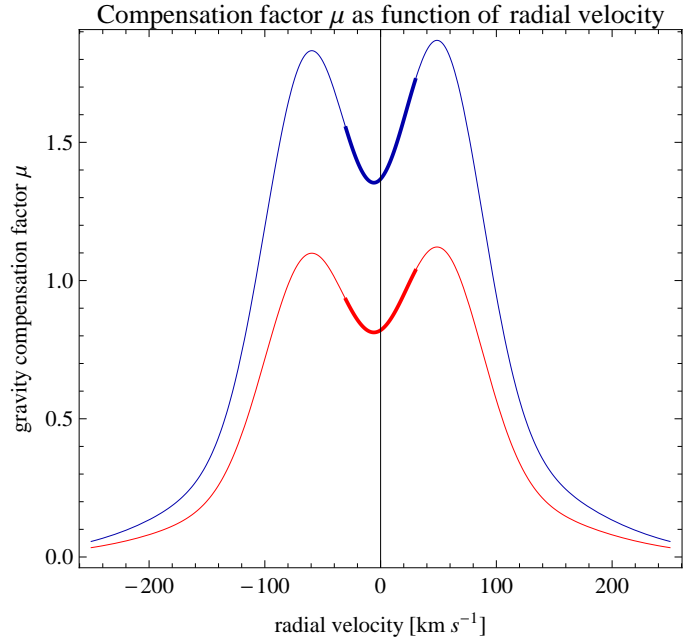}		
  	\caption{Factor $\mu$ of solar radiation pressure force to solar gravity based on the model specified in Eq.~\ref{eqSolProfileFit} shown as a function of radial velocity of a H atom relative to the Sun for the total solar flux values corresponding to the minimum (red) and maximum (blue) of solar activity. Thick lines indicate the spectral region $\pm 30$~km s$^{-1}$ around 0 Doppler shift, relevant for the neutral interstellar hydrogen gas in the heliosphere.}
 		\label{figSolProf}
		\end{figure}

With this formula, one can calculate the $\mu$ factor for an arbitrary radial velocity $v_r$ providing that the total solar Lyman-alpha flux $I_{\mathrm{tot}}$ is known. The dependence of the $\mu$ factor on radial velocity for the total flux values representative for solar minimum and maximum conditions are shown in Fig.~\ref{figSolProf}, adapted from \citet{tarnopolski_bzowski:09}. 

The spectral region of the solar Lyman-alpha line most relevant for modeling of the heliospheric Lyman-alpha glow in the wavelength band straddling the central wavelength by approximately $\pm 30$~km s$^{-1}$. The spectral flux at the line center is closely correlated with the line-integrated flux $I_{\mathrm{tot}}$ \cite{vidal-madjar_phissamay:80}. This is beneficial to the modeling of the helioglow because it permits to easily calculate the $\mu$ factor based on measurements of the solar line-integrated Lyman-alpha flux.

\subsection{Approximate values of the $\mu$ factor}

\begin{figure}[t]
		\centering
		\includegraphics[scale=0.5]{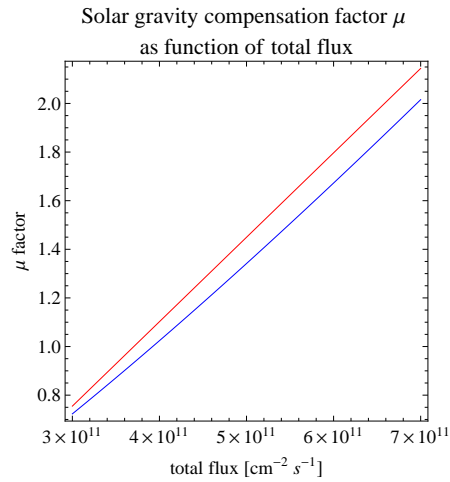}		
  	\caption{Solar radiation pressure factor $\mu$ shown as a function of the total flux in the Lyman-alpha line calculated  by \citet{emerich_etal:05} from Eq.~\ref{eqEmerichFormula}, which connects the total flux with the spectral flux precisely at line center (blue), and by \citet{bzowski_etal:08a} from Eq.~\ref{eqBzFormula} (red), where correlation between the total solar flux and spectral flux is adopted from averaged spectral flux over $\pm 30$~km~s$^{-1}$ around the line center.}
 		\label{figEmerichVsBz}
		\end{figure}

Since modeling of neutral interstellar hydrogen with the use of a model that takes the full solar line profile is computationally demanding, the approach where the $\mu$ factor is treated as independent of radial velocity are still widely used. To that purpose a formula to translate the line-integrated flux into the $\mu $ factor is needed. In the past, this issue was addressed just by taking
		\begin{equation}
		\label{eqIToMuSimple}
		\mu \left( I_{\mathrm{tot}} \right)=3.0303 \times 10^{10} a\, I_{\mathrm{tot}}
		\end{equation}
where $I_{\mathrm{tot}}$ is the disk- and line-integrated solar Lyman-alpha flux and $a$ is a constant usually adopted between 0.85 and 1. With increasing accuracy of measurements, more sophisticated formulae became available. \citet{emerich_etal:05} fitted the following relation between the spectral flux at line centre to the total flux:	
		\begin{equation}
		\label{eqEmerichFormula}	\frac{F_\lambda}{10^{12}\mathrm{cm}^{-2}\mathrm{s}^{-1}\mathrm{nm}^{-1}}=0.64\left(\frac{I_{\mathrm{tot}}}{10^{11}\mathrm{s}^{-1}\mathrm{cm}^{-2}}\right)^{1.21}\pm0.08
		\end{equation}	
while \citet{bzowski_etal:08a} found a linear relation between the spectral flux averaged by the spectral range $\pm 30$~km s$^{-1}$ about the line center and the line- and disk-integrated flux:
		\begin{equation}
		\label{eqBzFormula}		
		\mu \left( I_{\mathrm{tot}} \right) = 3.473 10^{-12} I_{\mathrm{tot}}-0.287
		\end{equation}
A comparison of the two predicted $\mu$ values as a function of the solar total Lyman-alpha flux is shown in Fig.~\ref{figEmerichVsBz} and suggests that if one decides not to use the model of radiation pressure force dependent on the radial velocity of the atom, then calculation of a good effective $\mu$ factor is not a straightforward task. Use of the model with $\mu$ being a function of $v_r$ is obligatory when one wants to model interstellar deuterium \citep{tarnopolski_bzowski:09}.

\begin{figure}[t]
		\centering
		\includegraphics[scale=0.7]{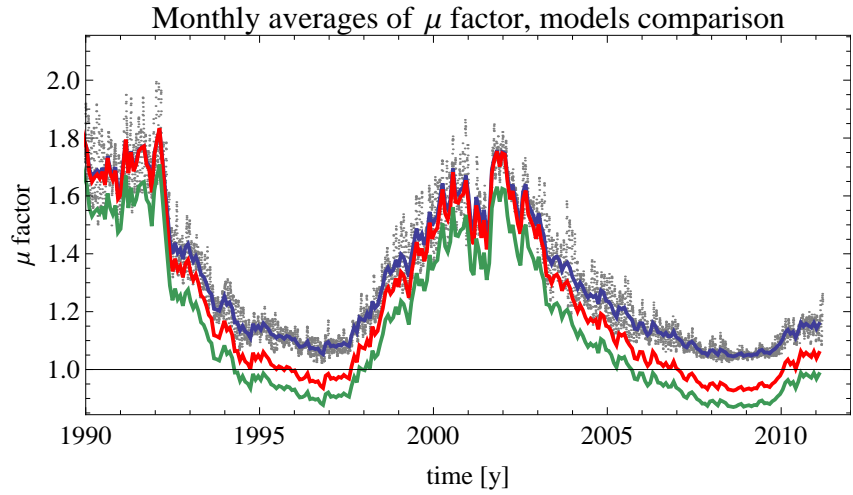}		
  	\caption{Comparison of the solar radiation pressure factors $\mu$ that approximate the compensation of solar gravity by the resonant radiation pressure force acting on neutral H atoms. Gray dots are daily values from the composite Lyman-alpha flux series presented in Fig.~\ref{figCompoLyaDaily}, blue line is the monthly (Carrington rotation averaged) values of the flux scaled using Eq.~\ref{eqIToMuSimple}, green line are monthly values obtained from Eq.~\ref{eqEmerichFormula} \citep{emerich_etal:05} and red line represents monthly values obtained from Eq.~\ref{eqBzFormula} \citep{bzowski_etal:08a}.}
 		\label{figMuValues}
		\end{figure}

The $\mu$ values returned by the formulae by \citet{emerich_etal:05} and \citet{tarnopolski_bzowski:09} calculated from the monthly values of the LASP composite Lyman-alpha flux shown in Fig.~\ref{figCompoLyaDaily} are presented in Fig.~\ref{figMuValues}. Differences between the $\mu$ values obtained from equations (\ref{eqEmerichFormula}) and (\ref{eqBzFormula}) are on the order of 10\%, i.e. on the order of the uncertainty of the total flux. 

\section{Ionization processes}

The three main ionization processes of neutral interstellar hydrogen atoms in the heliosphere are the following:\\
charge exchange with solar wind charged particles (mostly protons):
		\begin{displaymath}
		\mathrm{H} + \mathrm{p} \rightarrow \mathrm{p_{PUI}} + \mathrm{H_{ENA}}
		\end{displaymath}
photoionization by solar EUV radiation:
 		\begin{displaymath}
		\mathrm{H} + \nu \rightarrow \mathrm{p_{PUI}} + \mathrm{e}
		\end{displaymath}
and ionization by impact of solar wind electrons: 
 		\begin{displaymath}
		\mathrm{H} + \mathrm{e} \rightarrow \mathrm{p_{PUI}} + 2\mathrm{e}
		\end{displaymath}
The reaction product is always a pickup proton p$_{\mathrm{PUI}}$ and either an energetic neutral atom H$_{\mathrm{ENA}}$ or one or two electrons. There are no recombination processes that would inject neutralized protons to the region of phase space occupied by the population of interstellar neutral atoms.

As a result of the charge exchange reaction, a solar wind proton captures the electron from the neutral interstellar H atom and becomes an energetic neutral atom (ENA), but maintains its momentum and thus does not enter the interstellar population. In this respect, even though one neutral H atom is replaced by another, such a reaction is still a loss process despite the total number of H atoms in the system  does not change -- but it does change in the population under study. 

The newly-created protons regardless of the reaction they originate from, are picked up by the solar wind flow \citep{fahr:73, vasyliunas_siscoe:76}, creating a distinct population that can be measured \citep{mobius_etal:85a, gloeckler_etal:93a} and analyzed \citep[e.g.][]{gloeckler_geiss:01a, gloeckler_etal:04b, bzowski_etal:08a}, but this problem is outside the scope of this text. The solar wind is not depleted in protons due to the charge exchange reaction since eventually a solar wind core proton is replaced with a pickup proton -- but the concentration of protons per unit volume is not changed even though the distribution function of solar wind protons is modified.

The two remaining ionization reactions cause genuine losses, the H atom enters the reaction and is not replaced with another at a different velocity: eventually, a proton--electron pair is created. 
 
\subsection{Charge exchange}
\subsubsection{General formula}

The process of resonant charge exchange between H atoms and protons is of crucial importance for the physics of the heliosphere. It contributes to the pressure balance between the solar wind and interstellar gas and makes possible energy and momentum transport between solar wind and interstellar gas across the heliopause. Charge exchange losses of neutral interstellar gas in the supersonic solar wind, which will be discussed in this section, are only a small piece of a large picture.

The rate of charge exchange between neutral H atoms and solar wind protons can be regarded as probability of a charge exchange act per unit time in a given location in space. For a H atom traveling with velocity $\vec{v}_H$ and a local proton distribution function $f_p(\vec{v}_p)$, where $\vec{v}_p$ is the velocity vector of an individual proton, the rate of charge exchange can be calculated from the formula:
		\begin{equation}
		\label{eqBetaCXGeneral}
		\beta_{\mathrm{CX}}=\int\sigma_{\mathrm{CX}}\left(\left|\vec{v}_H-\vec{v}_p\right| \right) \left|\vec{v}_H-\vec{v}_p\right|f_p\left(\vec{v}_p
\right)\mathrm{d}\vec{v}_p,
		\end{equation}
where $\sigma_{\mathrm{CX}}\left(\left|\vec{v}_{H}-\vec{v}_p\right|\right)$ is the reaction cross section and $\vec{v}_H-\vec{v}_p \equiv \vec{v}_{\mathrm{rel}}$ is the relative velocity between the H atom and an individual proton. The integration covers the entire proton velocity space. This formula can be put into an equivalent form:
		\begin{equation}
		\label{eqBetaCXVRel}		
		\beta_{\mathrm{CX}}=\int\sigma_{\mathrm{CX}}\left(\left|\vec{v}_{\mathrm{rel}}\right| \right) \left|\vec{v}_{\mathrm{rel}}\right|f_p\left(\vec{v}_H-\vec{v}_{\mathrm{rel}}
\right)\mathrm{d}\vec{v}_{\mathrm{rel}}
		\end{equation}
Depending on the underlying plasma regime and on the velocity of the H atom, various simplifications can be made. When the kinetic spread of the plasma $u_{\mathrm{T},p}$ is small compared with the plasma flow velocity $\vec{v}_{\mathrm{SW}}$: $u_{\mathrm{T},p} = \sqrt{2 k T_p/m_p} \ll \vec{v}_{\mathrm SW} = \int \vec{v_p} \,f\left(\vec{v}_p\right) \mathrm{d} \vec{v}_p/n_p$, we can approximate the proton distribution function by delta-function centered at the solar wind speed and the formula for charge exchange rate simplifies to:
		\begin{equation}
		\label{eqBetaCXSolWind}		
		\beta_{\mathrm{CX}}=\sigma_{\mathrm{CX}}\left(v_{\mathrm{rel}}\right)n_p v_{\mathrm{rel}},
		\end{equation}
where $n_p$ is the local proton density and $v_{\mathrm{rel}}$ becomes $v_{\mathrm{rel}} \equiv \left| \vec{v}_H  - \vec{v}_{\mathrm SW}\right|$. This is the case for the scenario of H$_\mathrm{ENA}$ in the supersonic solar wind for the atoms that travel at $v_{\mathrm{H}}\sim 50$~km~s~$^{-1}$ or faster. 

For $v_H \ll v_{\mathrm SW}$, as in the case of the thermal interstellar H populations in the supersonic solar wind, we have $v_{\mathrm{rel}}=v_p$. Then the rate of charge exchange between H atoms and solar wind protons will be given by
		\begin{equation}
		\label{eqBetaCXSWIsGas}		
		\beta_{\mathrm{CX}}=\sigma_{\mathrm{CX}}\left(v_{\mathrm{SW}}\right)n_p v_{\mathrm{SW}}
		\end{equation}
This is not the case, however, for H$_\mathrm{ENA}$ which travel at $v_H\geq$50~km~s$^{-1}$, which is not negligible compared with the typical solar wind speed of $v_{\mathrm{SW}}\sim$440~km~s$^{-1}$, in the subsonic solar wind (inner heliosheath), and within the Baranov wall (outer heliosheath) thermal speeds of local plasmas are comparable to specific velocities of H atoms and the full form of the charge exchange formula must be applied. 

\subsubsection{Charge exchange cross section}		
		
The collision speed range most relevant for the heliospheric physics is from $\sim 1$~km~s$^{-1}$ to $\sim 1000$~km~s$^{-1}$, which is equivalent to 0.005~eV and 5.2~keV. Relative speed between interstellar neutral H atoms and protons in the supersonic solar wind range from $\sim 300$~km~s$^{-1}$ to $\sim 1000$~km~s$^{-1}$. 

A detailed discussion of the charge exchange process and of the cross sections for this reaction can be found in \citet{fahr_etal:07a} and will not be repeated here. For the purpose of this work it is important to point out that there were basically 4 cross section formulae used in the heliospheric physics: from \citet{fite_etal:62, maher_tinsley:77, barnett_etal:90, lindsay_stebbings:05a}. 

\citet{fite_etal:62} and \citet{maher_tinsley:77} proposed to approximate the charge exchange cross section as a function of relative velocity $v_{\mathrm{rel}}$ between the colliding partners by a formula
   \begin{equation}
   \label{eqMTsigCX}
   \sigma_{\mathrm{CX}}\left( v_{\mathrm{rel}}\right) = \left(a + b \ln v_{\mathrm{rel}}\right)^2
   \end{equation}
The validity of the formula by \citet{fite_etal:62} was claimed to be between 20 and 2000~eV and of \citet{maher_tinsley:77} between 0.005 and 1 keV. 

\begin{figure}[t]
		\centering
		\includegraphics[scale=0.7]{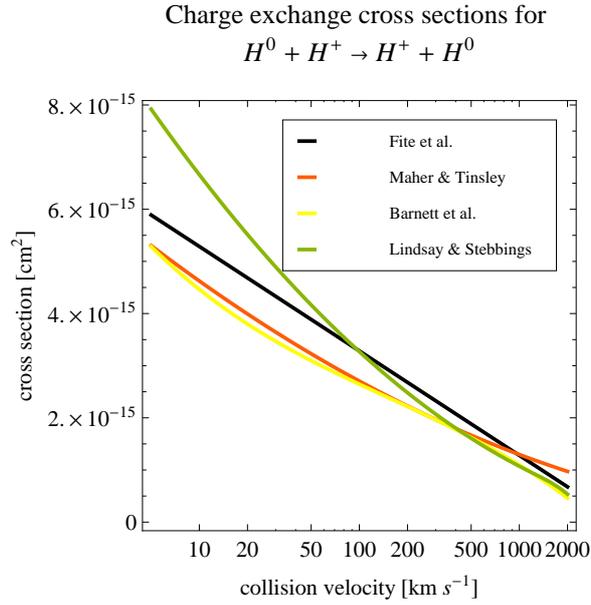}		
  	\caption{Cross sections for charge exchange reaction between protons and H atoms in an energy range most important in the heliospheric physics. The recommended relation from \citet{lindsay_stebbings:05a} is compared with the formulae used in the past by \citet{fite_etal:62, maher_tinsley:77} and \citet{barnett_etal:90}.}
 		\label{figSigCX}
		\end{figure}

\citet{barnett_etal:90} proposed and involved fit in a form of Chebyshev polynomials valid in a broad energy range. \citet{bzowski:01b} approximated the data used by \citet{barnett_etal:90} restricted to $v_{\mathrm{rel}}< 800$~km~s~$^{-1}$ by a formula
   \begin{equation}
   \label{eqMBsigCX}
    \sigma_{\mathrm{CX}}\left(v_{\mathrm{rel}} \right) = a_0+\sum\limits_{i=1}^{3}a_i\,\left(\ln v_{\mathrm{rel}}\right)^i
   \end{equation}
The most recent and authoritative compilation of measurements and calculations was presented by \citet{lindsay_stebbings:05a}, who suggested the following formula for the cross section expressed in cm$^2$, valid for collision energies $E$ between 0.005 and 600~keV:
   \begin{equation}
   \label{eqLSsigCX}
   \sigma_{\mathrm{CX}}\left(E\right) = 10^{-16}\left(1 - \exp\left[-67.3/E \right] \right)^{4.5}\,\left(4.15-0.531\ln E \right)^2
   \end{equation}
A comparison of the cross sections from the 4 formulae is presented in Fig. \ref{figSigCX}. 

It is important to note that while all four formulae return similar results for the collision speeds relevant for the supersonic solar wind, the recommended formula by \citet{lindsay_stebbings:05a} returns a significantly larger cross sections for low energies relevant for the outer heliosheath. Thus adoption of the older formulae may result in a significant underestimation of the coupling strength between the neutral interstellar gas and the plasma in the outer heliosheath, where the secondary population of interstellar H atoms is formed, with consequences for the results of heliospheric modeling, as elaborated on in this book in the chapter by Izmodenov et al. 

\subsection{Photoionization}
\subsubsection{From solar spectrum to photoionization rate}

\begin{figure}[t]
		\centering
		\includegraphics[scale=0.7]{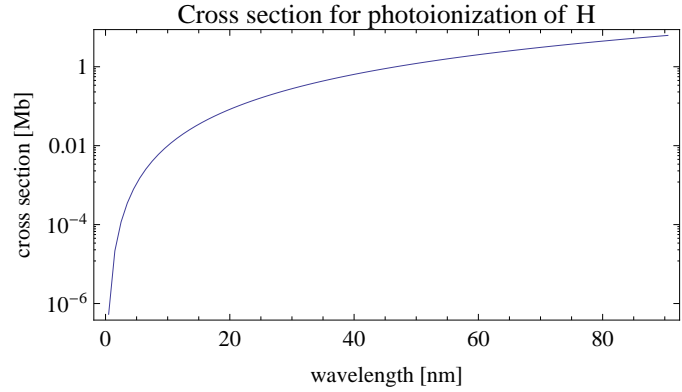}		
  	\caption{Cross section for photoionization of hydrogen based on Eq.~\ref{eqPhCrossSec} adapted from \citet{verner_etal:96}.}
 		\label{figBetaPhCrossSec}
		\end{figure}

\begin{figure}[t]
		\centering
		\includegraphics[scale=0.7]{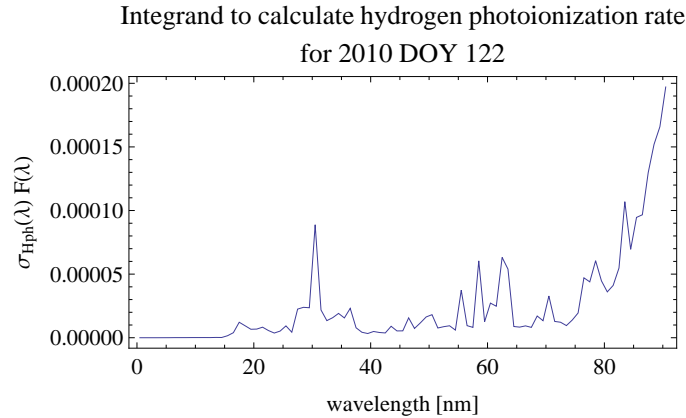}		
  	\caption{Integrand function from Eq.~\ref{eqPhotoIon} for the flux obtained from TIMED/SEE for 2010 DOY 122 and photoionization cross section defined in Eq.~\ref{eqPhCrossSec} and shown in Fig.~\ref{figBetaPhCrossSec}.}
 		\label{figPhIntegrand}
		\end{figure}

Photoionization is a secondary ionization factor of neutral interstellar H, but its significance has recently increased because of the decrease in the average solar wind flux observed since the last solar maximum, resulting in a decrease in the intensity of the dominant charge exchange ionization rate. The rate of photoionization $ \beta _{\mathrm{ph}}\left( t \right)$ at a time $t$ can be calculated from the formula
		\begin{equation}
		\label{eqPhotoIon}		
		\beta_{\mathrm{ph}}\left( t \right)=\int\limits_{0}^{\lambda_{\mathrm{ion}}}\sigma_{\mathrm{ph}}\left(\lambda \right) F_{\lambda} \left( \lambda, t \right) \mathrm{d}\lambda
		\end{equation}		
where $\sigma_{\mathrm{ph}}\left(\lambda\right)$ is the cross section for photoionization for wavelength $\lambda$ and $F_{\lambda}\left(\lambda, t\right)$ is the solar spectral flux at the time $t$ and wavelength $\lambda$; $\lambda_\mathrm{ion}$ is the wavelength for the threshold ionization energy. In the case of hydrogen, the spectral range of the radiation capable of knocking out electrons from H atoms is entirely in the EUV range. The cross section for photoionization of H can be expressed by the following formula \citep{verner_etal:96}:
		\begin{equation}
		\label{eqPhCrossSec}		
		\sigma_{\mathrm{Hph}}\left( \lambda \right)=6.82297 \times 10^{-10}\left( \frac{9.36664}{\sqrt{\lambda}}+1\right)^{-2.963}\left( \lambda - 2884.69 \right)^2 \lambda^{2.0185},
		\end{equation}	
where the cross section is expressed in megabarns (Mb), wavelength in nm and $\lambda \le 91.18$~nm. The cross section is shown in Fig.~\ref{figBetaPhCrossSec}, and the importance of various portions of the spectrum for the photoionization is illustrated in Fig.~\ref{figPhIntegrand}, which shows the integrand function from Eq.~\ref{eqPhotoIon} for DOY 122 in 2001 \citep[in preparation]{bochsler_etal:11b}. 		

It is clear from Fig.~\ref{figPhIntegrand} that the most important for the  photoionization of hydrogen is the portion of the spectrum immediately next to the photoionization threshold. The photoionization rate varies because the solar EUV spectral flux varies. Direct measurements of the solar EUV spectrum in the entire relevant energy range have been available on a 2-hourly basis since 2002, when the TIMED/SEE experiment \citep{woods_etal:05a} was launched. Before, the spectral coverage was intermittent and one had to resort to indirect methods. 

Basically, these methods can be grouped into two classes: (1) direct integration using Eq.~\ref{eqPhotoIon} with the spectrum $F_{\lambda}\left(\lambda, t \right)$ calculated from proxies, and (2) using correlation formulae between selected proxies and photoionization rates obtained for the times when the spectrum measurements are available. Since measurements covering only a portion of the spectrum with a relatively low spectral resolution are also available (SOHO/CELIAS/SEM) \citep{hovestadt_etal:95a, judge_etal:98}, a variant of method (2) would require finding a correlation between these partial direct measurements and the photoionization rate. None of the method gives perfect results; furthermore, a reliable application of method (2) became possible only after a sufficiently rich database of precise measurements of the solar spectrum became available.

Inspection of Fig. \ref{figPhIntegrand} shows that the band from 50 to 30~nm, which is the most relevant for helium and has been observed by CELIAS/SEM, is of secondary significance for hydrogen. Since, however, variations in various portions of the solar EUV spectrum are correlated to some extent, the evolution of the EUV flux in the CELIAS/SEM bands is a reasonable indicator of the evolution of the photoionization rates of H as well. 

Employment of method (1) requires a proxy model of the solar spectrum in the relevant photon energy range. A number of such models was developed in the past, including SERF1 (also known as HFG or EUV81) \citep{hinteregger_etal:81a}, EUVAC \citep{richards_etal:94a}, SOLAR2000 \citep{tobiska_etal:00c}, and NRLEUV \citep{warren_etal:98a, warren_etal:98b, lean_etal:03a, warren:06a}. The models are based on empirical correlations found between various portions of the solar spectrum and selected time series of available measurements, including typically the F$_{10.7}$ radio flux, the MgII$_{\mathrm{c/w}}$ index, the solar Ca II K index, and even the sunspot time series. Methodology and problems of creation of such models have been recently reviewed by \citet{floyd_etal:02a, floyd_etal:03a, floyd_etal:05a, lean_etal:11a} and the question of finding suitable proxies by \citet{kretzschmar_etal:06a, dudokdewit_etal:05a, dudokdewit_etal:08a, dudokdewit_etal:09a}. Typically, linear correlations have been sought for, what does not seem to be an optimum solutions at all cases \citep{bochsler_etal:11b}.

\subsubsection{Variation in the photoionization rate of H in the ecliptic plane}
The process of photoionization of heliospheric hydrogen was extensively discussed by \citet{ogawa_etal:95} and we will not repeat such a discussion in this Chapter. We will briefly present current views on the rate of photoionization of heliospheric hydrogen based on actual measurements and models. 

\begin{figure}[t]
		\centering
		\includegraphics[scale=0.5]{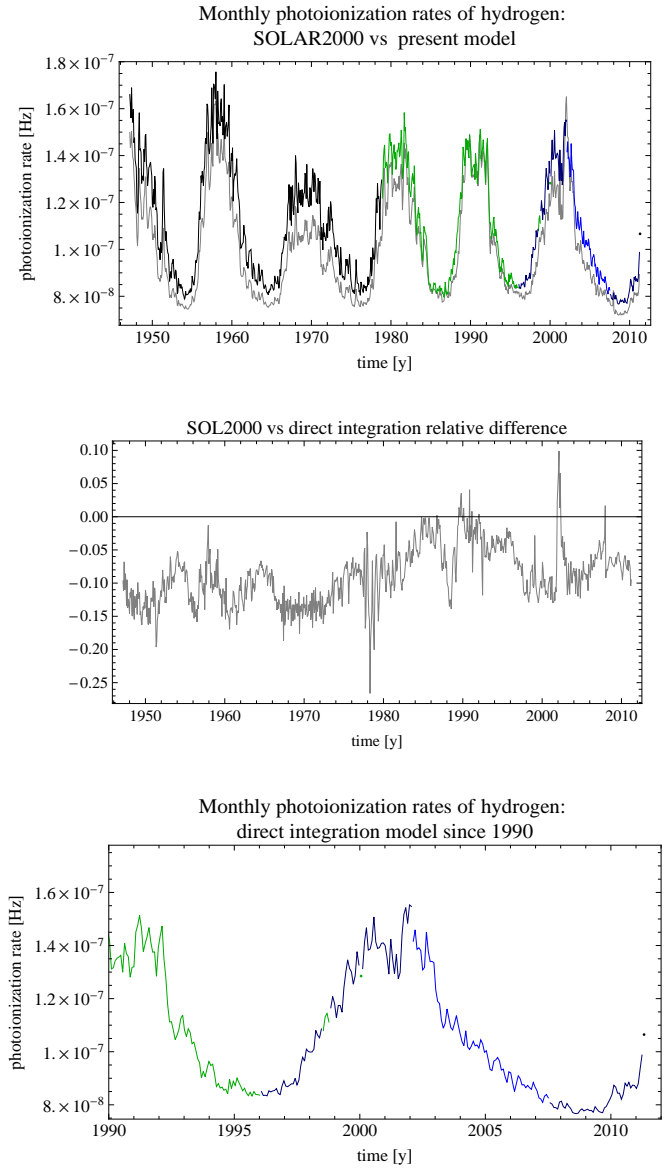}		
  	\caption{Comparison of the monthly averages of hydrogen photoionization rate obtained from the SOLAR2000 \citep[gray]{tobiska_etal:00c} and direct-integration model (blue), extended using the SEM/Ly-alpha (dark blue), Mg~II$_\mathrm{c/w}$ (green), and $F_{10.7}$ proxies (black) \citep[in preparation]{bochsler_etal:11b} in the upper panel; relative difference between the SOLAR 2000 and present model (middle panel), and exploded view of the photoionization rate from the present model during the two past solar cycles (lower panel).}
 		\label{figSOL2vsOurPh}
		\end{figure}

We have developed a model of evolution of the Carrington rotation averages of the hydrogen photoionization rates that is based on directly measured solar spectra from TIMED/SEE and a hierarchy of proxies \citep[in preparation]{bochsler_etal:11b}. From the spectra at full time resolution available from 2002 until present we calculated the photoionization rates using Eq.~\ref{eqPhotoIon}. Since the time series obtained showed clear signatures of flares and local particle-background contamination, we filtered it against the outliers beyond two sigmas, which show up in the time series of change rates $\left( \beta\left( t_{i+1} \right) - \beta\left( t_i \right)\right)/\left(t_{i+1} - t_i \right)$. From the filtered time series we calculated monthly averages, which are shown in the blue color in ther lower panel of Fig.~\ref{figSOL2vsOurPh}.  

Since the coverage by TIMED/SEE is limited in time and intercalibration or comparison of various measurements in the heliosphere taken at different times requires a knowledge of homogeneously-derived time series of ionization rates, we used a hierarchy of proxies to extend the directly-obtained photoionization rates backwards in time until the end of 1947, when the measurements of the F$_{10.7}$ flux became available. 

We started from directly integrated photoionization rates calculated from the filtered TIMED spectra, which now cover the full interval from solar maximum to solar minimum. We calculated a time series of monthly averages and we found a correlation formula between these values and Carrington rotation averaged measurements in Channel 1 and Channel 2 from CELIAS/SEM and the time series of Lyman-alpha flux from LASP, which is the following:
		\begin{eqnarray}
		\label{eqSEMLyaToTimed}		
		\beta_{\mathrm{Hph}}&=&5.39758 \times 10^{-20}\,I_{\mathrm{tot}}+2.36415 \times 10^{-16} \,\mathrm{SEM}_\mathrm{Ch1}^{0.765549}+\nonumber\\
                                    &+& 3.98461 \times 10^{-16}\,\mathrm{SEM}_\mathrm{Ch2}^{0.765549}+2.05152 \times 10^{-8}
		\end{eqnarray}
Using this formula, we calculated the Carrington averages of photoionization rates for the entire interval for which the SEM data are available. For the times when they are unavailable, but the MgII$_{\mathrm{c/w}}$ index from LASP is available, we use another correlation formula:
		\begin{equation}
		\label{eqMgIIToTimed}		
		\beta_{\mathrm{Hph}}=3.56348 \times 10^{-6}\,\mathrm{MgII_{c/w}}-8.5947 \times10^{-7}
		\end{equation}
When the MgII$_{\mathrm{c/w}}$ index is unavailable, we use the following correlation formula with the F$_{10.7}$ flux expressed in the sfu units (i.e. $10^{-22}$~W m$^{-2}$~Hz$^{-1}$):
		\begin{equation}
		\label{eqF107ToTimed}		
		\beta_{\mathrm{Hph}}=1.31864 \times 10^{-8}\,\mathrm{F_{10.7}^{0.474344}}-1.7745 \times 10^{-8}
		\end{equation}
It is worth pointing out that the exponent at the F$_{10.7}$ flux is close to $1/2$, not 1, as frequently adopted.

The results of the model with color coding of the sources used is shown in the lower panel of Fig.~\ref{figSOL2vsOurPh} for the time interval since 1990 until present. 

The photoionization rate obtained in the way described above can be compared with the rate from the SOLAR2000 model, used extensively in the previous studies \citep[e.g.][]{bzowski_etal:09a}. The comparison is shown in the upper panel of Fig.~\ref{figSOL2vsOurPh}. The two models agree to about $10 - 15\%$, with the direct integration model giving almost always higher values. Such an accuracy is basically equal to the present accuracy of the EUV measurements, especially in the low-energy portion of the spectrum which contributes most to the ionization. 

		\clearpage

\subsubsection{Latitude variation of photoionization rate}

Similarly as in the case of solar Lyman-alpha radiation, also radiation in the photoionization spectral region is expected to vary with heliolatitude. \citet{auchere_etal:05c} constructed a model of equatorial and polar flux in the solar 30.4~nm line, mostly responsible for ionization of helium, and demonstrated that a $\sim 0.8$ pole-to-equator ratio (fluctuating) can be expected. A similar modeling for the spectral range relevant for photoionization of hydrogen is not available, but coronal observations by \citet{auchere_etal:05b} suggest that such an anisotropy can indeed be expected and that some north-south asymmetry cannot be ruled out. 

In lack of a complete model and sufficient data we surmise that a latitude variation of hydrogen photoionization rate approximated by a formula similar to Eq.~\ref{eqLyaFlat} may be tentatively adopted. The subject certainly needs further studies.

\subsection{Electron ionization}

Significance of the electron-impact ionization reaction for the distribution of neutral interstellar gas in the inner heliosphere was pointed out by \citet{rucinski_fahr:89}, who developed a model of the electron ionization rate based on the local electron temperature and density. Further insight into the problem of electron-impact ionization of neutral interstellar hydrogen inside the heliosphere can be found in \citet{bzowski_etal:08a, bzowski:08a}. 

The ionization rate in the electron-impact reaction at a location described by the radius-vector $\vec{r}$ can be calculated from the formula \citep{owocki_scudder:83a}
   \begin{equation}
   \label{eqElImpRate}
   \beta_{\mathrm{el}}\left(\vec{r}\right) = \frac{8 \pi}{m^2_{\mathrm{e}}} \int\limits^{\infty}_{E_{\mathrm{ion}}}\sigma_{\mathrm{el}}\left(E\right) f_{\mathrm{e}}\left(E, \vec{r}\right) E\,\mathrm{d}E,
   \end{equation}
where $\sigma_{\mathrm{el}}$ is the energy-dependent reaction cross section, $E$ collision energy and $E_{\mathrm{ion}}$ the ionization threshold energy, for hydrogen equal to $\sim 13.6$~eV. Ionization occurs for H atom -- electron collisions with energies exceeding the limiting energy $E_{\mathrm{ion}}$. Practically, almost entire energy of electrons in the solar wind at a few AU from the Sun is in the thermal motions. Kinetic energy of an electron moving at a typical solar wind expansion speed of 440~km~s~$^{-1}$ is about 0.5~eV, which is much less than the ionization threshold. Since the temperature of the core electron distribution function at 1~AU is $\sim 10^5$~K and of the halo component on the order of $\sim 10^6$~K, the thermal speeds of the solar wind electrons are on the order of a few thousand of km~s~$^{-1}$, which is much more than the expansion speed. Thus, the expansion speed of the electron fluid can be neglected in the calculation of the electron-impact rate. 

\begin{figure}[t]
		\centering
		\includegraphics[scale=0.7]{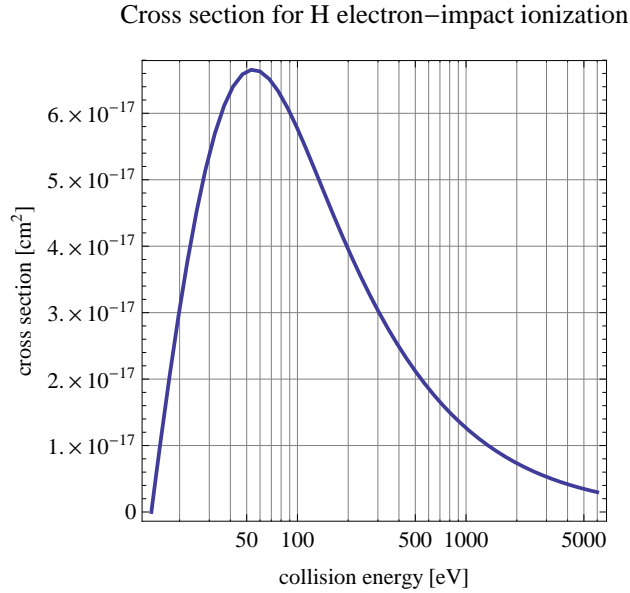}		
  	\caption{Cross section for electron-impact ionization of hydrogen, defined in  Eq.~\ref{eqElImpRate} from \citet{lotz:67}, as a function of H atom--electron collision energy in eV. }
 		\label{figElImpSigma}
		\end{figure}

The formula for the cross section for electron ionization was proposed by \cite{lotz:67a} and simplified for H by \cite{lotz:67}:
   \begin{equation}
   \label{eqBetaPhGeneralCrossSec}
   \sigma _{\mathrm{el}}\left(E\right)=\sum _{i=1}^{N_A} \frac{a_i q_i \ln \left(\frac{E}{P_i}\right)    \left(1-b_i \exp \left[-c_i
   \left(\frac{E}{P_i}-1\right)\right]\right)}{E P_i}
   \end{equation}
where $N_A$ is the number of electrons in the ion and the summation goes over the partial cross sections for knocking out all individual electrons from the ion. $P_i$ is the ionization potential for a given charge state of the ion, $E$ is the impacting electron energy, $a_i,b_i, c_i$ are parameters specific to a given ion and its charge state and $q_i$ is the statistical weight. For hydrogen, there is only one electron to be knocked out and Eq.~\ref{eqBetaPhGeneralCrossSec} takes the form:
   \begin{equation}
   \label{eqLotzHCrossSection}
   \sigma_{\mathrm{el}}\left(E\right) = 4.0 \left(1 - 0.60 \exp\left[-0.56 \left(\frac{E}{13.6} - 1\right)\right]\right)\frac{\ln\left(E/13.6\right)}{13.6 E} 
   \end{equation}
where $E \geq 13.6$ is expressed in eV. It is claimed by \citet{lotz:67} to be accurate to $\sim 10$\% and shown in Fig. \ref{figElImpSigma}.

Electron density can be adopted from quasi-neutrality and continuity conditions in the solar wind and calculated as equal to the local proton density + 2 $\times$ local alpha-particle concentration:
   \begin{equation}
   \label{eqElDens}
   n_e = n_p\left(1 + 2 \xi_{\alpha}\right),
   \end{equation}
where $\xi_{\alpha}$ is the local alpha-particle abundance relative to solar-wind protons.

The temperature behavior is more complex. Since electrons are subsonic while protons are supersonic on one hand, and since collisions are infrequent on the other hand, the electrons cool off faster than protons, and not adiabatically. The distribution function of electrons in the solar wind and its evolution with solar distance is fairly complex and requires further studies, especially that measurements performed on Ulysses using two different techniques: Quasi Thermal Noise \citep{issautier_etal:01a} and particle-measurements \citep{salem_etal:01a} return somewhat discrepant results \citep{issautier:09a}. Basically, the electron distribution function can be decomposed into three components: a warm core, a hot halo (both approximated by the Maxwellian function:
    \begin{equation}
    \label{eqElDiFu}
    f_e\left(\vec{r}, T_e \right) = n_e\,\left(\frac{m_e}{2 \pi \,k_B \,T_e\left(\vec{r}\right)} \right)^{3/2}\, \exp\left[-\frac{E}{k_B\, T_e\left(\vec{r}\right)} \right],
    \end{equation}
where $k_B$ is the Boltzmann constant, $m_e$ electron mass, $E$ kinetic energy of electron, $T_e$ temperature of the electron fluid, and $\vec{r}$ radius-vector of the location in space), and a fluctuating strahl, stretched along the local magnetic-field direction \citep{pilipp_etal:87c, pilipp_etal:87b}\footnote{Note that the electron distribution function was recently approximated by \citet{leChat_etal:10a, leChat_etal:11a} by the kappa function, which naturally covers both the core and halo components}. The contribution of the halo population to the net rate is on the level of a few percent and is an increasing function of the heliocentric distance \citep{maksimovic_etal:05a,stverak_etal:09a}. Estimates by \citet{bzowski_etal:08a} show that at 1~AU, the ionization rate due to the core population of the solar wind electrons is equal to about $0.4\times 10^{-7}$~s$^{-1}$ and to the halo population to less than $0.04\times 10^{-7}$~s$^{-1}$. The amplitude of fluctuations in the electron ionization rate may reach an order of magnitude, which is much more than the long-time variations related to variations in solar activity. On the other hand, the electron data from Wind \citep{salem_etal:03a} imply an in-ecliptic solar minimum (1995) rate of $\sim 0.68\times 10^{-7}$~s$^{-1}$ and a solar maximum (2000) rate of $\sim 0.73\times 10^{-7}$~s$^{-1}$. Thus assuming a constant rate over the solar cycle is a reasonable approximation.

Observations done with Ulysses \citep{phillips_etal:95c, issautier_etal:98, leChat_etal:11a} suggest that the electron ionization rate is a 3D, time dependent function of the solar cycle phase. Both the temperature magnitude and the cooling rate differ between the fast and slow solar wind. \citet{bzowski:08a} adopted the following radial profiles of the core $T_c$ and halo $T_h$ temperatures and the halo-to-core density ratios $\xi_{hc}=n_h/n_c$ for the slow \citep[after][]{scime_etal:94}: 
\begin{eqnarray}
\label{eqElSlowWind}
T_c & = & 1.3\cdot 10^5\,r^{-0.85} \nonumber\\
T_h & = & 9.2\cdot 10^5\,r^{-0.38} \\
\xi_{hc} = 0.06\,r^{-0.25} \nonumber 
\end{eqnarray}
and fast \citep[after][]{issautier_etal:98, maksimovic_etal:00a} solar wind: 
\begin{eqnarray}
 \label{eqElFastWind}
T_c & = & 7.5\cdot 10^4\,r^{-0.64} \nonumber \\
T_h/T_c & = & 13.57 \\
\xi_{hc}  & = & 0.03 \nonumber
\end{eqnarray}
In both solar wind regimes, the core $(n_c)$ and halo $(n_h)$ densities are calculated from the equations:
\begin{eqnarray}
\label{elCoreDens}
 n_c & = & \frac{1 + 2 \xi_{\alpha}}{1 + \xi_{hc}}\, n_p \nonumber \\
n_h & = & \xi_{hc}\, n_c
\end{eqnarray}
with the solar wind alpha abundance $\xi_{\alpha} = 0.04$, identical in both fast and slow wind regimes on average, and $n_p$ the proton density.

\citet{rucinski_fahr:89} inserted the formulae from equations \ref{eqBetaPhGeneralCrossSec},  \ref{eqElDens}, and \ref{eqElDiFu} to the integrand function in Eq.~\ref{eqElImpRate} and obtained a formula for each of the terms $i$ contributing to the total electron-impact cross section, which we present here in a slightly modified form:
   \begin{equation}
   \label{eqBelRateGeneral}
   \beta_{\mathrm{el},i}\left(T_e\right) = n_e \,\frac{ a_i\, q_i}{m_e\, P_i }\sqrt{\frac{8\, m_e}{\pi\,k_B T_e }}\, \left(\Gamma \left(0,\frac{P_i}{k_B T_e}\right)-\frac{\Gamma
   \left(0,c_i+\frac{P_i}{k_B T_e}\right)\, \exp\left[c_i\right]\, b_i P_i}{P_i+c_i\, k_B T_e}\right),
   \end{equation}
where $\Gamma(a,z)$ is the incomplete gamma function and for hydrogen $i = 1$, $q_i = 1$, $P_i = 13.6$~eV, $a_i = 4.0 \times 10^{-14}$, $b_i = 0.60$, $c_i = 0.56$ . 

\begin{figure}[t]
		\centering
		\includegraphics[scale=0.7]{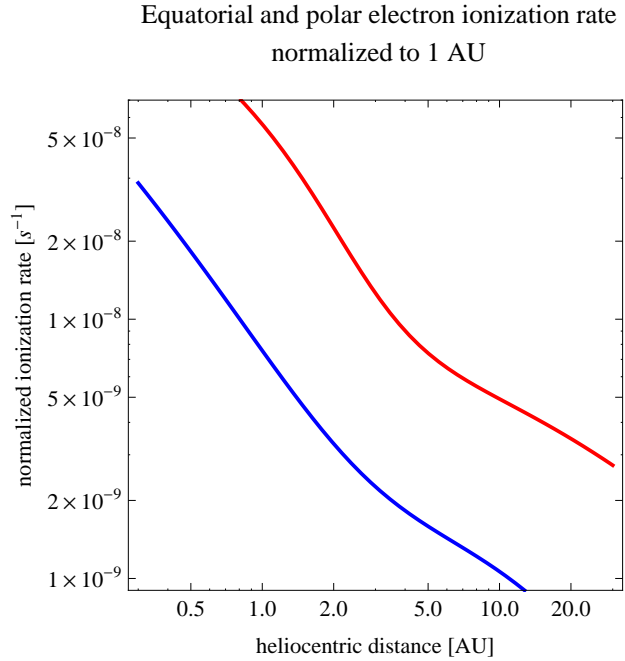}		
  	\caption{Normalized radial profiles $r^2\, \beta_{\mathrm{el}}\left(r\right)$ of the equatorial (red) and polar (blue) rates of electron-impact ionization of neutral interstellar H atoms in the supersonic solar wind, defined by Eq.\ref{eqBelSlowRate} and (\ref{eqBelFastRate}) \citep[adapted from][]{bzowski:08a} respectively.}
 		\label{figBelPolarEqtr}
		\end{figure}

Using these relations, \citet{bzowski:08a} employed the approach proposed by  \citet{rucinski_fahr:89}, assuming that the core and halo temperatures are isotropic (which is not exactly the case, as shown by \citet{stverak_etal:08a}), and calculated radial profiles of the electron-impact ionization rates separately for the fast and slow solar wind and approximated the results by the following phenomenological formulae:
\begin{eqnarray}
\label{eqBelSlowRate}
\beta_{\mathrm{el,s}}\left(r, n_p\right) & = & \frac{n_p}{r^2}\,\exp\left[\frac{\displaystyle \ln r \left(541.69 \ln r - 1061.32 \right) + 1584.32}{\displaystyle \left(\ln r- 29.17\right)\left(\left(\ln r - 2.02\right) \ln r + 2.91 \right)}\right] \\
\label{eqBelFastRate}
\beta_{\mathrm{el,f}}\left(r, n_p\right) & = & \frac{n_p}{r^2}\,\exp\left[\frac{\displaystyle \ln r \left(348.73 \ln r -917.39\right) + 2138.05}{\displaystyle \left(\ln r - 18.97 \right)\left(\left( \ln r - 2.53\right) \ln r + 5.74\right)}\right]
\end{eqnarray}
As is evident in these formulae, the electron-impact ionization rates are parametrized by local proton densities $n_p$ normalized to 1~AU and are fixed functions of heliocentric distance, which differ appreciably from the $1/r^2$ profiles that are typical of the solar wind flux and photoionization rate, as shown in Fig.~\ref{figBelPolarEqtr}. 
The behavior of electron temperature in the intermediate solar wind resulting from stream-stream interactions is unknown. \citet{bzowski:08a} approximated the electron temperature in the solar wind of intermediate velocity as a mean value between the fast and slow wind specific temperatures, weighted by the percentage of the solar wind density in transition from the slow to fast wind conditions. 

A consequence of the different magnitudes and radial behavior of the solar wind electron temperature between the slow and fast solar wind is a pronounced latitude anisotropy of the electron ionization rate throughout the solar cycle with an exception for a brief interval at solar maximum, when solar wind becomes almost spherically symmetric. The significance of the electron impact ionization in the overall balance of the contributing ionization reactions increases towards the Sun. Eventually, the role of electron ionization and its anisotropy is greater in the downwind hemisphere, where neutral interstellar gas has already passed the Sun (i.e., the streamlines of the H gas flow have passed their perihelia). Because of the fast cooling of the electron fluid with solar distance, the significance of the electron ionization reaction for the distribution of neutral interstellar hydrogen in the upwind hemisphere at the distances above $\sim 2$~AU (i.e., at and beyond the distance of the Maximum Emissivity Region (MER) of the heliospheric Lyman-alpha backscatter glow) is almost negligible. 

		\begin{figure}[t]
		\centering
		\includegraphics[scale=0.7]{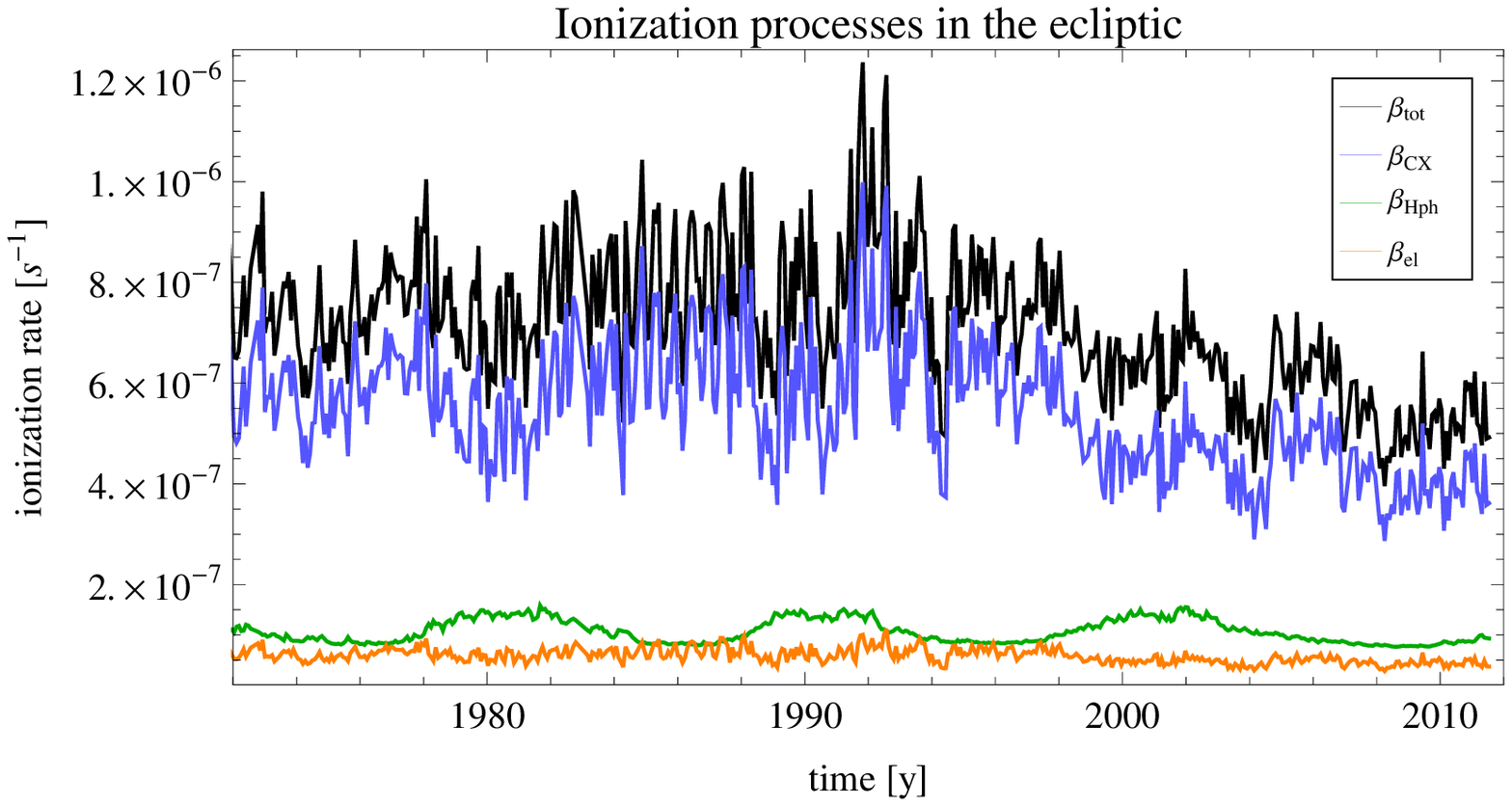}		
  	\caption{Ionization rates in the ecliptic plane from all relevant processes (black). Blue marks the charge exchange rate (calculated from Eq.~\ref{eqBetaCXSWIsGas} using the cross section from Eq.~\ref{eqLSsigCX} \citep{lindsay_stebbings:05a}), green the photoionization rate, orange the  electron-impact ionization rate.}
 		\label{figOMNIIonProcess}
		\end{figure} 

\section{Evolution of solar wind in the ecliptic plane}

The existence of solar wind was predicted on a theoretical basis by \citet{parker:57a} and discovered experimentally once the spacecraft Lunnik~II and Mariner~2 left the magnetosphere at the very beginning of the space age \citep{gringauz_etal:60a, neugebauer_snyder:62a}. Regular measurements of its parameters begun in the beginning of 1960s and up to now, data from more than 20 spacecraft are available, obtained using various techniques of observations and data processing. 

The highly supersonic solar wind in the ecliptic plane consists in fact of a sequence of various interleaved components: a ``genuine'' slow solar wind, fast solar wind, solar wind plasma from stream-stream interaction regions, and (intermittently) interplanetary coronal mass ejections (CME). The Mach number of the flow varies from $\sim 3$ to $\sim 10$. The balance between the populations changes with solar activity. 

Solar wind expands almost radially (i.e., its velocity has very small non-radial components in comparison with the radial component), with the speed basically invariable with solar distance between the outer boundary of the acceleration region near the Sun (located inside a few solar radii ) and approximately 10 AU, where the effects of mass loading due to the ionization of neutral interstellar gas \citep{fahr_rucinski:99, fahr_rucinski:01a, fahr_rucinski:02a} become measurable \citep{richardson_etal:95a, richardson_etal:08a}. The overall slowdown continues up to a few AU upstream from the termination shock, where the flow speed, by then already reduced by about 60 km~s~$^{-1}$ relative to the speed at 1~AU, is additionally slowed down by the component of protons reflected at the termination shock \citep{liewer_etal:93a, richardson_etal:08b}. For the purposes of modeling of the neutral gas distribution observed from 1~AU the reduction in solar wind speed with distance may be neglected. 

Outside the acceleration region of a few solar radii, solar wind density falls off with solar distance as $1/r^2$ and this relation, stemming directly from the continuity  equation, is not significantly altered by the interaction of the solar wind plasma with neutral interstellar gas. Indeed, with the main reaction being charge exchange, a reaction act results in a shift of the proton  in phase space (a core solar wind proton is replaced with a pickup proton moving at a different velocity), but not in a change in the local density. Only the two secondary reactions, photoionization and ionization by electron impact, actually inject new protons into the pickup ion region in phase space, thus increasing the local concentration of protons. Since the two secondary ionization processes contribute only $\sim 20$\% of the total ionization rate at 1~AU, as it will be clear later in this chapter, and since the PUI content  in the solar wind at the termination shock is on a level about 20\%, the net  increase in the absolute density of solar wind above the level dictated by the $1/r^2$ relation can be estimated as an excess of $20\% \times 20\% = 4$\% , which is negligible in comparison  with the uncertainty in the solar wind density, which will be discussed in a further part of this chapter.

\begin{figure}[t]
		\centering
		\includegraphics[scale=0.7]{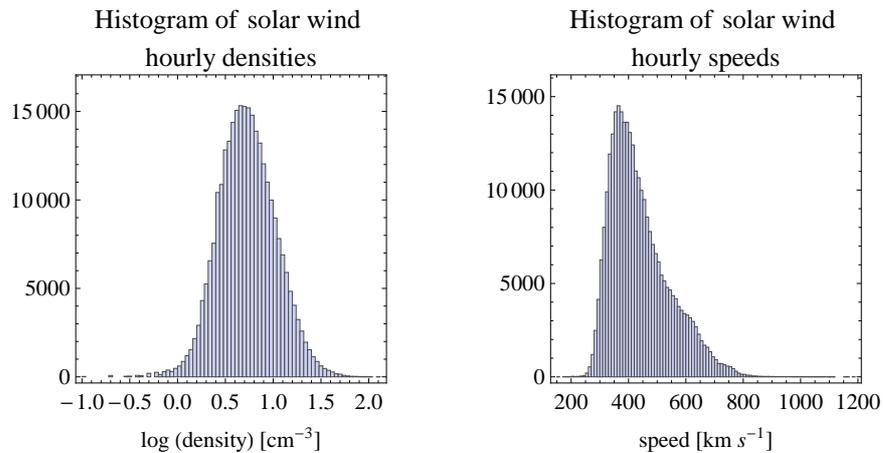}		
  	\caption{Histograms of hourly averages of solar wind density (adjusted to 1~AU) (left) and speed (right) from the entire OMNI-2 database, which currently spans an interval from 1963 until 2011.}
 		\label{figOMNIHistograms}
		\end{figure}

The probability distribution function (PDF) of hourly averages of densities is log-normal (i.e., logarithms of densities are distributed normally), while velovity features a complex PDF which can be approximately described as a sum of at least two normal distributions with differently located peaks of different widths. They are illustrated in Fig.~\ref{figOMNIHistograms} \citep[see also][]{deToma:11a}. 

From the viewpoint of charge exchange rate, details of the PDF of solar wind parameters are not as important as their mean values at appropriate time scales. Since the parameters of solar wind vary with the solar cycle and, as it was recently discovered by \citet{mccomas_etal:08a}, also secularly, and since neutral interstellar hydrogen, which is the background for the heliospheric Lyman-alpha backscatter glow, features variations on time scales comparable to the solar cycle length \citep{rucinski_bzowski:95b}, long and well-calibrated series of solar wind parameter measurements is a very important asset for heliospheric studies. Historically, measurements of solar wind speed obtained from various experiments generally agreed among themselves with an accuracy of $\sim 5$\%, but systematic differences between density values from different experiments up to $\sim 30$\% existed. Hence any study of a long-term behavior of solar wind required intercalibration of the results from different experiments. Such an initiative brought the OMNI data collection [King and Papitashvili available at at \url{http://omniweb.gsfc.nasa.gov/}] \citep[see also][]{king_papitashvili:05}, where historical and present measurements of the solar wind density, velocity, temperature, alpha abundance, and magnetic field vector were brought to a common calibration. 

Originally, the OMNI collection was created in the 1970s by the National Space Science Data Center (NSSDC)  at the NASA Goddard Space Flight Center. In 2003, a successor database OMNI-2 was made available, which has been maintained until present. The OMNI-2 data collection was basis for the present study of solar wind parameters in the ecliptic  and is used in this chapter to construct the charge exchange ionization rates.

The early period of the OMNI databases (1963 to 1971) includes data from multiple spacecraft \citet{bonetti_etal:69a, neugebauer:70a}, the middle-period data are mostly from IMP-8 and span from  1971 to 1994, and the later periods, from 1994 until present, include mostly data from IMP-8, Wind/SWE \citep{kasper:02a}, and ACE/SWEPAM \citep{mccomas_etal:98a}. Since there was no overlap between the early and middle period, the data from the early period in the OMNI-2 database are adopted unchanged from the original OMNI collection. The data from the early period were extensively intercalibrated between various spacecraft doing observations during that time, but no intercalibration with the middle and recent periods was possible. Still, owing to the overlap of the data between the middle and late periods, it was possible to perform an intercalibration between the data from these two periods. 
		
Most of the solar wind plasma data used in the OMNI collection were obtained from the MIT Faraday Cups \citep{bridge_etal:65a, lyon_etal:68a, lyon_etal:67a, lazarus_paularena:98a} and LANL electrostatic analyzers \citep{bame_etal:71a, hundhausen_etal:67a, ogilvie_etal:68a, feldman_etal:73a, asbridge_etal:76a, bame_etal:78a, bame_etal:78b, mccomas_etal:98a}. In the OMNI-2 series, \citet{king_papitashvili:05} adopted measurements of solar wind parameters performed by the Wind spacecraft as basis for the common calibration. They followed in this respect an analysis performed by \cite{kasper:02a, kasper_etal:06a}. The preparation of the data published in the OMNI collection involves removing from the original high-resolution data supplied by the Principal Investigators of the experiments possible Earth bow shock contamination and incomplete records, and subsequently time-shifting the data from the spacecraft location to the Earth location. The primary source of the solar wind data are nowadays Wind measurements, which are ultimately replaced with measurements from ACE, which unlike the data from Wind, are free from possible bow shock contamination because the ACE spacecraft operates near the Earth at a Lissajous orbit close to the L1 Lagrange point about a million kilometers upstream of the Earth bow shock. Because of different time scales of the processing of the data from various spacecraft, typically an interim data product becomes available once the first data are obtained, which are superseded with the final product when all the data needed become available or are declared as unavailable. This results in some changes in the published time series during the time. Our experience shows, however, that the changes are seldom significant.

The data from different experiments are scaled to a common calibration using linear fits based on results of linear regression analysis. It is also important to mention the significance of the correlation work that the OMNI team performed on the data from the IMP-8 spacecraft and early Wind and ACE measurements. IMP-8 operated for 28 years in 1973--2001 interval and provides a bridge between the early and present observations, enabling presentation of a more or less homogeneous series of solar wind parameters shown further down in this chapter. 

The result of the intercalibration process is a time series of hourly-averaged solar wind parameters. Because of the varying quality of individual records, the time coverage of the parameters is not uniform and gaps may exist in some parameters, while correct data for the same time interval may be available for other.  Since the distribution function of solar wind is inhomogeneous and varies rapidly in time, the values of solar wind parameters retrieved from observations depend on the method used to process the data. Typically, the LANL team take moments of the observed distribution function to calculate the density, speed and temperature of the solar wind, while the MIT team fit the measurements to an anisotropic Maxwellian or bi-Maxwellian function using a nonlinear fit method. To assess differences resulting from the two aforementioned approaches, the MIT team calculated the density, speed and temperature from the Wind/SWE distributions using both mentioned methods. 

King and Papitashvili at http://omniweb.gsfc.nasa.gov/ extensively discuss the differences and correlations between data from various sources. They show that the velocities are very well linearly correlated, with the coefficients of the relation $v_{\mathrm{SWE}} = a + b\, v_{\mathrm{ACE}}$  equal to $a = -2.135 \pm 0.387$ , $b = 1.010 \pm 0.001$. In the case of densities, it is the logarithms of density which are linearly correlated and the coefficients of the formula $\log n_{\mathrm{SWE}} = a + b\, \log n_{\mathrm{ACE}}$ slowly vary with speed, $a$ changing from 0.006 for $v < 350$~km~s~$^{-1}$ to 0.091 for $350 < v < 450$~km~s~$^{-1}$ to 0.082 for $v > 450$~km~s~$^{-1}$ and $b$ changing from 1 for $v < 350$~km~s~$^{-1}$ to 1.036 for $350 < v < 450$~km~s~$^{-1}$ to 1 for $v > 450$~km~s~$^{-1}$. This brings up differences on the order up to 20\% between ACE and Wind, which are comparable to the uncertainty in density coming up from the application of various methods of parameter derivation discussed earlier. In a nutshell, while a very good correlation of speeds is obtained, the correlation between logarithms of densities is close to linear, but with a scatter of approximately 30\% around the fit line. This is probably a good measure of inherent uncertainty of the densities even without uncertainties in the absolute calibrations.

\begin{figure}
		\centering
		\includegraphics[scale=0.7]{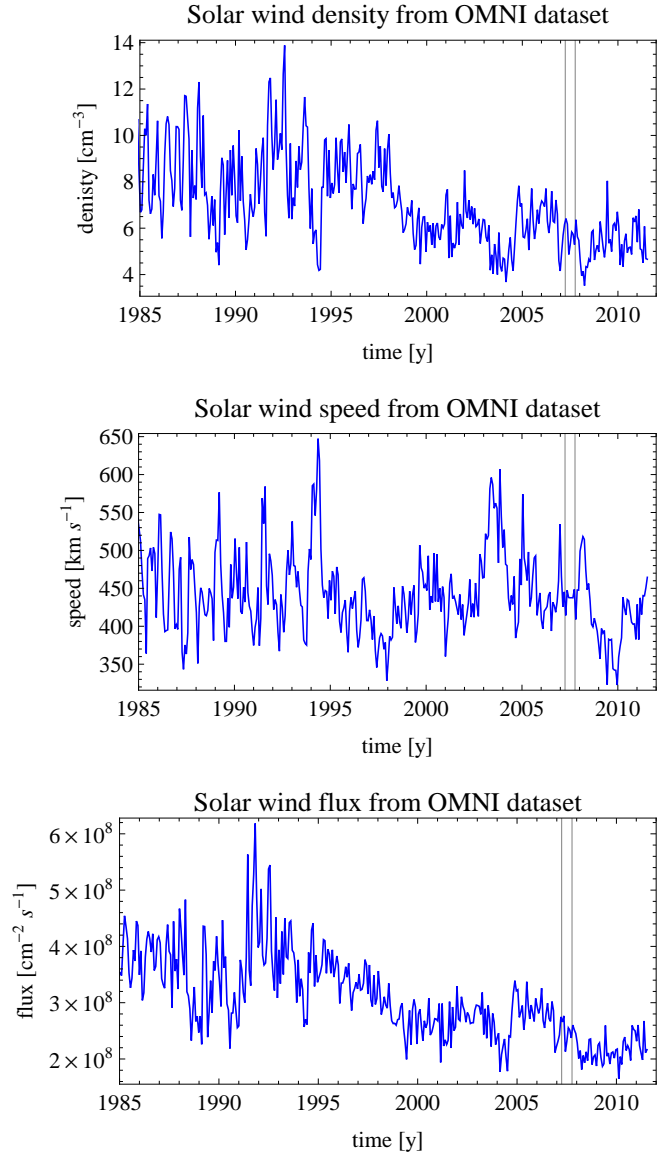}		
  	\caption{Carrington rotation averages of solar wind density adjusted to 1~AU (upper panel), speed  (middle panel), and flux (lower panel) calculated from hourly averages from the OMNI-2 database \citep{king_papitashvili:05}. The thin vertical lines mark the time interval shown at daily resolution in Fig. \ref{figOMNIDailyParams}.}
 		\label{figOmniDensSpeed}
		\end{figure}

\begin{figure}
		\centering
		\includegraphics[scale=0.7]{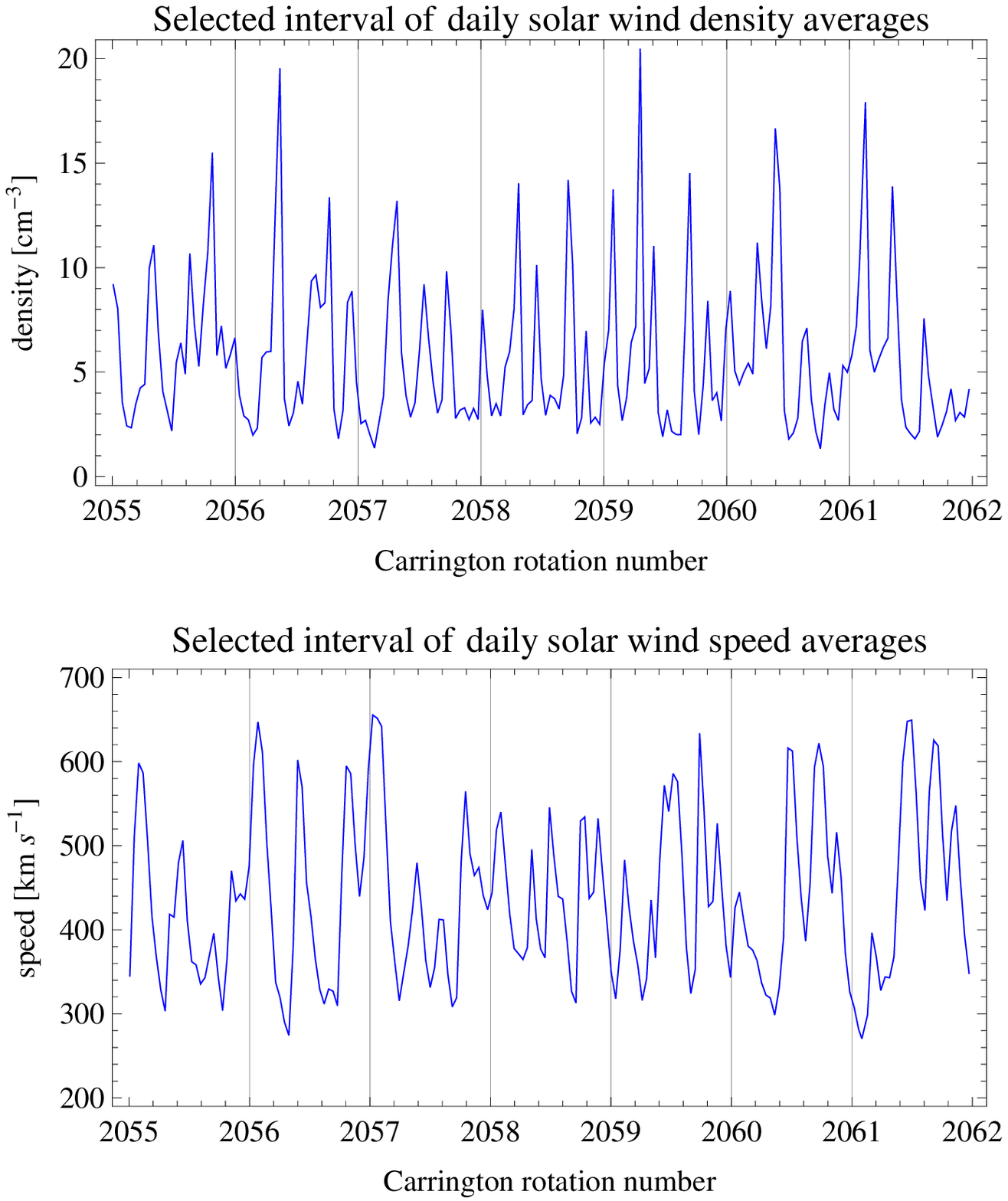}		
  	\caption{Daily averages of solar wind density (adjusted to 1~AU) (upper panel) and speed  (lower panel) calculated from hourly averages obtained from the OMNI-2 database \citep{king_papitashvili:05} for an example time interval covering 7 full Carrington rotation periods, presented to illustrate the complex structure of solar wind parameter evolution at different time scales and the approximate anticorrelation of density and speed.}
 		\label{figOMNIDailyParams}
		\end{figure}

Based on the OMNI database, we constructed a time series of Carrington period-averaged parameters of the solar wind normalized to 1 AU, with the grid points set precisely at halves of the Carrington rotation intervals. Small deviations of the times from the halves of the rotation periods were linearly interpolated. Averaging over the Carrington rotation enables constructing of a consistent model of the ionization rate, where axial symmetry of the solar parameters is safeguarded. Such a procedure seems to be the only reasonable one at present. Solar wind parameters show considerable variations during one solar rotation period, with quasi-periodic changes from slow to fast solar wind speeds and related changes in density (see Fig.~\ref{figOMNIDailyParams}). The time scale of the changes of the fast wind streams is comparable to the solar rotation period and thus constructing of a full and accurate model of the solar wind variation as a function of time and heliolongitude is presently not feasible because of the lack of sufficient data. Thus, we developed an equally-spaced time series of the solar wind in-ecliptic densities and speeds, which are well suited for the purpose of global heliospheric modeling and to calculate the charge exchange rates of heliospheric neutral atoms. The time series of density, speed, and charge exchange rate in the approximation of neutral H atom stationary relative to the Sun is presented in Fig.~\ref{figOmniDensSpeed}.

The time interval shown in this figure starts before the solar activity minimum in 1986 and includes the solar minima of 1995 and 2007, as well as the two recent maxima of 1990 and 2001. One observes a striking difference in the appearance of the solar wind equatorial parameters in comparison with the behavior of solar EUV radiation. Neither density nor speed seem to be correlated with the level of solar activity. There is no clear minimum--maximum--minimum variation, which was clearly seen in the EUV-related time series. Speed shows multi-timescale variations, but its mean value is basically unchanged in time. By contrast, density features a secular change, which begun just before the last solar maximum and leveled off shortly before the present minimum. The overall drop in the average solar wind density is on the order of 30\% between 1998 and 2005; it seems that the drop in the charge exchange rate is even more dramatic, namely by a factor of two. The present rate of charge exchange is at a level similar to the charge exchange rate observed by Ulysses at the poles during its first fast latitude scan \citep{phillips_etal:95c, mccomas_etal:99}. Thus the solar wind density features a ``plateau'' until 1998, then a ``cliff'' and a ``foot'' starting from 2002. Within the ``foot'' there seem to exist fluctuations of density and anticorrelated fluctuations of speed. These can be associated with the persistence of coronal holes at equatorial latitudes, as convincingly illustrated by \citet{deToma:11a}. This overall long-standing drop in the solar wind flux results in an overall enhancement in the neutral interstellar gas density near the Sun. The secular variations in the solar wind, on time scales significantly longer than the solar cycle time scale, suggest that the heliosphere does not evolve strictly periodically and that monitoring of these variations is an essential element of any effort aimed at quantitative analysis of all kinds of heliospheric~observations. 

\section{Latitudinal structure of the solar wind and its evolution during the solar cycle}
Shortly after the discovery of solar wind, a question whether or not it is spherically symmetric appeared. It is currently known to feature a latitudinal structure which varies with the solar activity cycle. While direct observations of the solar wind in the ecliptic plane have been conducted since the early 1960s, information on its latitudinal structure was available only from indirect and mixed observations of the cometary ion tails \citep{brandt_etal:72a, brandt_etal:75a}. The situation changed when radio-astronomy observations of interplanetary scintillation and appropriate interpretation of observations of the Lyman-alpha helioglow became available. Apart from the in situ measurements obtained from Ulysses, these two techniques remain the only source of global, time-resolved information on the solar wind structure. 

The launch of the Ulysses spacecraft \citep{wenzel_etal:89a}, the first and up to now the only interplanetary probe to leave the ecliptic plane and sample interplanetary space in polar regions, improved our understanding of the 3D behavior of solar wind by offering a very high resolution in latitude, but a poor resolution in time (the same latitudes were visited only a few times during the $\sim 20$-year mission). Hence the studies of solar wind parameters as a function of time and heliolatitude are still a work in progress and therefore are discussed in a separate section in this chapter.

\subsection{Historical perspective: insight from interplanetary scintillation}

Interplanetary scintillation measurements involve radiotelescope observations of remote compact radio sources (like quasars) and searching for fluctuations of the signal that come up because of diffraction of radio waves on electron density fluctuations $\sim 200$~km in diameter occurring in the line of sight. Specifically, the observable is the scintillation index, defined as a quotient of the  r.m.s. of the observed intensity fluctuation to the mean intensity of a source \citep{manoharan:93b}. Owing to a correlation between the fluctuations amplitude and solar wind speed it is possible to deduce the solar wind speed from careful analysis of registered diffraction patterns in the observed radio source signal \citep{hewish_etal:64a}. The scintillation signal is the sum of waves scattered along the line of sight to the observed radio source. Most of the scattering occurs at the closest distances to the Sun along the sightline because the strength of the electron density fluctuations rapidly decreases with solar distance.  Had the solar wind been spherically symmetric, it would be possible to define a weighting function and the IPS ``midpoint'' speed would be the spatial average of the solar wind speed centered at the closest point of the sightline \citep{coles_maagoe:72a}. But solar wind is not spherically symmetric and features streams, which result in discrepancies of the measured speed from the actual one \citep{houminer:71a}. 

An improvement in the accuracy of measurements of solar wind speed can be achieved owing to a better identification of the scintillation patterns in the sightlines when they are simultaneously observed by multiple stations displaced longitudinally on the Earth surface, and by using the solar wind tomography technique \citep{jackson_etal:97a, jackson_etal:98a, kojima_etal:98, asai_etal:98a}.

Despite the high sophistication of the tomography technique, its accuracy inevitably depends on purely geometrical considerations (the telescopes are located in the north hemisphere of the Earth, whose orbit is tilted at an angle of 7.25 deg to the solar equator), on the number of observations available, and ultimately on the fidelity of the adopted correlation formula between the scintillation index, which is directly measured, and the solar wind speed. While correlating these quantities was possible early enough for the equatorial solar wind \citep{harmon:75a}, the out-of-ecliptic IPS measurements could only be calibrated once in situ data from Ulysses became available \citep{kojima_etal:01}. Even then one has to keep in mind that ideally, such a calibration should be repeated separately for each solar cycle because, as discussed earlier in this chapter, solar wind features secular changes.

Early measurements of solar wind speed using the IPS technique brought mixed conclusions: while \citet{dennison_hewish:67a} discovered an increase in solar wind speed outside the ecliptic plane, \citet{hewish_symonds:69a} did not find such an increase. Further observations, however, reported by \citet{coles_rickett:76a}, clearly showed that the solar wind is structured, with a band of slow speed around the solar equator and polar caps of a much faster wind. 

An extensive program of IPS observations of solar wind, initiated in 1980s in the Solar-Terrestrial Environmental Laboratory at the Nagoya University (Japan) \citep{kojima_kakinuma:90a}, resulted in a homogeneous dataset that spans almost three solar cycles and enables studies of the evolution of solar wind speed profile with changes in solar activity \citep{kojima_kakinuma:87a}. Even before the introduction of the Computer Assisted Tomography technique, they suggested that the solar wind structure varies with solar activity, with the slow wind reaching polar regions when the activity is high. Supported and interpreted by the tomography technique \citep{hayashi_etal:03a}, it enables detailed studies of the structure of solar wind in the varying solar activity conditions \citep{kojima_etal:99a, kojima_etal:01, ohmi_etal:01a, ohmi_etal:03a, fujiki_etal:03a, fujiki_etal:03b, fujiki_etal:03c, ohmi_etal:03a, kojima_etal:07a, tokumaru_etal:09a, tokumaru_etal:10a}. The solar wind speed data obtained from these observations will be discussed later in this chapter. 

\subsection{Historical perspective: insight from heliospheric backscatter glow}

Observations of the Lyman-alpha heliospheric backscatter glow, carried out since the beginning of 1970s \citep{bertaux_blamont:71}, have also been used as a tool to discover the 3D structure of solar wind and its evolution with solar cycle. The bi-modal structure of the solar wind, with the slow and dense wind in an equatorial band and rarefied, fast wind at the polar caps, results in a distinctly structured charge exchange ionization rate. Since the direction of inflow of neutral interstellar gas is very close to solar equator, the inflowing atoms whose orbits happen to be in a plane close to solar equator spend their entire travel through the heliosphere in the region of increased probability of ionization. In contrast, those traveling in planes inclined at greater angles to solar equator spend relatively little time at low heliolatitudes or even do not get close to the region of increased ionization at all and hence more of them are able to survive the travel towards the Sun \citep[see Fig.~1 in ][for an illustrative sketch]{lallement_etal:85b}. As a result, a region of reduced density of neutral interstellar hydrogen gas is created close to solar equator. At heliolatitudes farther from the equator the density of the gas at similar distances is higher. The gas is illuminated by an approximately spherically symmetric solar Lyman-alpha radiation, which is backscattered by resonance fluorescence. Since at equatorial latitudes the density of the gas is reduced and the illuminating flux is almost homogeneous in heliolatitude, the intensity of the backscattered radiation is lower at equatorial latitudes than at the polar caps. This equatorial dimming of the helioglow, referred to as the heliospheric groove, was observed already in 1970s and 1980s \citep{kumar_broadfoot:78, kumar_broadfoot:79, lallement_etal:85a, lallement_etal:86, lallement_stewart:90} and correctly interpreted as due to the enhanced ionization level at equatorial latitudes because of the anisotropy of solar wind. 

It is important to stress, however, that analysis of the heliospheric Lyman-alpha backscatter glow is only able to yield the latitude structure of the total ionization rate of neutral interstellar hydrogen. From the view point of density structure of neutral interstellar hydrogen near the Sun the nature of the ionization processes is not important, only the result they produce, i.e. a decrease in the total density at an equatorial latitude band. Hence no differentiation between the charge exchange, photoionization, and electron-impact ionization can be made based solely on the heliospheric glow analysis. Since, however, charge exchange is the dominant process and since photoionization is only slightly anisotropic in latitude, in the first approximation the latitude variation in the total rate can be regarded as latitude variation in the charge exchange rate, which is proportional to the latitude variation in the total solar wind flux, modulated by the dependence of the charge exchange rate on solar wind speed (see Eqs (\ref{eqBetaCXSWIsGas} and (\ref{eqLSsigCX})).

\citep{lallement_etal:85b} proposed a model of the charge exchange rate between the solar wind protons and neutral interstellar H atoms as a function of heliolatitude $\phi$ defined by the formula
\begin{equation}
\label{eqCXLatiSin}
\beta_{\mathrm{CX}}\left(\phi\right)  = \beta_0\left(1 - A \sin^2 \phi \right)
\end{equation}
This is a two-parameter relation, where $\beta_0$ corresponds to the rate at the equator and $A$ is a pole-to-equator amplitude, which can be fitted from observations of the helioglow. This formula reproduced approximately the limited observations of the groove obtained at the early stage of the research. It is able to reproduce various pole-to-equator contrasts in the ionization rate, also the situation when the contrast virtually disappears and solar wind becomes approximately spherically symmetric. However, the profile of the ionization rate obtained from this formula has a full width at half maximum $\sim 45 \deg$ and is perfectly symmetric about solar equator. Consequently, is able to reproduce neither the north-south asymmetries in the solar wind nor the situations when the range of the slow solar wind significantly differs from $\sim 45 \deg$. 

Despite these deficiencies, Eq.~(\ref{eqCXLatiSin}) was successfully used to infer qualitatively the solar wind structure by a number of authors \citep[e.g.]{lallement_etal:85b, pryor_etal:98b, pryor_etal:03a}. The conclusions were similar to those obtained from the IPS analysis: during solar minimum solar wind is latitudinally structured, with a band of enhanced flux at the equator and two polar caps of a rarefied and fast wind. During solar maximum the ordered structure changes and the polar caps become almost fully covered with the slow wind. However it became clear that the simple model given by Eq.~\ref{eqCXLatiSin} is not fully adequate to describe the reality and a need for more observations became evident. \citet{bertaux_etal:95} proposed an experiment to study solar wind anisotropies using the technique of analysis of the heliospheric Lyman-alpha backscatter glow, which was implemented in the French/Finnish project SWAN onboard the Solar and Heliospheric Observatory mission (SOHO). Already shortly after launch it became evident that the early conclusions on the evolution of the heliospheric groove, and thus the solar wind, with the solar activity cycle are supported \citep{bertaux_etal:96a, bertaux_etal:97a, bertaux_etal:99,  kyrola_etal:98}, but the formula used to describe the latitude profile of the ionization rate needs modification. Thus in the latter work, Eq.~\ref{eqCXLatiSin} was modified to describe separately the northern and southern hemispheres:
\begin{equation}
\label{eqCXLatiSin1}
\beta_{\mathrm{CX}}\left(\phi\right)  = \beta_0\left[\Theta\left(\phi\right)\left(1 - A_N \sin^2 \phi \right) + \Theta\left(-\phi\right)\left(1 - A_S \sin^2 \phi \right) + B\left( \phi \right)\right],
\end{equation} 
where $\Theta$ is the Heavyside step function, $A_N, A_S$ are the separate anisotropy parameters for the northern and southern hemispheres, and $B\left(\phi\right)$ is used to narrow the width of the equatorial band of enhanced solar wind flux. Even more sophisticated approach was proposed by \citet{summanen:96}, who suggested to approximate the equatorial band in the latitudinal profile of the total ionization rate by 
\begin{equation}
\label{eqTulaAnalyt}
\beta_{\mathrm{CX}}\left(\phi\right)  = \beta_0\left[1 - A\left(t\right) \sin^2 \left(c\, \phi \right) \right]
\end{equation}
for heliolatitudes $-40\deg \le \phi \le 40 \deg$, where $c = 9/4$ limits the equatorial band to $\pm 40\deg$,  and by 
\begin{equation}
\label{eqTuulaAnalytPolar}
\beta_{\mathrm{CX}}\left(\phi\right)  = \beta_0\, \exp \left[ -\left(\frac{1}{0.2 P}\left(t - P/2 \right)\right)^2 \right],
\end{equation}
where $P$ is the solar cycle length and $t$ time. In this formula, there was no north-south asymmetry allowed, but it was possible to homogeneously reproduce variations of the anisotropy parameters in tact with the solar activity cycle. 

Since the north-south anisotropies in the ionization rate and their evolution with solar activity were evident on one hand, and on the other hand the first fast latitude scan by Ulysses \citep{phillips_etal:95c} suggested that the profile of solar wind parameters can be approximated by an equatorial plateau (with a ``rough surface'' of the gusty slow wind) standing out from a flat ``foot'' of the fast polar wind, \citet{bzowski:01a, bzowski_etal:02, bzowski:03} suggested to approximate it by the formula
\begin{eqnarray}
\label{eqMBCXForm}
\beta_{\mathrm{CX}}\left(\phi ,t\right)&=&\left(\beta_{\mathrm{CX,pol}} + \delta_{CX}\, \phi \right) + \left(\beta _{\mathrm{CX,eqtr}}\left(t\right)-\beta_{\mathrm{CX,pol}}\right)\\  
& \times &\exp \left[-\ln  2 \left(\frac{2 \phi  - \phi_{N}\left(t\right)- \phi_{S}\left(t\right)}{\phi_{N}\left(t\right)-\phi_{S}\left(t\right)}\right)^N\right],\nonumber
\end{eqnarray}
where $\phi $ is heliographic latitude and $N$ is a shape factor; $\beta_{\mathrm{CX,pol}}$ is the average ionization rate at the poles and the term $\left(\beta_{\mathrm{CX,pol}}+ \delta_{CX}\, \phi \right)$ describes the north-south asymmetry of the polar ionization rates; the term $\left(\beta_{\mathrm{SW,pol}} + \delta_{CX}\, \phi \right)+\left(\beta_{\mathrm{SW,eqtr}}\left(t\right)-\beta_{\mathrm{CX,pol}}\right)$ for $\phi = 0$ corresponds to the charge exchange rate at solar equator; and the term $\exp\left[-\ln 2 \left(\frac{2 \phi - \phi_{N} - \phi_{S}}{\phi_{N}-\phi_{S}}\right)^N\right]$ describes the latitudinal dependence of the ionization rate \citep[see also][]{bzowski:08a}. The shape of the central bulge is controlled by the exponent $N$; for $N = 2$ the shape is Gaussian, for $N = 8$ it is close to rectangular.

The parameters in the model by \citet{bzowski_etal:03a} are the north and south boundaries of the equatorial slow wind band $\phi_N$ and $\phi_S$, the equator/north pole and equator/south pole ratios of the ionization rate, and the polar north/south asymmetry parameters. In order to obtain absolute values of the ionization rate, an independent assessment of the ionization rate at the equator is needed. \citet{bzowski_etal:03a} used the theory developed by \citet{bzowski:03} and interpreted a carefully selected subset of SWAN observations. To maintain as much symmetry in the data as possible, and to include in the data the full span of heliolatitudes with a well balanced presence of measurements from all heliolatitudes, they chose observations taken within a week of the passage of SWAN through the projection of inflow axis ion the ecliptic plane, i.e. at the beginning of June and December of each year, and they restricted the field of view to a narrow band going through the projections on sky of solar equator and poles. They eliminated ``searchlights'' \citep[discovered by][]{bertaux_etal:00}, i.e. reflections on sky of the point-like active regions on the solar disk, traveling across the sky with the angular velocity of solar rotations, and cleaned the data from the contamination by extraheliospheric ``chaff'' (Milky Way, stars etc.). To eliminate possible bias from an imperfect absolute calibration, they tried to fit the lightcurves normalized to equatorial values instead of attempting to fit the absolute intensity of the helioglow.

In agreement with other studies, \citet{bzowski_etal:03a} found that the ordered structure of the solar wind present during solar maximum disappears with the increasing solar activity. The boundaries between the fast and slow wind regions move polewards and ultimately at solar maximum the slow wind encompasses the whole space. The motion of the fast/slow wind boundaries in the north and south hemispheres were found to be shifted in phase by approximately a year. \citet{bzowski_etal:08a} discovered that the areas of the polar fast wind regions are linearly correlated with the areas of the polar holes observed by \citet{harvey_recely:02}, which enabled them to calculate the variation in the boundaries between the fast and slow solar wind $\phi_N, \phi_S$ for the time span of the polar coronal holes observations, i.e. from 1990 until 2002. \citet{bzowski_etal:08a} suggested also that the evolution of $\phi_N, \phi_S$ can be approximated by a formula
\begin{equation}
\label{eqSWBndrEvol}
\phi_{N,S}\left(t\right) = \phi_{0} + \phi_{1}\, \exp\left[-\cos^3\left(\omega_{\phi}\, t \right) \right],
\end{equation} 
where $\omega_p$ was obtained as $2 \pi /$Lyman-alpha main period. The validity range of this approximation is limited to the time interval from 1990 to 2002. 

Based on this model of ionization, it was possible to infer the evolution of solar wind speed and density as a function of time and latitude. This topic will be covered in a further part of this chapter.

The model from \citet{bzowski_etal:03a, bzowski_etal:08a} was able to reproduce more correctly the latitudinal span of the slow wind region and its evolution with solar activity, but was inadequate to correctly reproduce the boundaries themselves. Observations from SWAN showed that especially during the transition phases of solar activity, the photometric latitudinal profiles of the groove are complex and variable in time. To address this problem, it  turned out to be necessary to use an approach proposed by \citet{summanen_etal:93} and to model the ionization rate as a multi-step function, with different levels in fixed, though arbitrarily selected latitudinal bands. Such a model of the ionization rate was implemented already by \citet{lallement_etal:85b}, but then the values in the latitudinal ``slots'' were filled with the values obtained from the analytical models discussed earlier in this chapter. Now, as discussed by \citet{quemerais_etal:06b, lallement_etal:10b}, the ionization rates in the latitudinal bands are free parameters fitted to the maps of heliospheric backscatter glow, without any assumptions on  relations between the neighboring bands. In contrast to the approach exercised by \cite{bzowski_etal:03a}, \citet{quemerais_etal:06b, lallement_etal:10b} used all the data available, cleaned only with use of a mask intended to cut off the known extraheliospheric ``chaff''. Thus a time series of ionization rate profiles at a resolution of $\sim 10 \deg$ in heliolatitude and $\sim 2$ days in time  was obtained from fitting of the model to the filtered full-sky maps and subsequent scaling to the in-ecliptic ionization rates obtained from in situ measurements. An illustration of results of this analysis is shown in \citet{lallement_etal:10b}.

\subsection{Historical perspective: Ulysses measurements}

\begin{figure}
\centering
\includegraphics[scale=0.68]{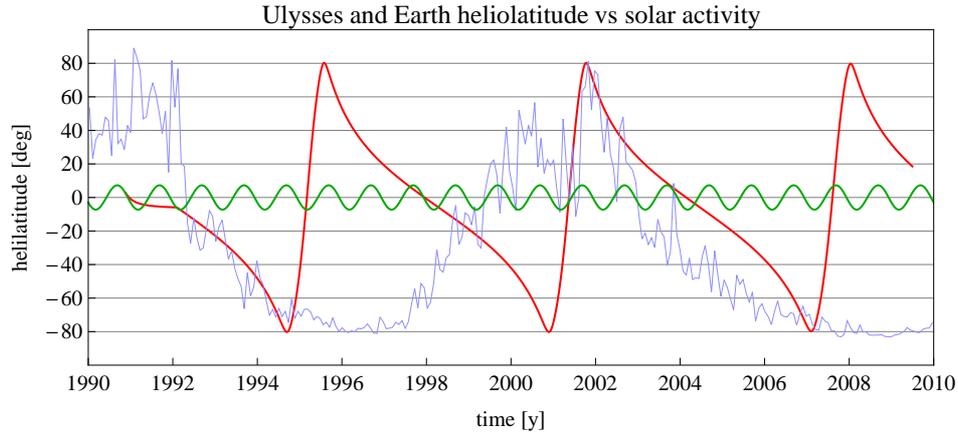}
\caption{Illustration of heliolatitude of Ulysses (red) and Earth (green) during the time span of Ulysses mission. The pale blue line is the F$_{10.7}$ solar radio flux, superimposed to correlate variations in solar activity with Ulysses heliolatitude during its more than 3 orbits in a polar plane almost perpendicular to the inflow direction of neutral interstellar gas.}
\label{figUlyssesComposite}
\end{figure}

Ulysses, launched in October 1990, has been the first spacecraft to traverse polar regions of the heliosphere and provide a unique view of solar wind \citep{wenzel_etal:89a, smith_etal:91a}. After a cruise to Jupiter carried out close to the ecliptic plane, it performed a Jupiter gravity assist maneuver which cast it away from the ecliptic plane on an elliptical nearly polar orbit with an aphelion of $\sim 5.5$~AU and perihelion $\sim 1.4$~AU in a plane nearly  perpendicular both to the ecliptic plane and solar equator as well as to the inflow direction of interstellar gas. The period of the orbit was about 6 years. The heliolatitude track of Ulysses is shown in Fig. \ref{figUlyssesComposite}. The spacecraft was launched at solar maximum and the radial cruise was completed just when solar activity was beginning to decrease. The first dive towards the south solar pole was carried out during an interval of decreasing activity, followed by the first so-called fast latitude scan, when the spacecraft coasted from south to north solar pole, covering almost full span of heliolatitudes during just about a year at solar minimum activity conditions. This fast latitude scan preceded by about a year the first observations of the  helioglow by SWAN/SOHO, which provided an opportunity to calibrate the model of ionization rate obtained from analysis of the heliospheric Lyman-alpha glow. Ulysses continued on its polar orbit, performing its first full slow latitude scan during an interval of increasing solar activity. The following second fast latitude scan occurred in a totally different solar cycle phase, namely during solar maximum. Again this scan took about a year and was performed during an interval of dynamically variable solar wind structure. The following slow latitude scan was, in contrast to the first slow scan, during decreasing solar activity and afterwards the spacecraft performed its last fast latitude scan, again during a solar minimum activity interval. 

The geometry and timing of Ulysses trajectory brought a unique dataset of direct in situ measurements of parameters of solar wind plasma, obtained from SWOOPS \citep{bame_etal:92a} and SWICS \citep{gloeckler_etal:92} experiments. The discoveries and findings from the plasma measurements were presented in dozens of scientific papers \citep[see, e.g.,][]{phillips_etal:95a, phillips_etal:95c, marsden_smith:97, mccomas_etal:98b, mccomas_etal:99, mccomas_etal:00a,mccomas_etal:00b, mccomas_etal:02a, mccomas_etal:02b, mccomas_etal:03a, mccomas_etal:06b, mccomas_etal:08a}. An additional benefit from this unique mission is the use of the Ulysses flight spare plasma instrument with only minor modifications on the ACE mission \citep{mccomas_etal:98a}, which facilitates intercalibrating the Ulysses measurements with the OMNI time series.

		\begin{figure}
		\centering
		\includegraphics[scale=0.7]{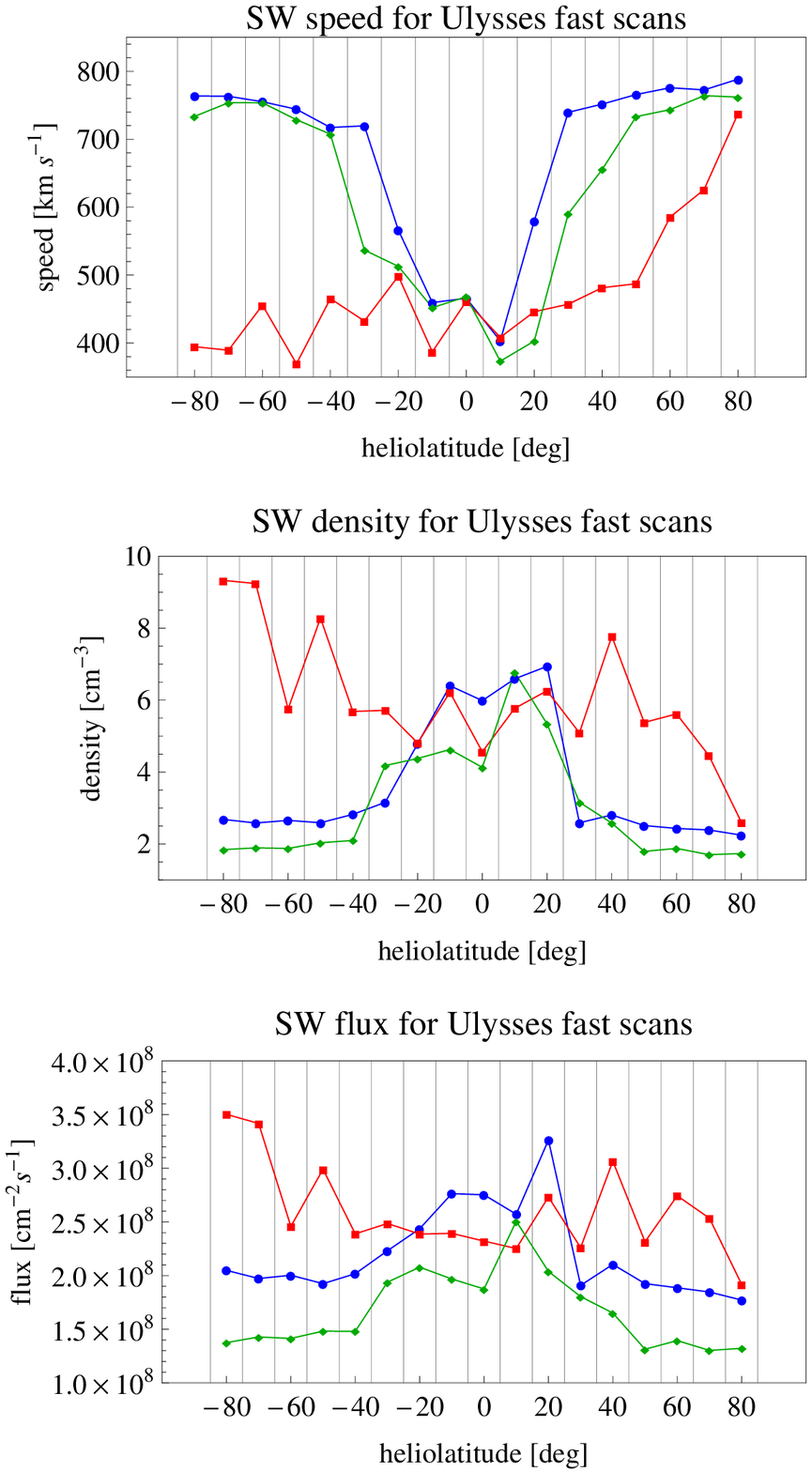}		
  	\caption{Solar wind speed (upper panel), adjusted density (middle panel), and adjusted flux (lower panel) as a function of heliolatitude for the first (blue), second (red) and third (green) Ulysses fast latitude scans. The parameters are averaged over 10-degree heliolatitude bins.}
 		\label{figUlyDensSpeedFlux}
		\end{figure} 

The Ulysses solar wind dataset is unique and invaluable because it is the first and only direct, in situ measurement of the solar wind parameters outside the ecliptic plane. The evolution of solar wind speed, adjusted density and flux during the previous and current solar minima and during the previous maximum are compiled in Fig. \ref{figUlyDensSpeedFlux}, where the parameter values are averged over 10-degree bins in heliolatitude. Adjustment throughout this text means scaling to 1~AU assuming an average dropoff with heliocentric distance as $1/r^2$. It can be seen that the heliolatitude structure during the two minima is basically similar, featuring an equatorial enhancement in density with the associated reduction in velocity (the slow wind region), and that during solar maximum the slow wind expands to all heliolatitudes (see also Fig. \ref{figUlysses2IPSfast} and discussion of time scales in the variation of the solar wind structure). However, the region of slow wind seems to reach farther in heliolatitude during the last solar minimum than during the minimum of 1995, which, interestingly, is much less conspicuous in density. A striking feature is a strong reduction in flux, reported already by \citet{mccomas_etal:08a, ebert_etal:09a}. The reduction in flux is visible as a continuous trend from the 1995 minimum through the 2002 maximum until present. Another interesting trend is a variation of $\sim 1$~km~s$^{-1}$~deg$^{-1}$ in the fast polar solar wind, discovered by \citet{mccomas_etal:00a} and expanded upon by \citet{ebert_etal:09a}.

\subsection{Retrieval of solar wind evolution: introduction}
In this section we present an attempt to retrieve the history of evolution of solar wind density and speed during the past solar cycle to provide the modelers working within the FONDUE group with a tool to enable them to develop a model of distribution of neutral interstellar hydrogen [see Izmodenov et al., this volume] that could be used as background for inter-calibration of various sets of UV observations. We attempt to use all relevant datasets, paying special attention to absolute calibrations and possible biasing. As it will be shown in the further part of this section, this is still the work in progress. Since an essential part of information must be drawn from a careful interpretation of the Lyman-alpha helioglow observations from SWAN/SOHO, which are subject to improved absolute calibration (which is one of the goals of the FONDUE group), the effort of constructing a homogeneous set of solar wind parameters must be an iterative one.

\begin{figure}
\centering
\includegraphics[scale=0.7]{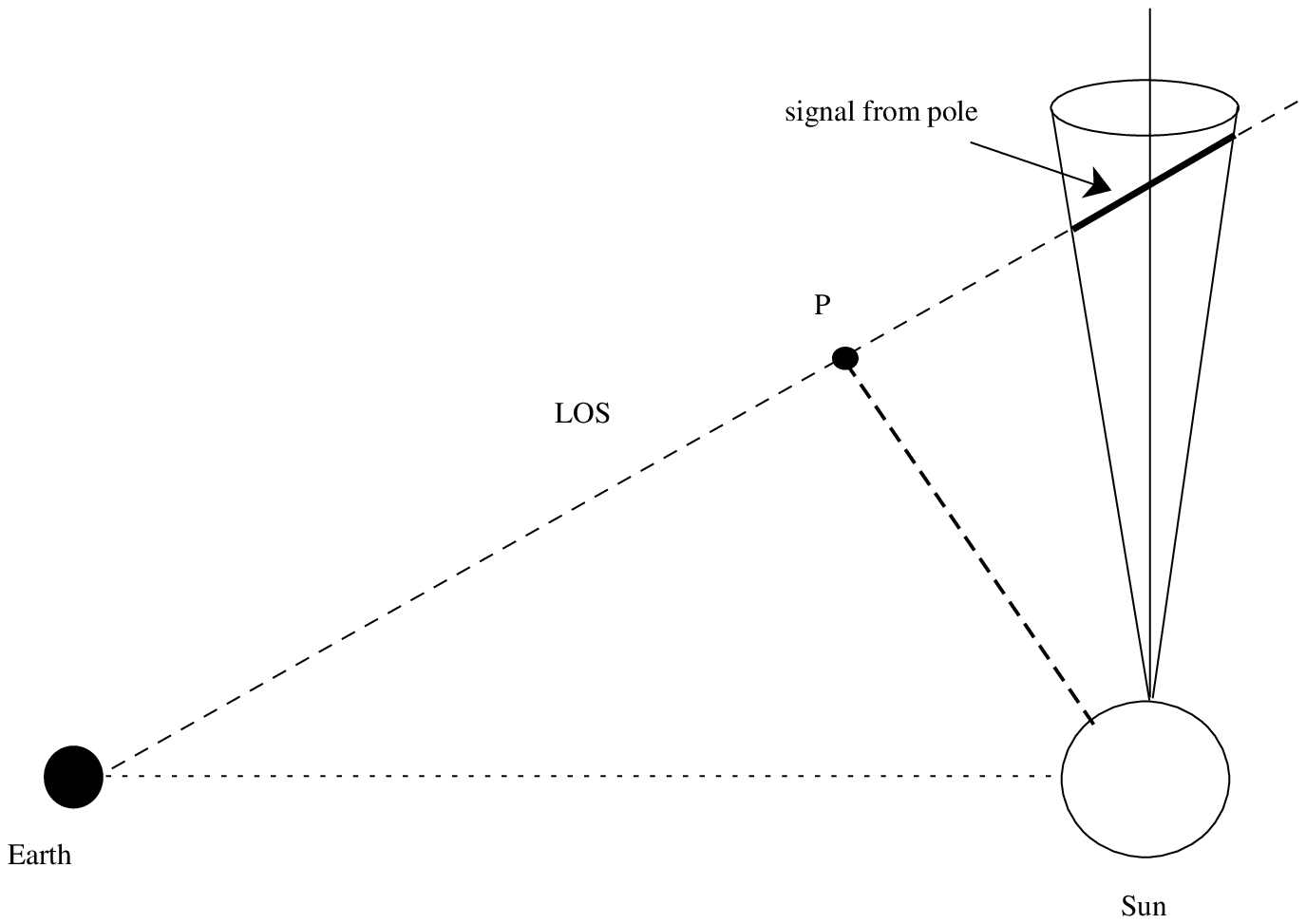}
\caption{Illustration of geometry of the line of sight (LOS) in an attempt of remote sensing observation of a solar polar region. The observer is close to the Earth in the ecliptic plane and aims its instrument (e.g., a radio telescope antenna or a Lyman-alpha photometer) at a target so that the line of sight crosses a cone with a small opening angle centered at the pole. The signal is collected from the full length of the line of sight, but the contributions from various parts are different and depend on the observations technique. In the case of IPS observations, the strongest contribution to the signal is from the point nearest to the Sun along the line of sight, marked with P, because the source function of the scintillation signal drops down with the square of solar distance. In the case of helioglow observations, the maximum of the signal comes from the so-called Maximum Emissivity Region, which is located within 1.5 -- 5~AU from the Sun, so it is important to carefully select the solar elongation of the line of sight. Note that the angular area of the polar region is quite small, so with an observations program that maps the entire sky only a small region in the maps indeed includes the signal from polar regions.}
\label{figLOSScheme}
\end{figure}

In the following, we will present a construction of  the procedure to retrieve the variability of solar wind speed and density in time and heliographic latitude with use of the available datasets. The procedure to retrieve the time and heliolatitude evolution of solar wind relies on the absolute calibration of the OMNI dataset both in speed and density in the ecliptic plane. Out of ecliptic, the baseline is the absolute calibration of Ulysses measurements and interplanetary scintillation observations (IPS). The prime source of information on the heliolatitude structure of solar wind are observations of IPS, interpreted by tomography modeling, that generally agree quite well with the Ulysses in situ measurements out of ecliptic and with the OMNI measurements in the ecliptic. Up to now, no continuous measurements of solar wind density as a function of ecliptic latitude have been available. 

Tests revealed that an appropriate balance between the latitudinal resolution of the coverage and fidelity of the results is obtained at a subdivision of the heliolatitudes' range into 10-degree bins. Concerning the time resolution, the most welcome would be Carrington rotation averages, identical with the resolution of the photoionization rate and Lyman-alpha flux. Regrettably, such a high resolution seems to be hard to achieve because (1) the time coverage in the data from IPS has gaps that typically occur during $\sim 4$ months at the beginning of each year, which would induce an artificial 1-year periodicity in the data, and (2) the fast latitude scans by Ulysses were about 12 months long and hence differentiating between time and latitude effects in its measurements is challenging. Thus a reliable latitude structure of solar wind can only be obtained at a time scale of 1~year and this is the time resolution of the model we are going to present. 

Concerning the global mapping of solar wind parameters from ecliptic it has to be pointed out that the accuracy of measurements of solar wind parameters decreases with latitude because of geometry reasons. The polar values are the most uncertain (and possibly biased) because the signal in the polar lines of sight is only partly formed in the actual polar region of space, which can be understood from a sketch presented in Fig. \ref{figLOSScheme}. 

In the following part of this chapter, we will present a procedure to retrieve the solar wind speed evolution in time and heliolatitude, and subsequently two procedures of retrieval of solar wind density. One is based on a correlation between the solar wind speed and density that was established from the three Ulysses fast latitude scans and must be regarded as an interim solution, to be used until the other one, based on the SWAN Lyman-alpha helioglow observations, will be available.

\subsection{Latitude profiles of solar wind velocity from Interplanetary Scintillation observations}

InterPlanetary Scintillations (IPS) observations carried out by the Solar-Terrestrial Environment Laboratory (STEL) of Nagoya University (Japan) enable us to derive the latitude structure of solar wind speed and its variations in time. We use in the analysis data from 1990 to 2009, obtained from 3 antennas (Toyokawa, Fuji and Sugadaira), and from another antenna (Kiso) since 1994. The 4-antenna system was operated until 2005, when the Toyokawa antenna was closed \citep{tokumaru_etal:10a}; since then the system again was operated in a 3-antenna setup. The IPS data from STEL are typically collected on a daily basis during 11 Carrington rotations per year: there is a break in winter because the antennas get covered with snow. The IPS observations are line of sight integration of solar wind speed weighted by density turbulances and Fresnel filter. Each day 30-40 lines of sight for selected scintillating radio sources are observed. The line of sight integration effect is deconvolved using the Computer Assisted Tomography (CAT) method developed by the STEL group \citep{kojima_etal:07a, kojima_etal:98}. The LOS's are projected on the source surface at 2.5 solar radii $\left(\mathrm{R_{\odot}}\right)$, which is used as a reference surface in time sequence tomography. The latitude coverage of the sky by IPS observations is not uniform and is strongly correlated with the Sun position on sky, which changes during the year, and with the target distribution on sky, with relatively few of them near the solar poles. Additionaly, the observations of the south pole are of lower quality than these of the north pole because of the low elevation of the Sun during winter in Japan. The original latitude coverage was improved owing to the new antenna added to the system in 1994 and by optimization of the choice of the targets. 

For our analysis we take solar wind speed from 1990 to 2009, mapped at the source surface on a grid of $11 \times 360 \times 180$ records per year, which correspond to a series of Carrington rotations. The data are organized in heliolatitude from $89.5\deg$ North to $89.5\deg$ South. 

A comparison of the tomography-derived solar wind speed with the in situ measurements by Ulysses we performed showed that the accuracy of the tomographic results depends on the number of IPS observations available for a given rotation. The intervals with a small number of data points clearly tend to underestimate the speed. Consequently, we removed from the data the Carrington rotations with the total number of points less than 30\,000. Small numbers of available observations typically happen at the beginning and at the end of the year and at the edges of data gaps. The selection of data by the total number of points per rotation constrainted the data mainly to the summer and autumn months, when all latitudes are well covered.

\begin{figure*}[t]
		\centering
		\begin{tabular}{ccc}		
\includegraphics[scale=0.4]{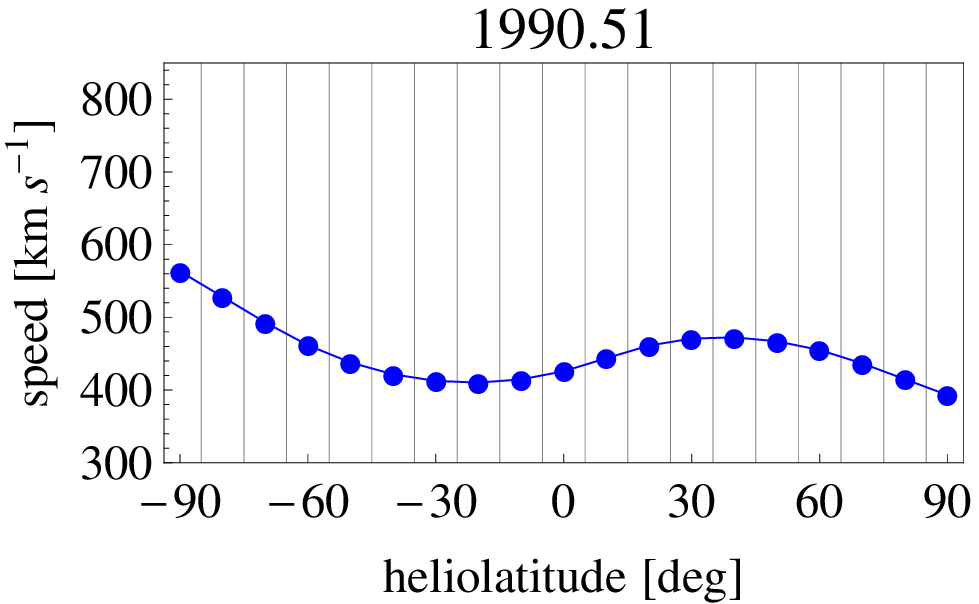}&\includegraphics[scale=0.4]{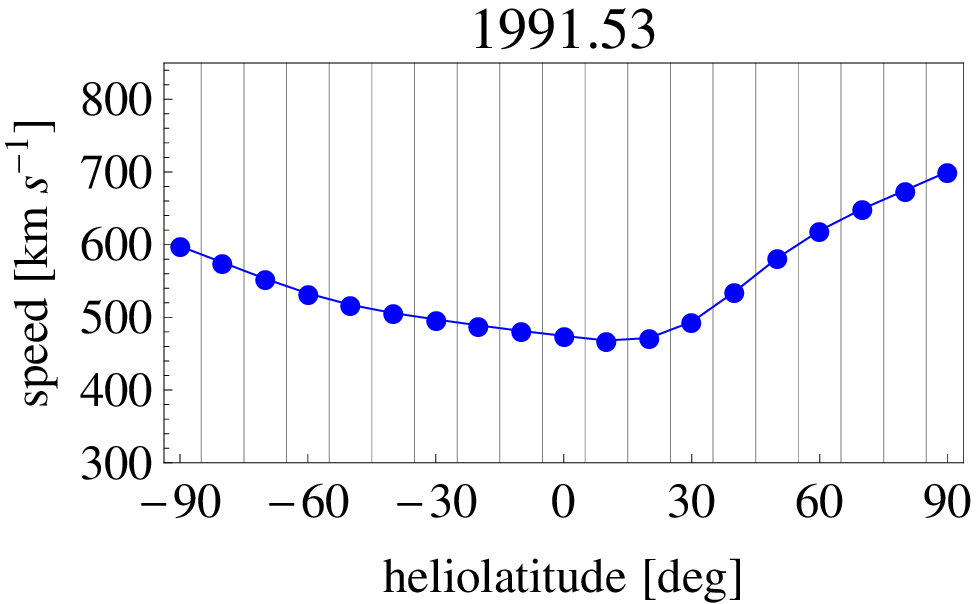}&\includegraphics[scale=0.4]{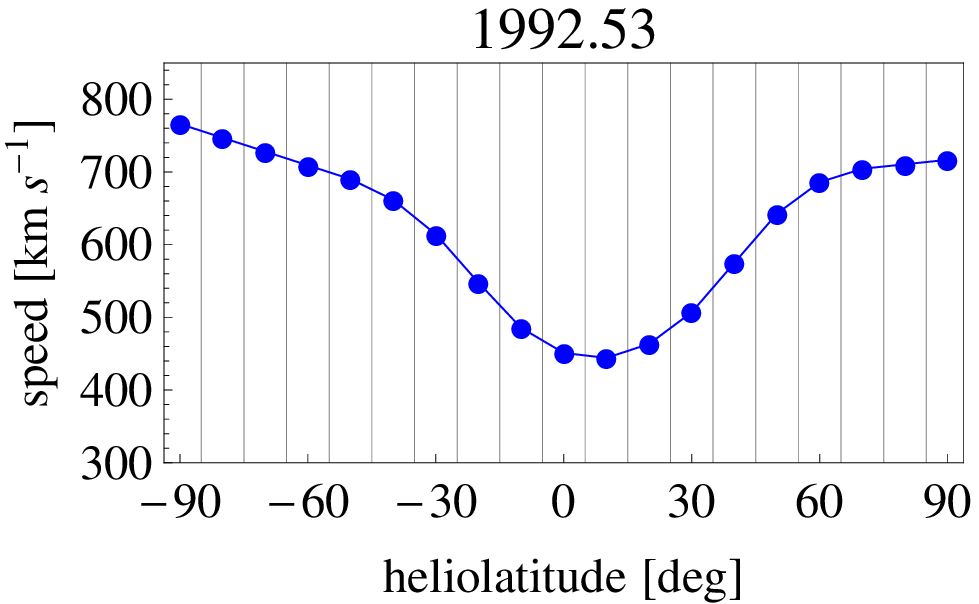}\\	\includegraphics[scale=0.4]{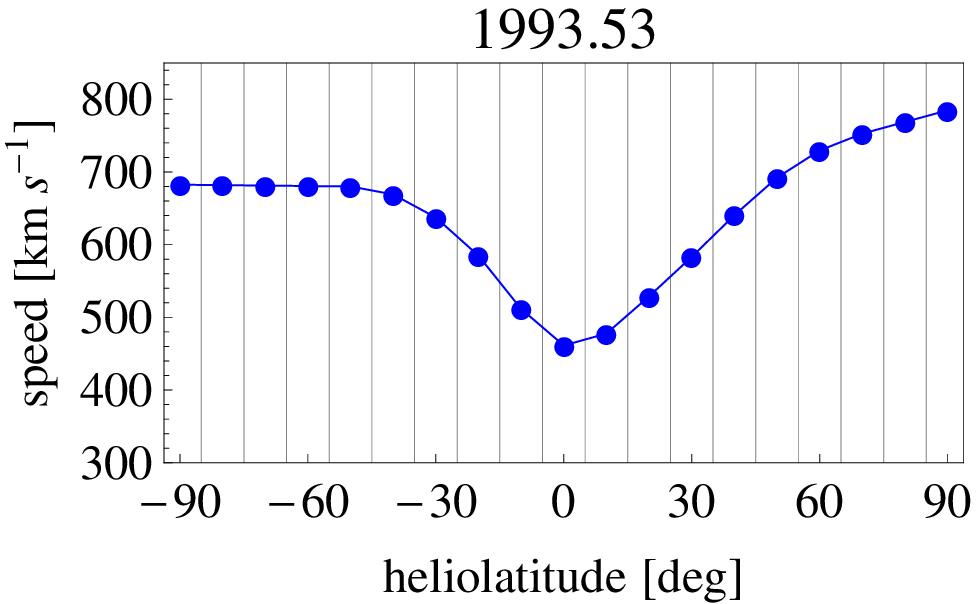}&\includegraphics[scale=0.4]{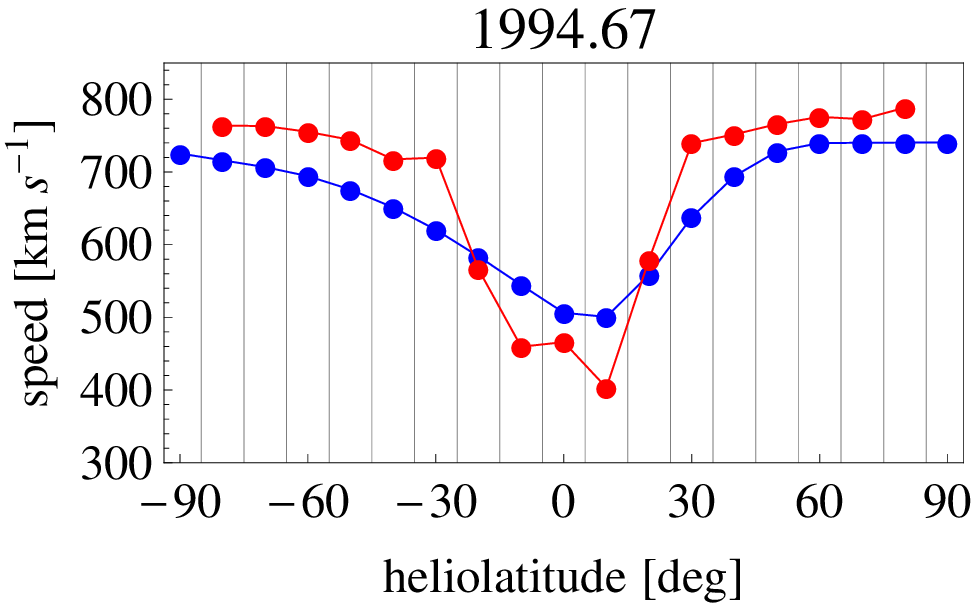}&\includegraphics[scale=0.4]{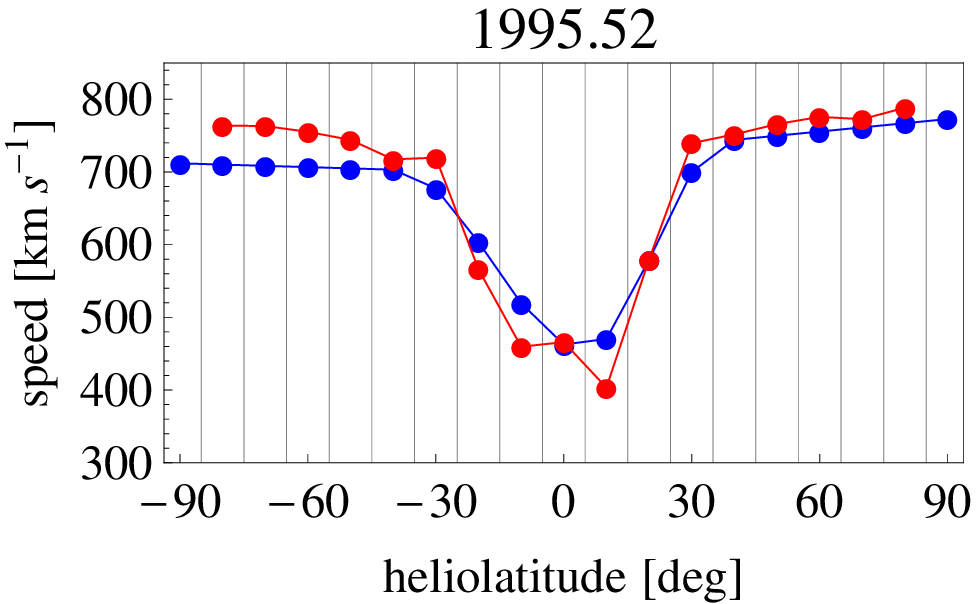}\\	\includegraphics[scale=0.4]{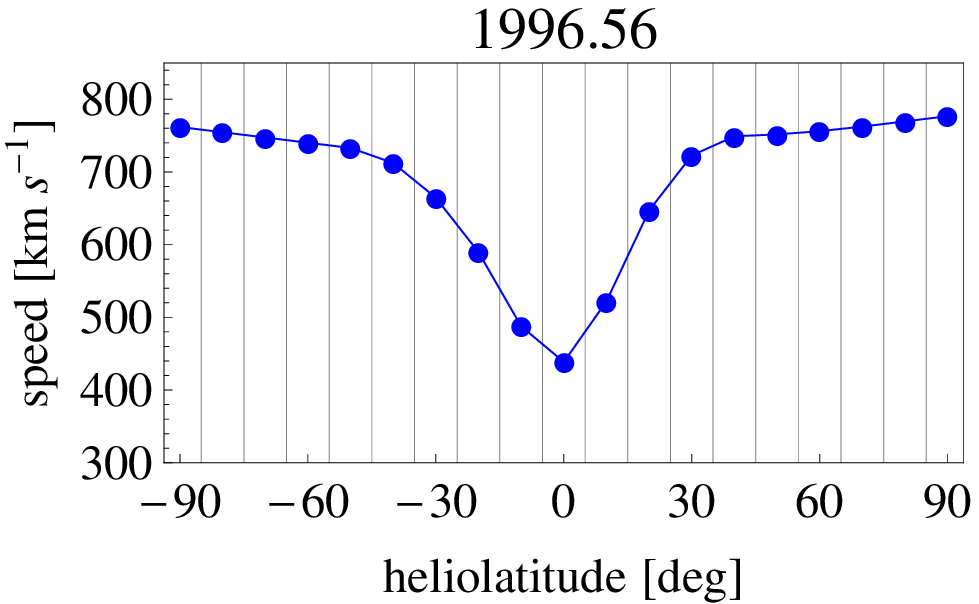}&\includegraphics[scale=0.4]{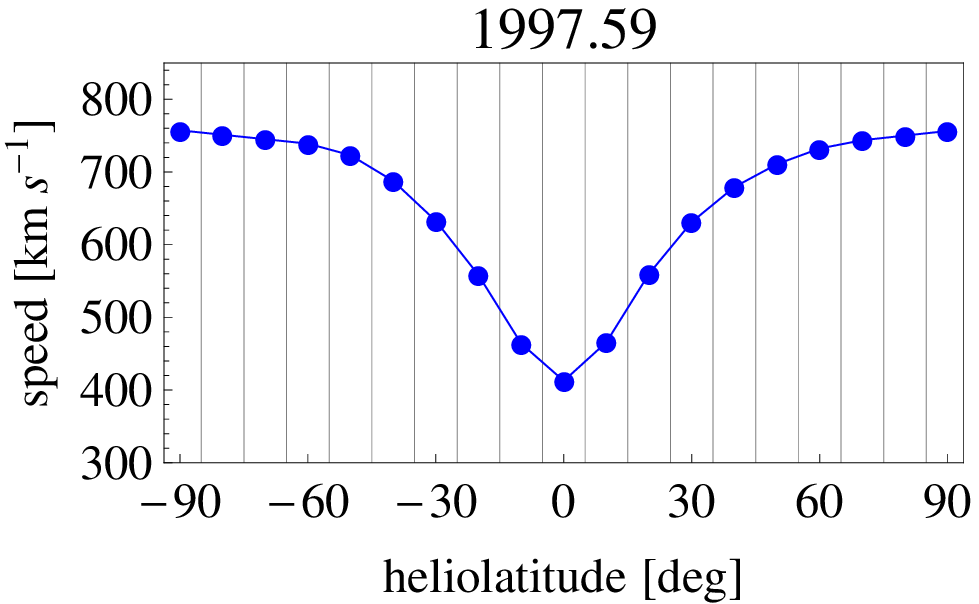}&\includegraphics[scale=0.4]{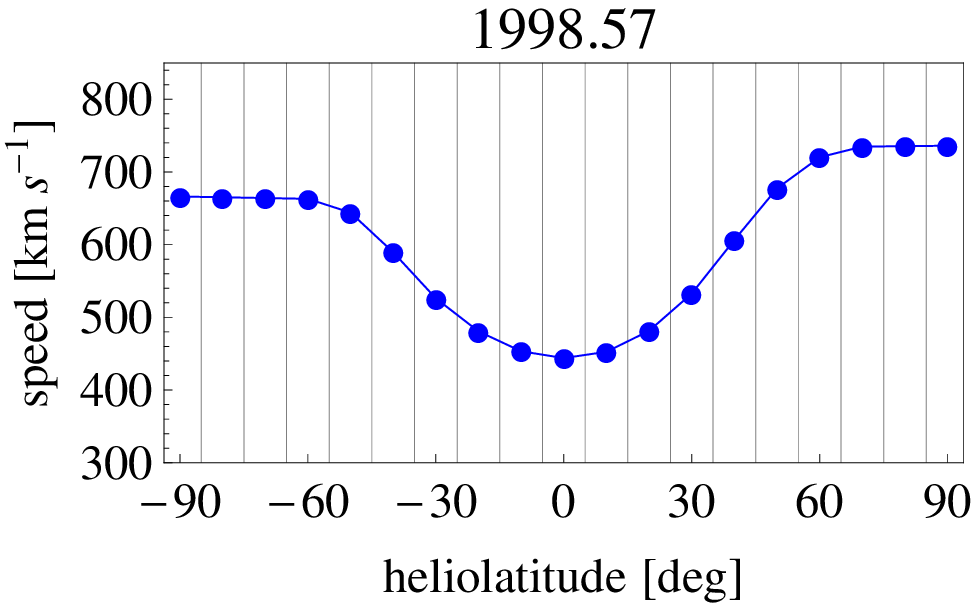}\\	\includegraphics[scale=0.4]{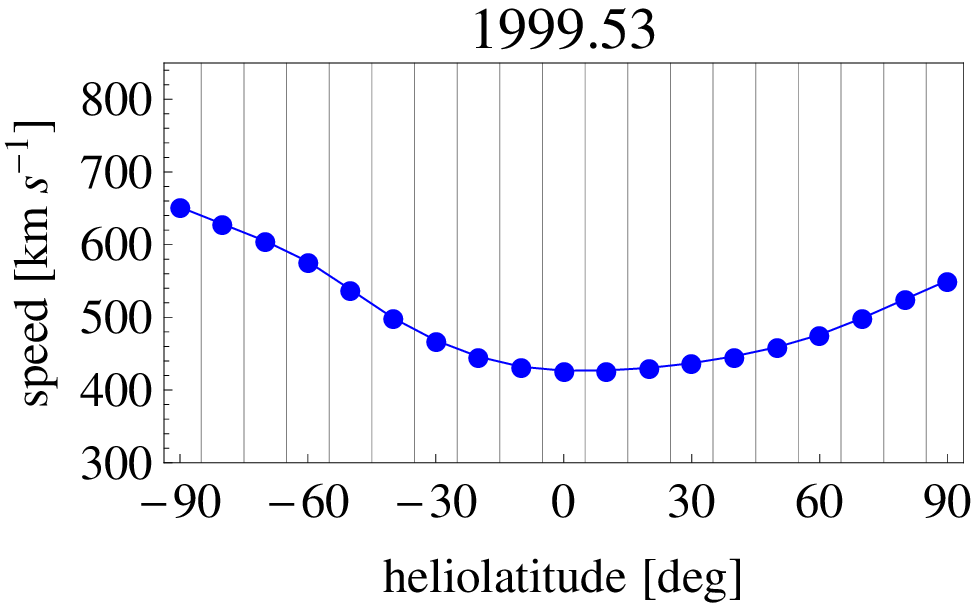}&\includegraphics[scale=0.4]{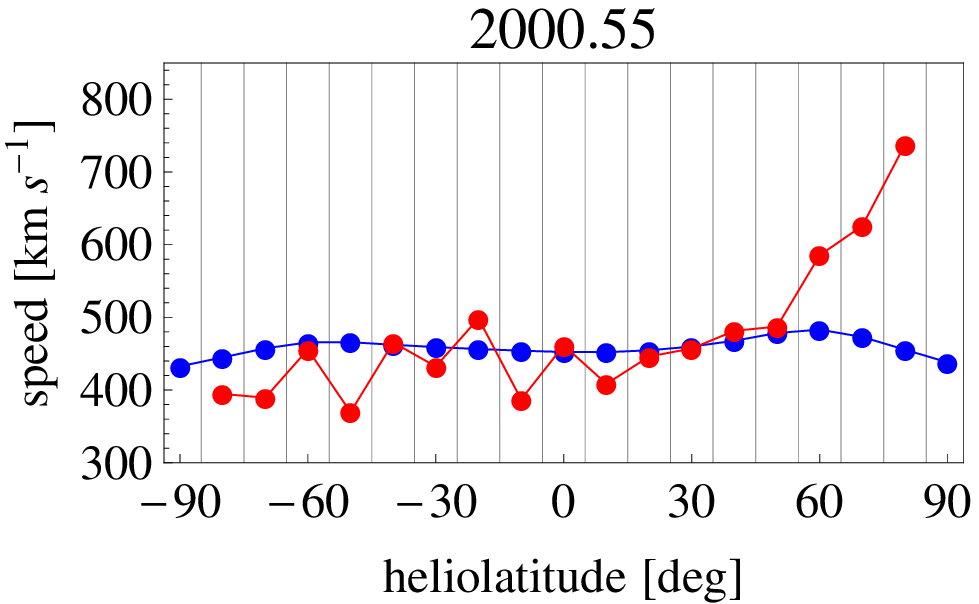}&\includegraphics[scale=0.4]{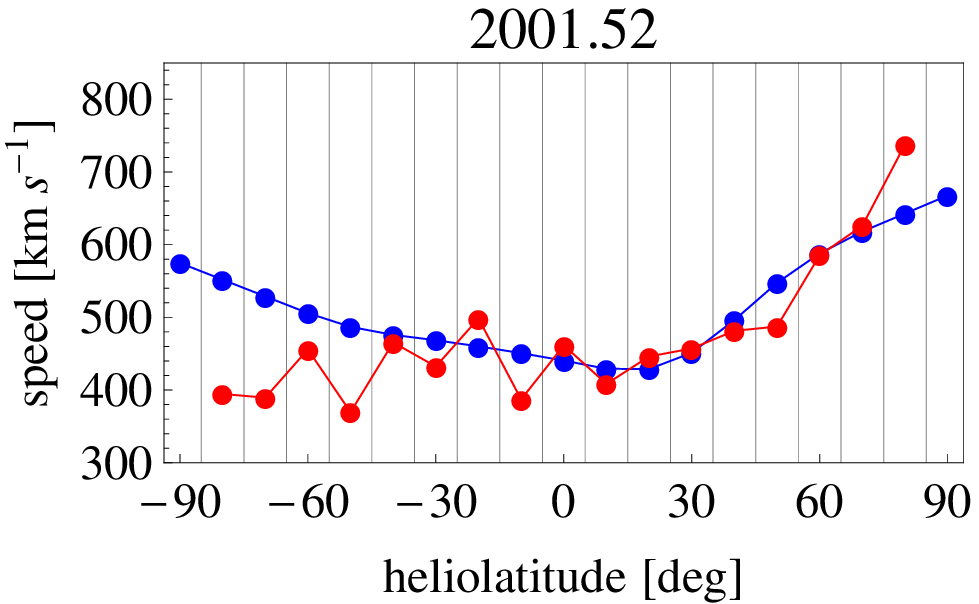}\\	\includegraphics[scale=0.4]{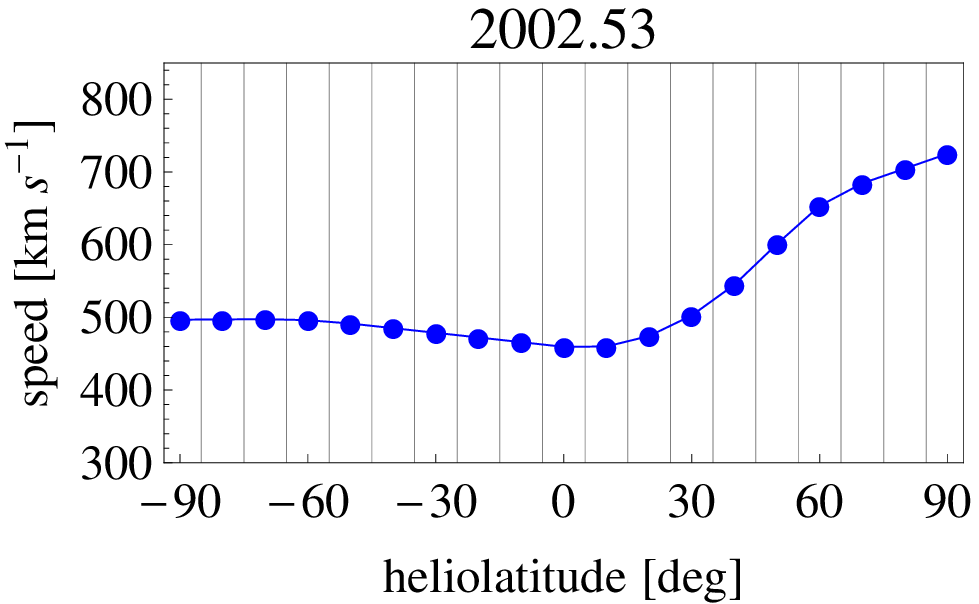}&\includegraphics[scale=0.4]{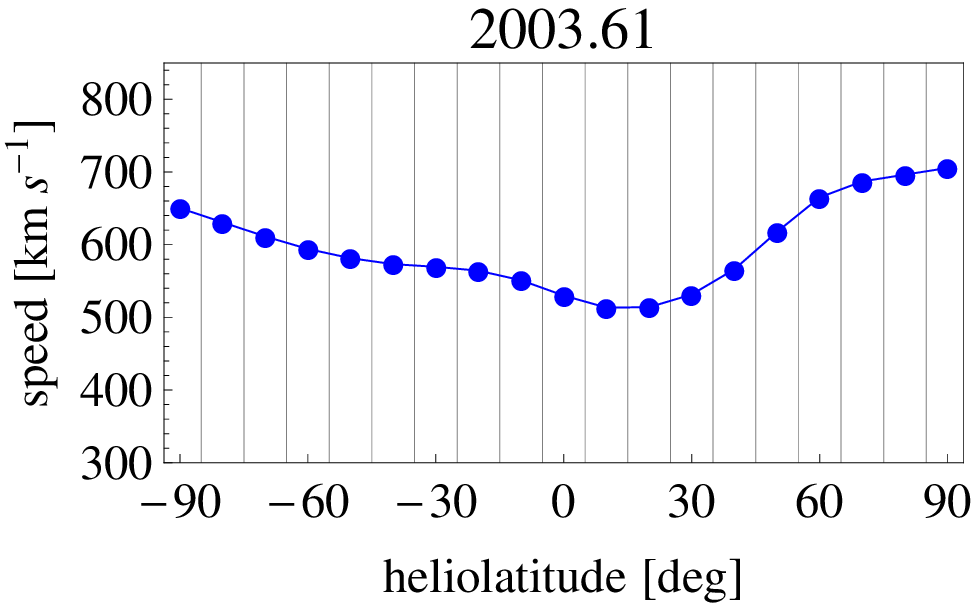}&\includegraphics[scale=0.4]{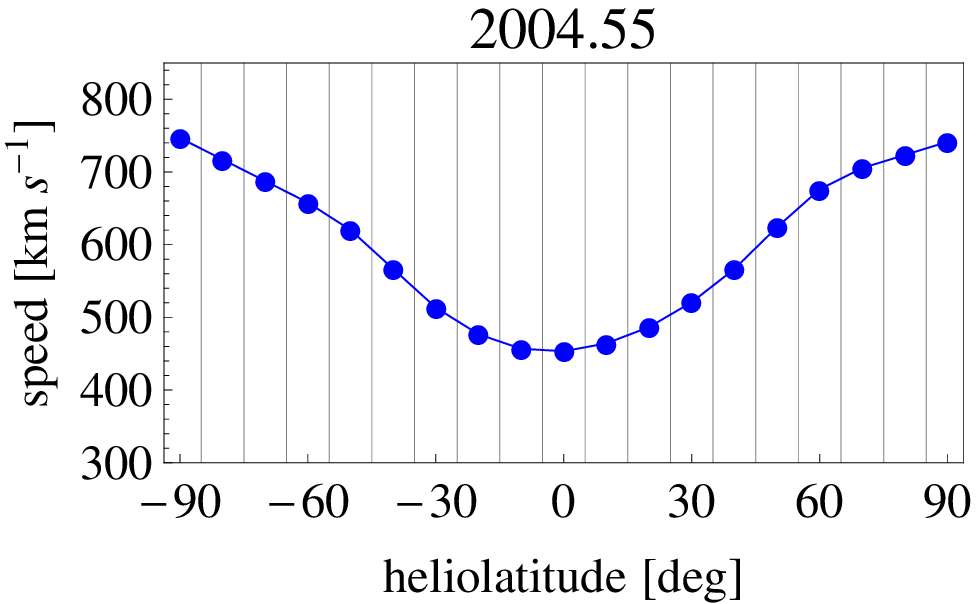}\\	\includegraphics[scale=0.4]{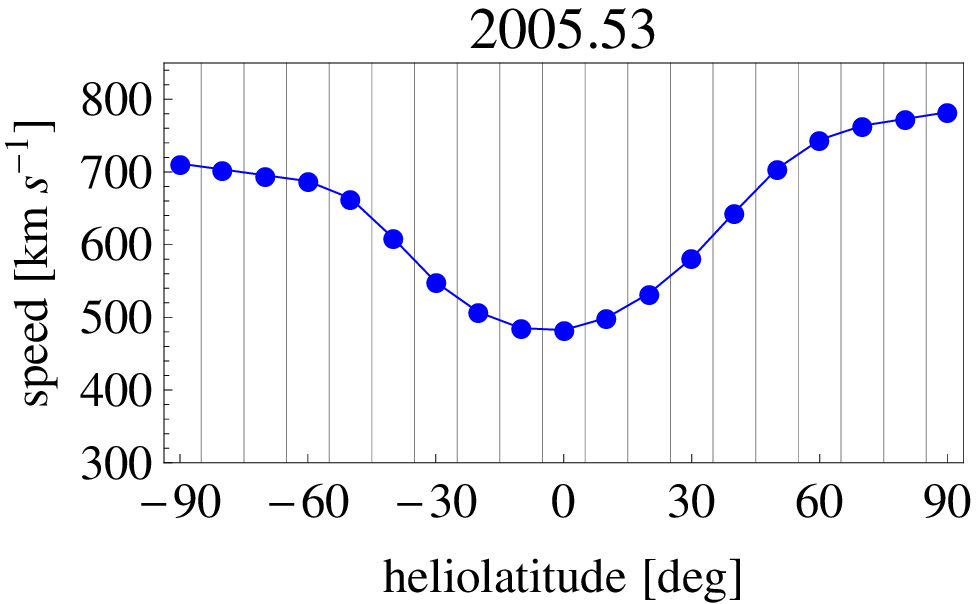}&\includegraphics[scale=0.4]{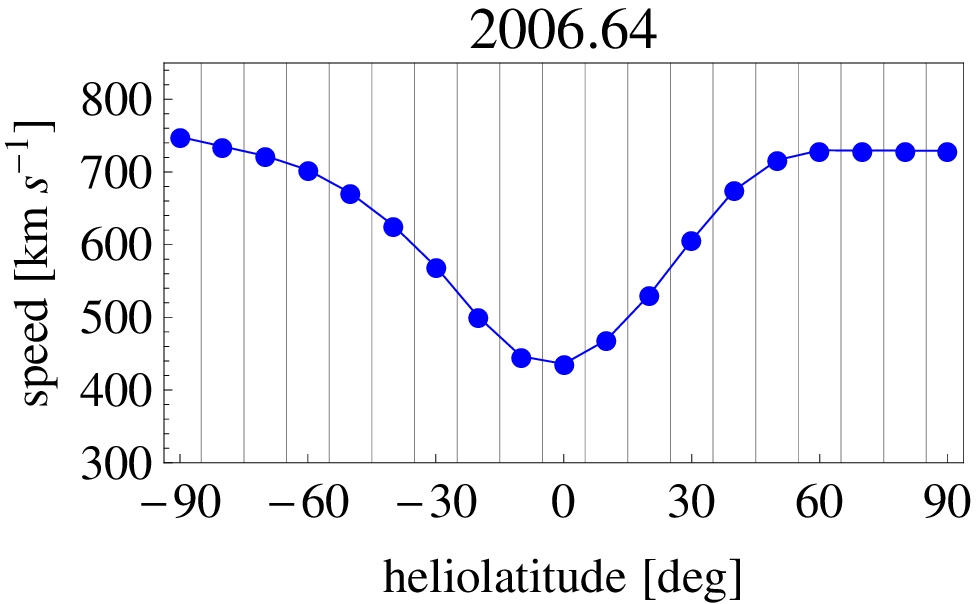}&\includegraphics[scale=0.4]{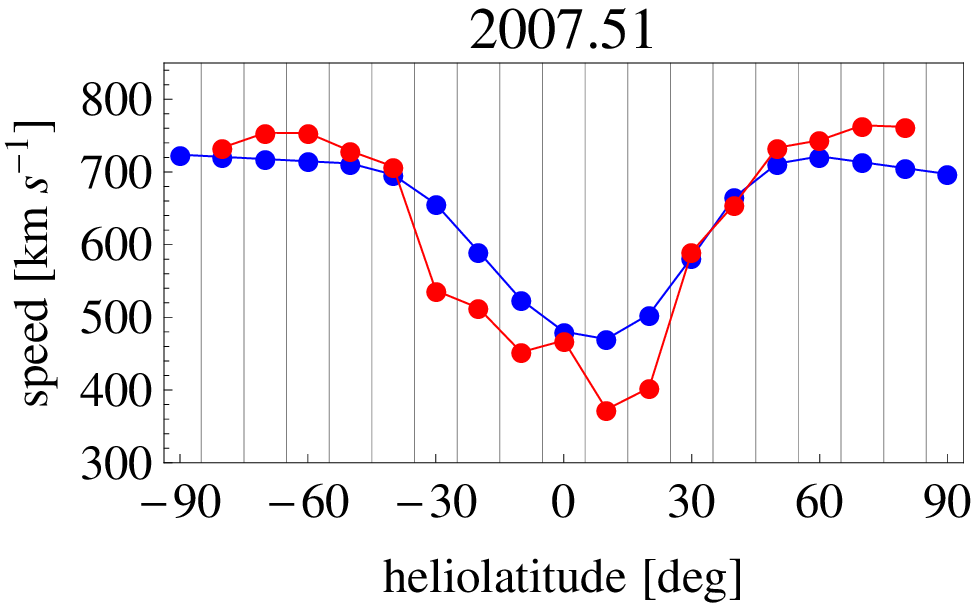}\\	
		\includegraphics[scale=0.4]{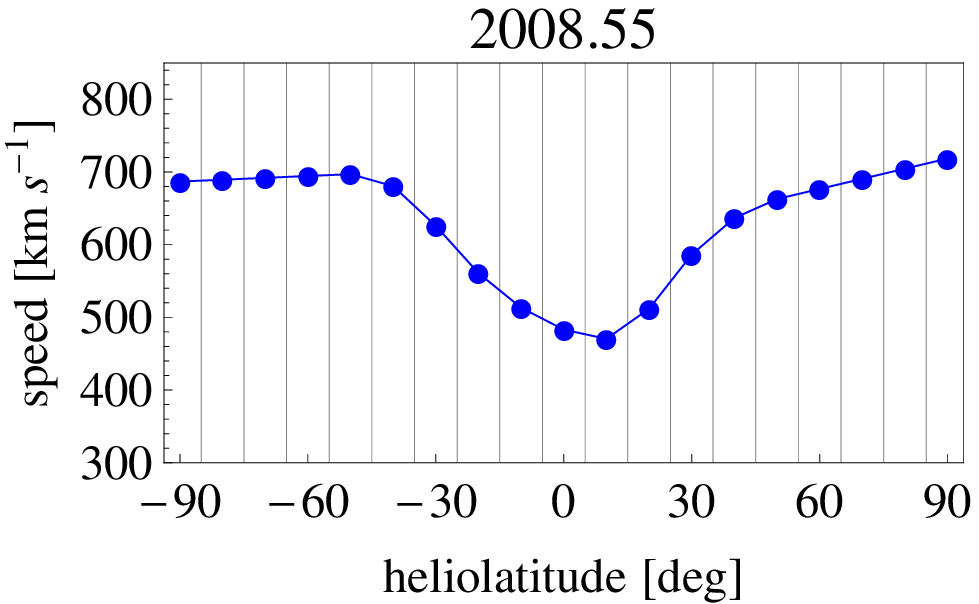}&\includegraphics[scale=0.4]{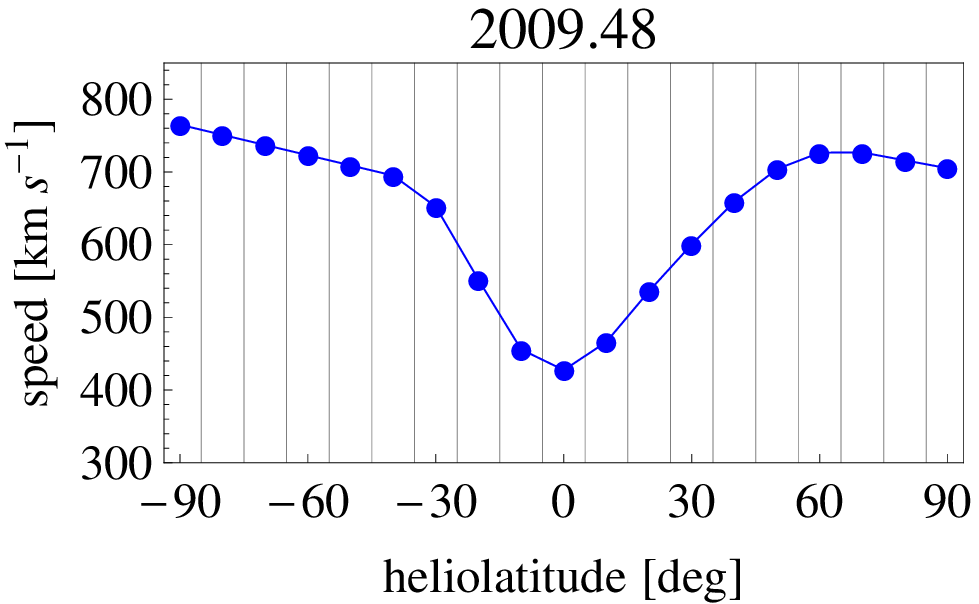}\\
		\end{tabular}
		\caption{Heliolatitude profiles of yearly averaged solar wind speed obtained from the Computer Assisted Tomography analysis of interplanetary scintillation observations \citep{tokumaru_etal:10a}.}
		\label{figIPSprofiles}
		\end{figure*}

The selected subset of data was split into years, and within each yearly subset into 19 heliolatitude bins, equally spaced from $-90\deg$ to $90\deg$. The speeds averaged over bins and over year for the latitudinal bins yield the yearly latitudinal profiles of solar wind speed, shown in Fig.\ref{figIPSprofiles}. They cover a half of solar cycle 22 and full solar cycle 23. In the analysis we use the two-step calculation, first Carrington rotation averaged values per bin and next the yearly averaged calculated from the monthly ones. It is worth to note that the solar wind profiles for specific Carrington rotations have a very similar shape to the yearly profiles, which suggests that the latitude structure is stable during a year and changes only on a time scale comparable with solar activity variations.   

		\clearpage
		
The results confirm that solar wind speed is bimodal during solar minimum, slow at latitudes close to solar equator (and thus the ecliptic plane) and fast at the poles. The latitude structure evolves with the solar acitivity cycle and becomes flatter when the activity is increasing. The structure is approximately homogeneous in latitude only during a short time interval during the peak of solar maximum, when at all latitudes solar wind is slow (see the panel for 2000 in Fig.~\ref{figIPSprofiles}). Shortly after the activity maximum the bimodal structure reappears and the fast wind at the poles is observed again, but switchovers from the slow to fast wind at the poles may still occur during the high activity period: compare the panels for 2001 and 2003 in the aforementioned figure and see the solar activity level depicted with the blue line in Fig.~\ref{figUlyssesComposite}. During the descending and ascending phases of solar activity there is a wide band of slow solar wind that straddles the equator and extends to midlatitudes; the fast wind is restricted to polar caps and upper midlatitudes. At solar minimum, the structure is sharp and stable during a few years straddling the turn of solar cycles, with high speed at the poles and at midlatitudes and a rapid decrease at the equator. Thus, apart from short time intervals at the maximum of solar activity, the solar wind structure close the poles is almost flat, with a steady fast speed value typical for wide polar coronal holes, which is in perfect agreement with the measurements from Ulysses \citep{phillips_etal:95a, mccomas_etal:00a, mccomas_etal:06b}.

To verify the results obtained from the IPS analysis, we compared them with the data from the three Ulysses fast latitude scans and with the OMNI measurements in the ecliptic.

\begin{figure}
		\centering
		\begin{tabular}{ccc}		
\includegraphics[scale=0.6]{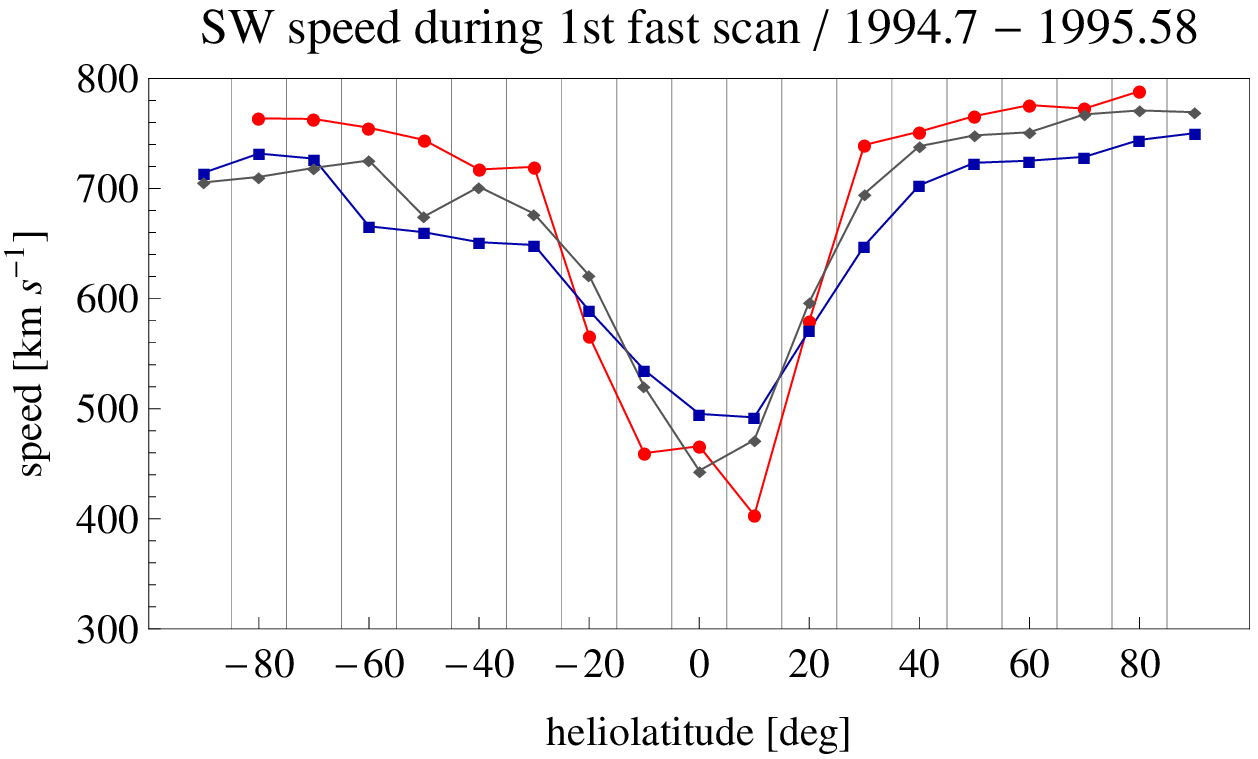}\\
\includegraphics[scale=0.6]{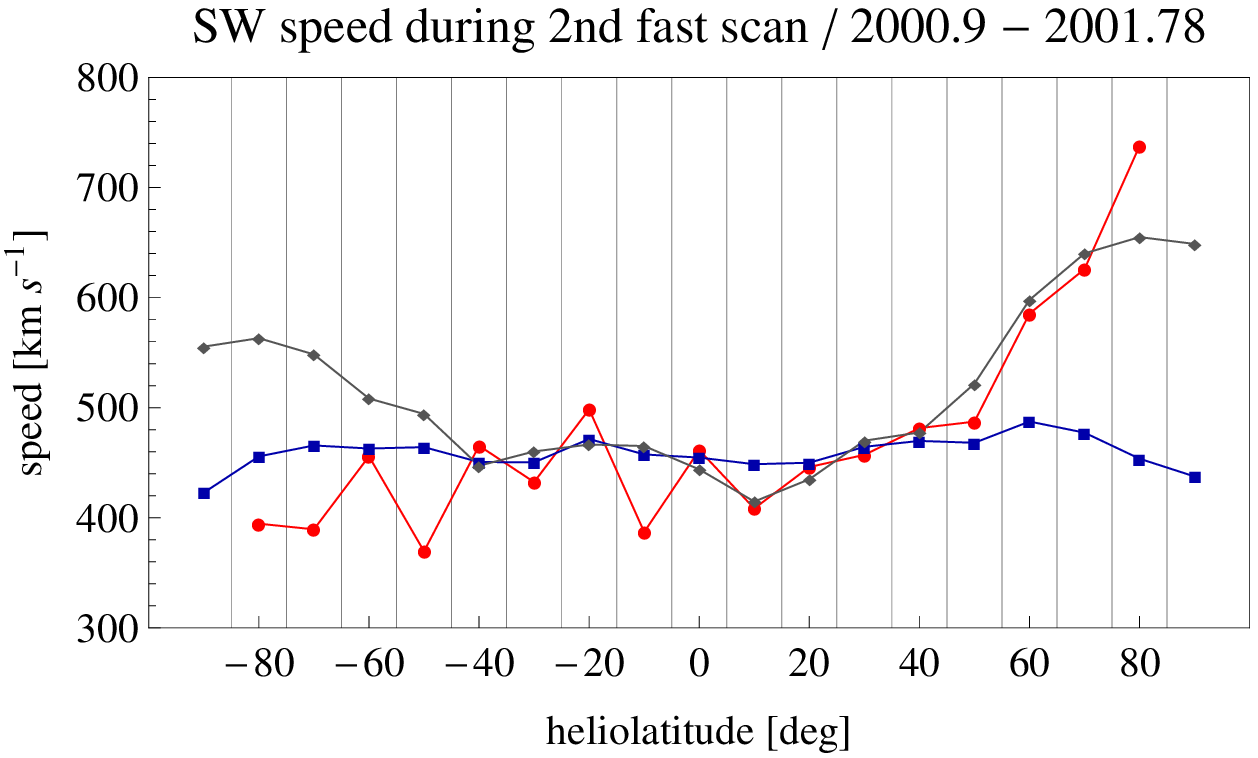}\\
\includegraphics[scale=0.6]{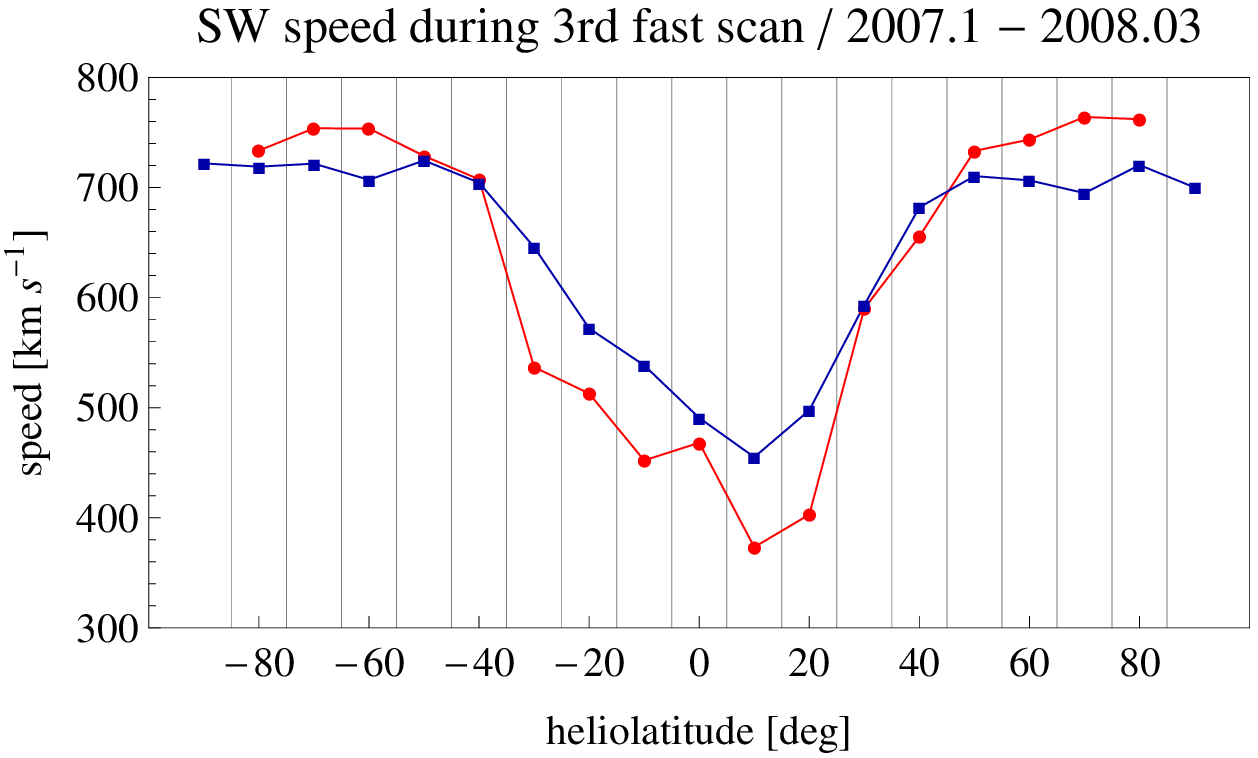}\\	
		\end{tabular}
		\caption{Solar wind speed profiles from Ulysses measurements and IPS observations for the 3 fast Ulysses scans. Red: Ulysses, blue: IPS during the year of beginning of a Ulysses fast scan, gray: IPS during the year of the end of Ulysses fast scan. Top panel: the first fast scan during solar minimum, middel panel: the second fast scan during solar maximum, bottom panel: the third fast scan during minimum.}
		\label{figUlysses2IPSfast}
		\end{figure}

The Ulysses velocity profiles used for this comparison were constructed from subsets of hourly averages available from NSSDC, split into identical heliolatitude bins as those used for the IPS data analysis and averaged. They are shown in Fig.~\ref{figUlysses2IPSfast} as red lines. Since the acquisition of the Ulysses profiles took one year each and the fast scans straddle the turn of the years, we show the IPS results for the years straddling the fast latitude scans; they are presented in blue and gray in Fig.~\ref{figUlysses2IPSfast}. 

The ``grass-line'' feature in the Ulysses profile for the second fast scan and in the equatorial part of the first and third fast scans is due to the time Ulysses was taking to scan the 10-degree bins. The fast scans were performed at the perihelion half of the Ulysses elliptical orbit, with the perihelion close to the solar equator plane. Hence, the angular speed of its motion was highest close to equator and traversing the 10-degree bin took it less than one solar rotation period. Thus the ``grass'' is an effect of incomplete Carrington longitude coverage of the bimodal solar wind, with slow wind interleaved with fast wind streams. Near the poles the angular speed was slower and it took more than 1 Carrington rotation to scan the 10-deg bin and during solar maximum, when the gusty slow wind engulfed the whole space, the ``grass effect'' expanded into the full latitude span. By contrast, during the low-activity scans the solar wind speed at high latitude was stable, which resulted in the lack of the small-scale latitude variations in the Carrington rotation averages at high latitudes.

		\clearpage

The profiles of solar wind speed from IPS and Ulysses are very similar, but some systematic differences exist. On one hand, it seems that a $\sim 50$~km~s$^{-1}$ is a typical difference between Ulysses and IPS values in the polar regions, with the northern region usually in  a better agreement than the southern. On the other hand, the agreement is sometimes almost perfect. 

The difference between the blue and gray lines in the top and middle panels of Fig. \ref{figUlysses2IPSfast} is a measure of true variation of the latitudinal profile of solar wind speed during one year. Ulysses was moving from south to north during the fast latitude scans, so the south limb of the profile from Ulysses ought to be closer to the south limb of the blue profile obtained from the IPS analysis, while the north limb of the Ulysses profile should agree better with the north limb of the gray IPS profile. Such a behavior is observed in the second panel, which corresponds to the solar maximum in 2001. In our opinion, it is a very interesting observation because it (1) shows how rapidly the latitude structure of solar wind varies during the maximum of solar activity, and (2) it confirms the credibility of both Ulysses and IPS results on solar wind structure. Originally, the interpretation of the speed profiles obtained from Ulysses was not clear, it was pondered whether the north-hemisphere increase in solar wind speed was a long-standing feature of the solar wind or was it just a time-variability of the wind at the north pole. Similarly, it was pondered whether IPS analysis is able to credibly reproduce the solar wind profiles given the fact that some of the profiles obtained approximately at the time of the fast scan seemed to disagree with the in situ data: our analysis suggests that the yearly-aveaged velocity profiles obtained from the tomography analysis of IPS observations agree with the in-situ observations from Ulysses even at times when the solar wind is restructuring rapidly at the peak of solar activity.

\begin{figure}
\centering
\includegraphics[scale=0.7]{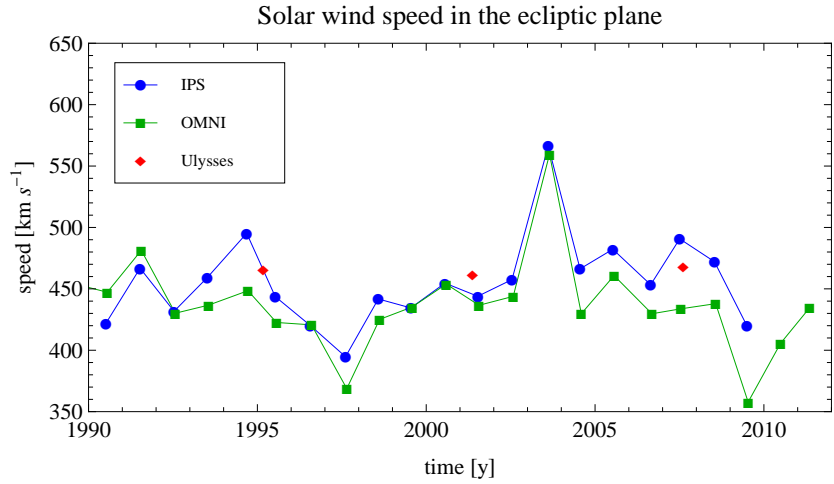}
\caption{Yearly averages of solar wind speed from the IPS analysis (blue) and in-situ measurements collected in the OMNI-2 database (green), compared with the solar wind speed measured by Ulysses during its passage through the ecliptic plane during the three fast latitude scans (red dots).}
\label{figEclSpeed}
\end{figure}

The IPS data are in a very good agreement with the OMNI data collected in the ecliptic plane (see Fig.~\ref{figEclSpeed}). Up to 2004 the agreement is almost perfect, afterwards small differences appear. The agreement is better than with the in-cliptic Ulysses measurements from the fast latitude scans. This, in our opinion, is because the measurements of solar wind parameters at the ecliptic plane obtained during the fast latitude scans from Ulysses are challenging to directly compare with the OMNI and IPS measurements because Ulysses was passing ecliptic 10-degree latitude bin in a time equal to about half of the Carrington rotation and thus a reliable longitude averaging of the solar wind parameters could not be obtained.

Thus the IPS solar wind speed profiles provide a solar wind latitude structure that can be adopted as an interim solution. They agree well both with the OMNI time series in the ecliptic and with the Ulysses measurements out of ecliptic for the time intervals when they can be compared directly. 
		\begin{figure*}[t]
		\centering
		\begin{tabular}{ccc}		
\includegraphics[scale=0.4]{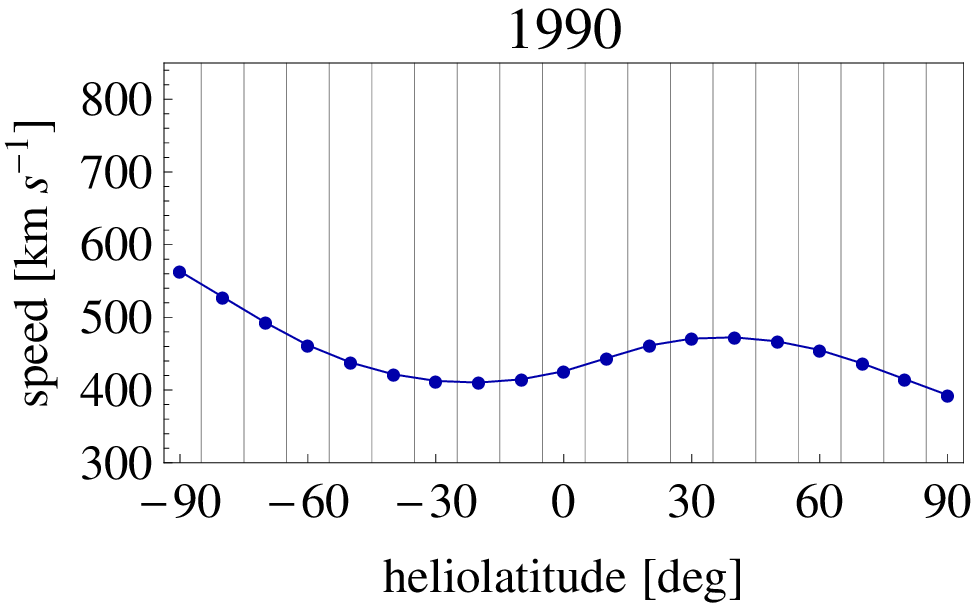}&\includegraphics[scale=0.4]{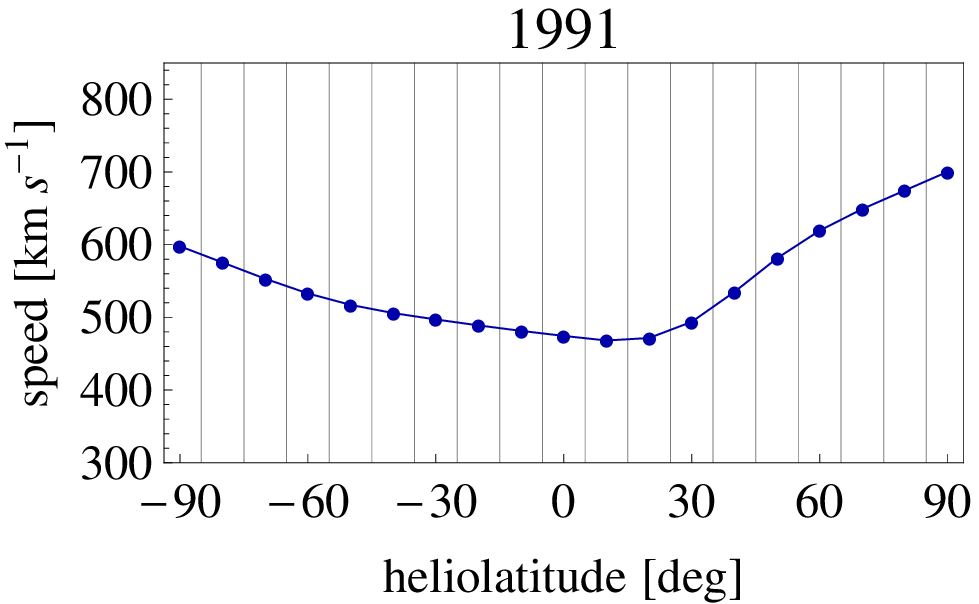}&\includegraphics[scale=0.4]{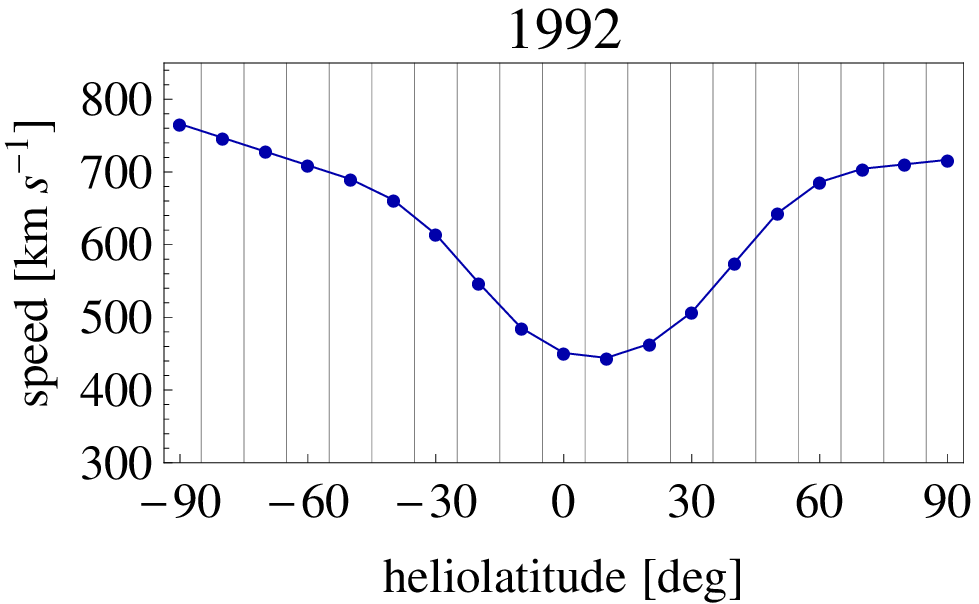}\\	\includegraphics[scale=0.4]{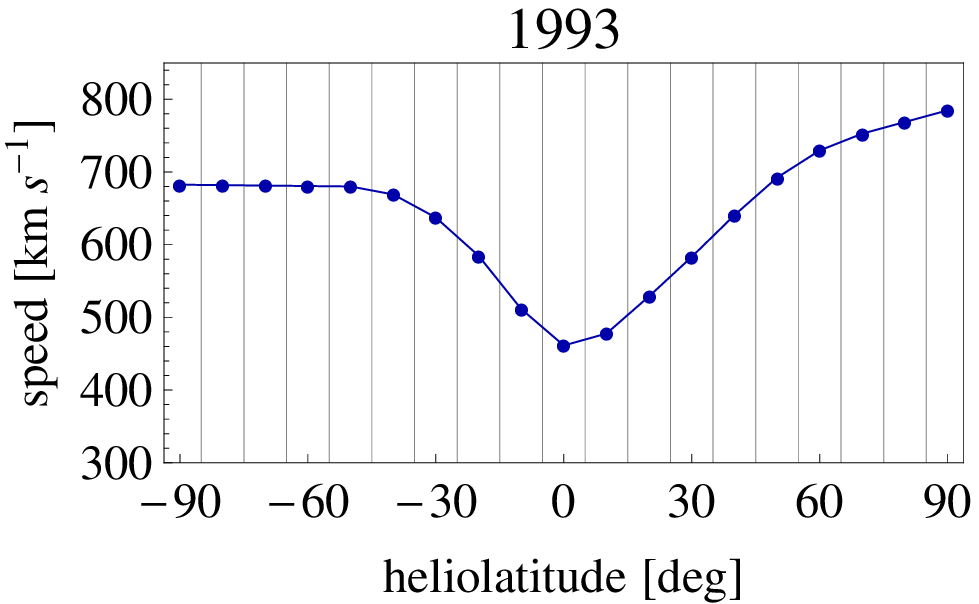}&\includegraphics[scale=0.4]{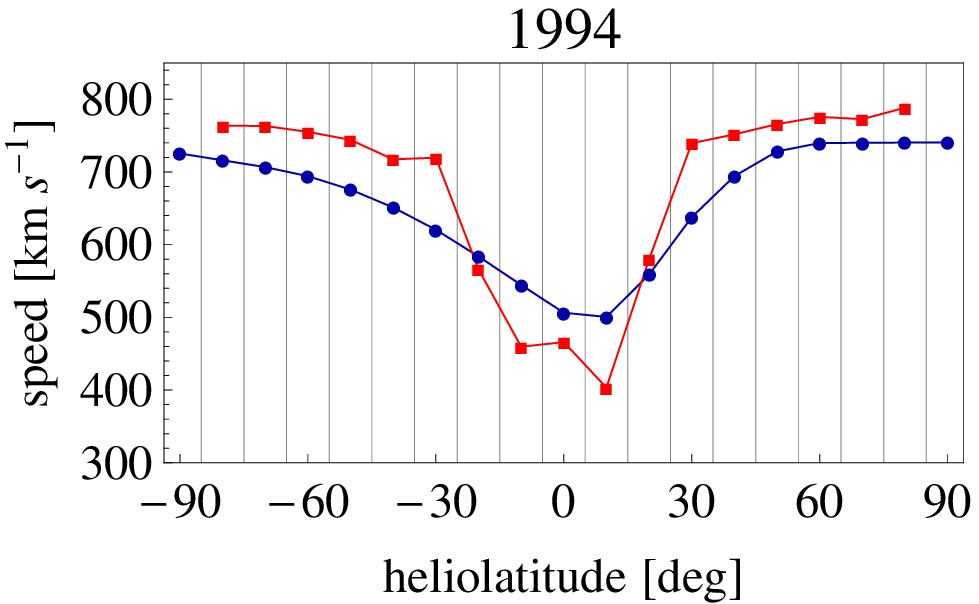}&\includegraphics[scale=0.4]{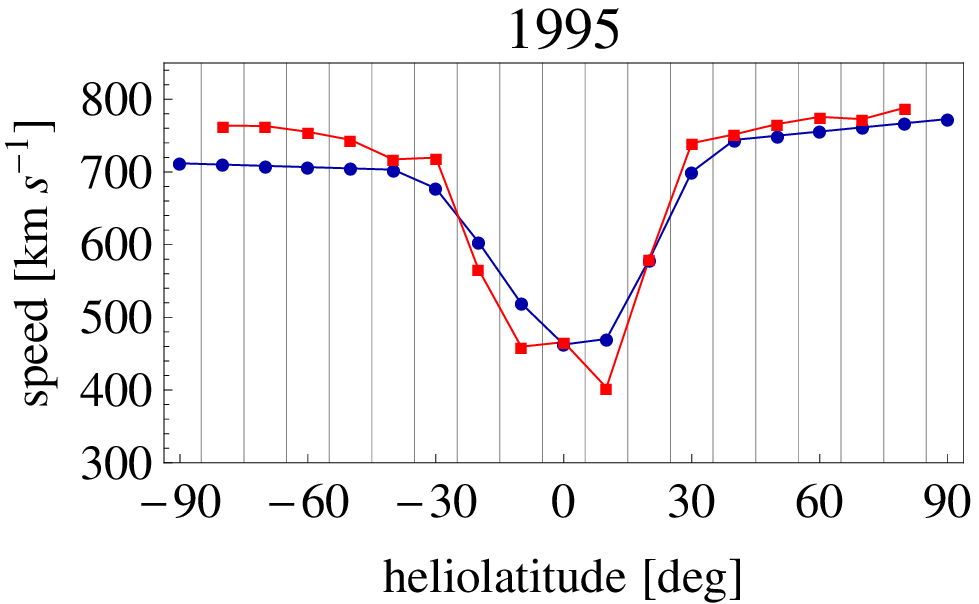}\\	\includegraphics[scale=0.4]{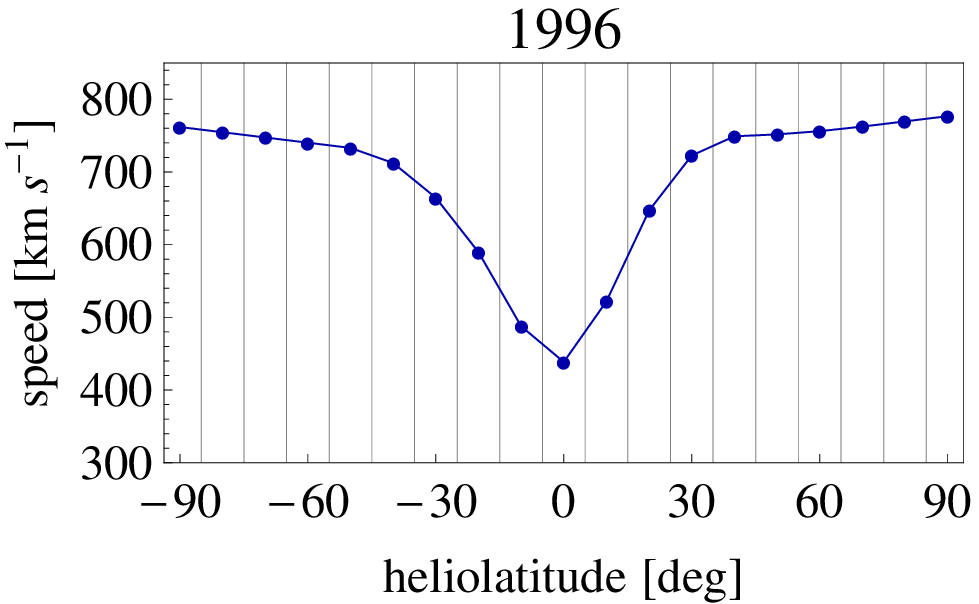}&\includegraphics[scale=0.4]{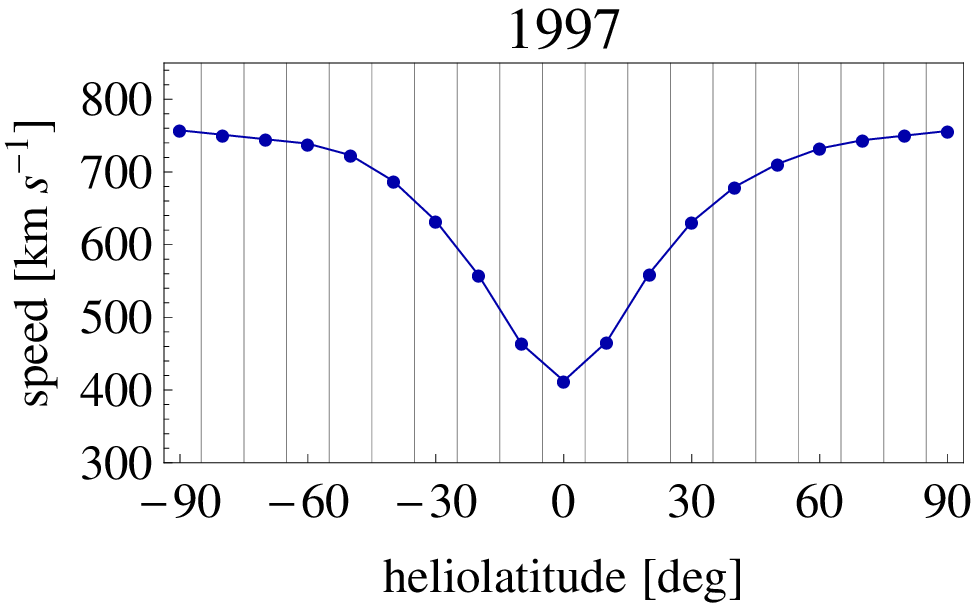}&\includegraphics[scale=0.4]{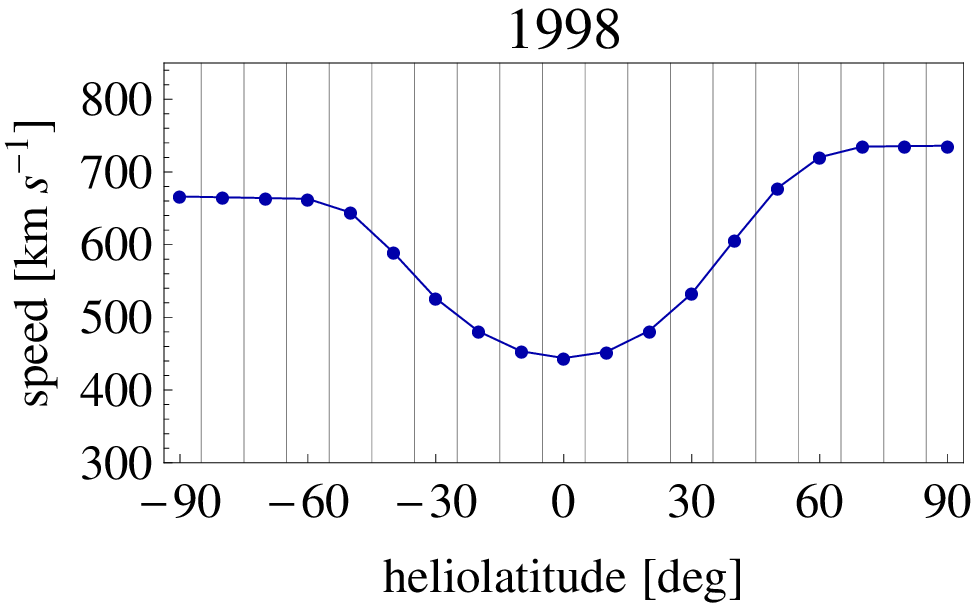}\\	\includegraphics[scale=0.4]{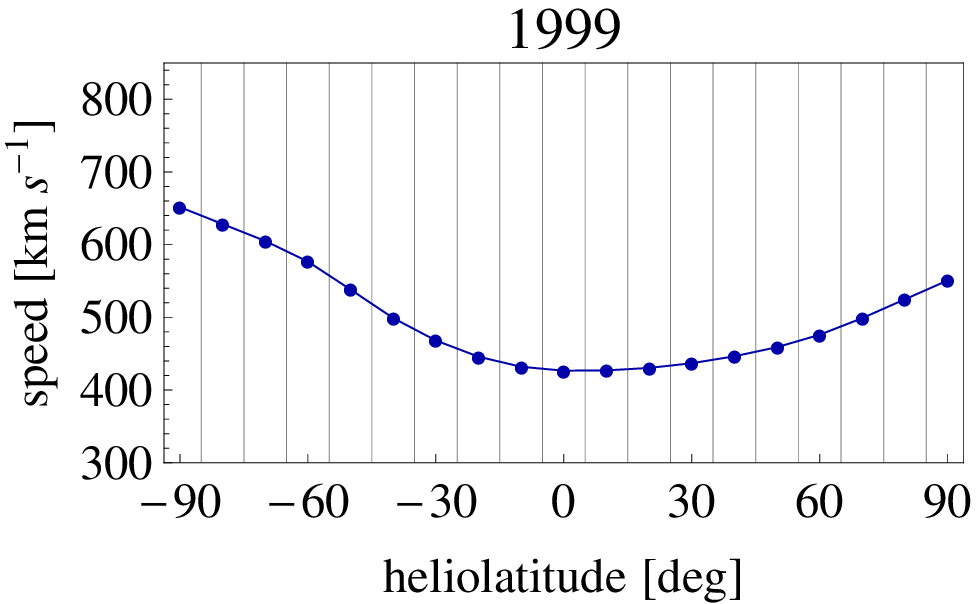}&\includegraphics[scale=0.4]{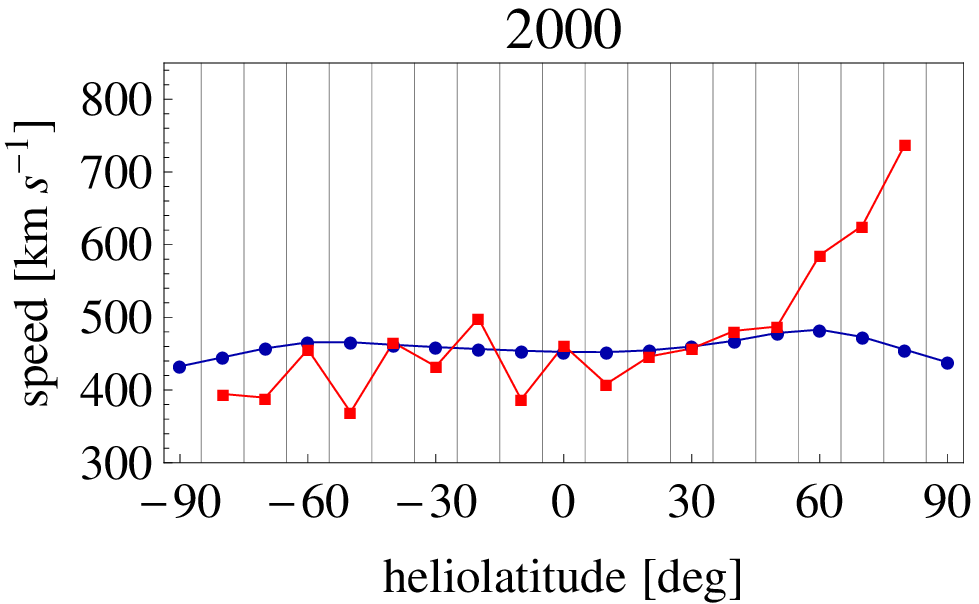}&\includegraphics[scale=0.4]{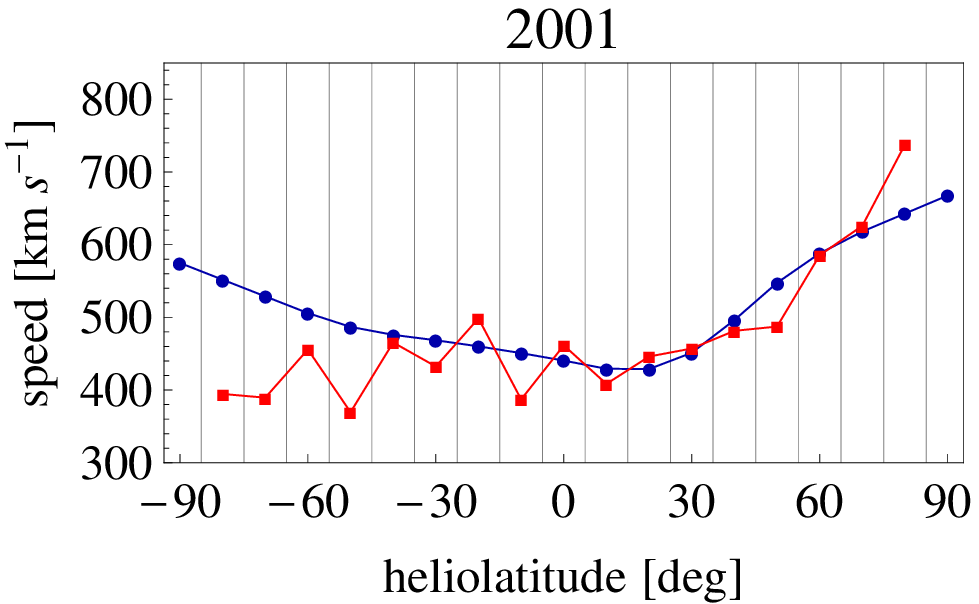}\\	\includegraphics[scale=0.4]{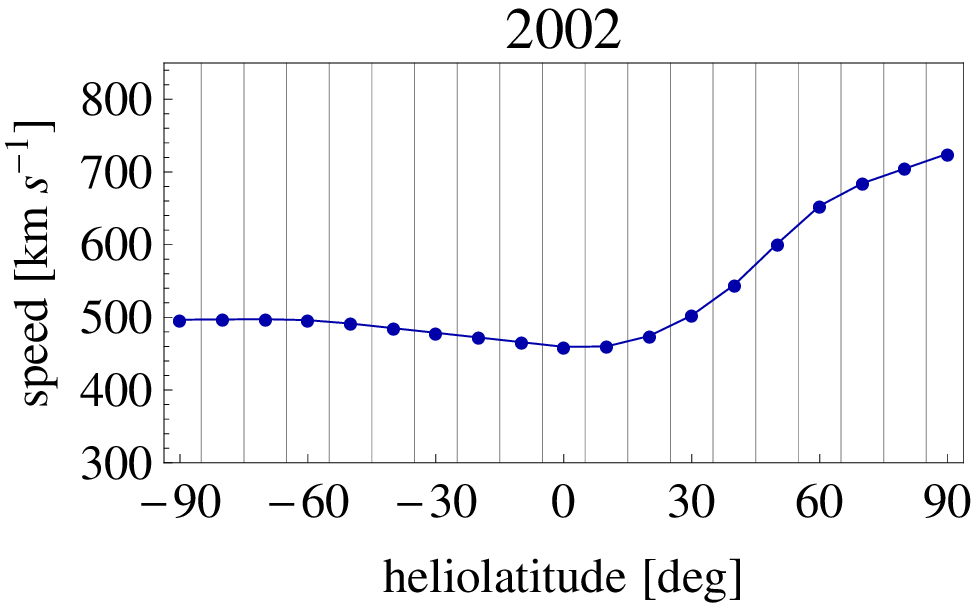}&\includegraphics[scale=0.4]{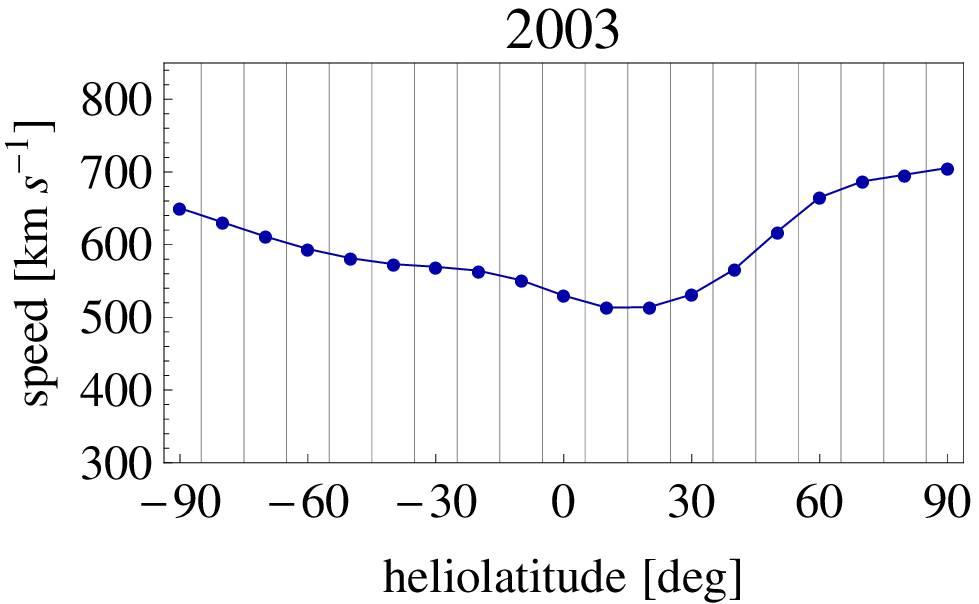}&\includegraphics[scale=0.4]{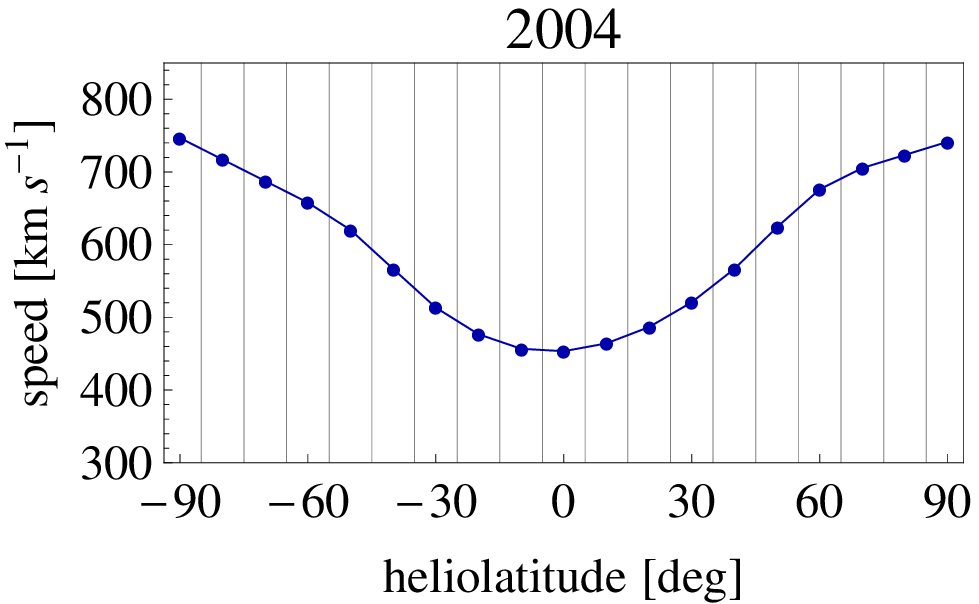}\\	\includegraphics[scale=0.4]{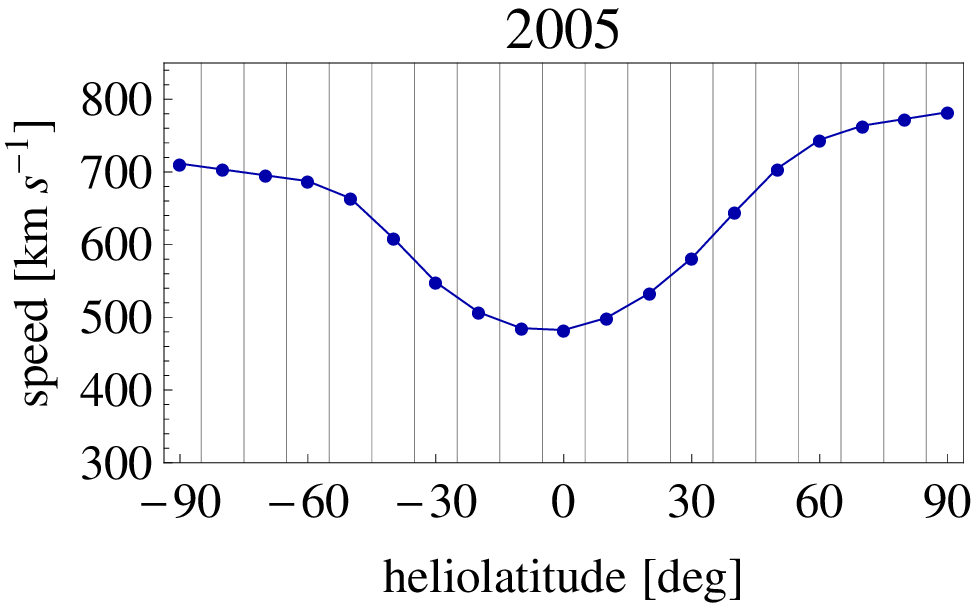}&\includegraphics[scale=0.4]{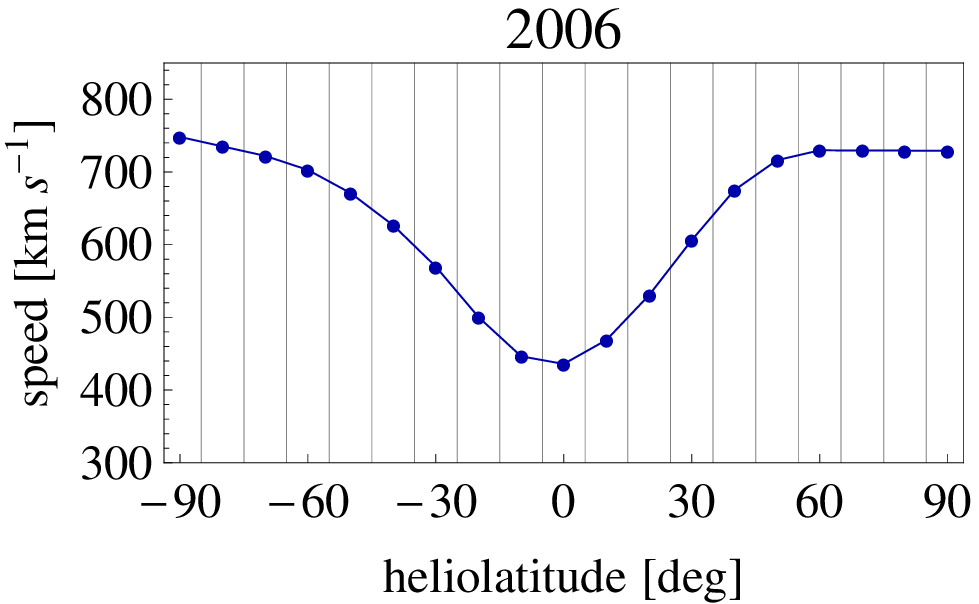}&\includegraphics[scale=0.4]{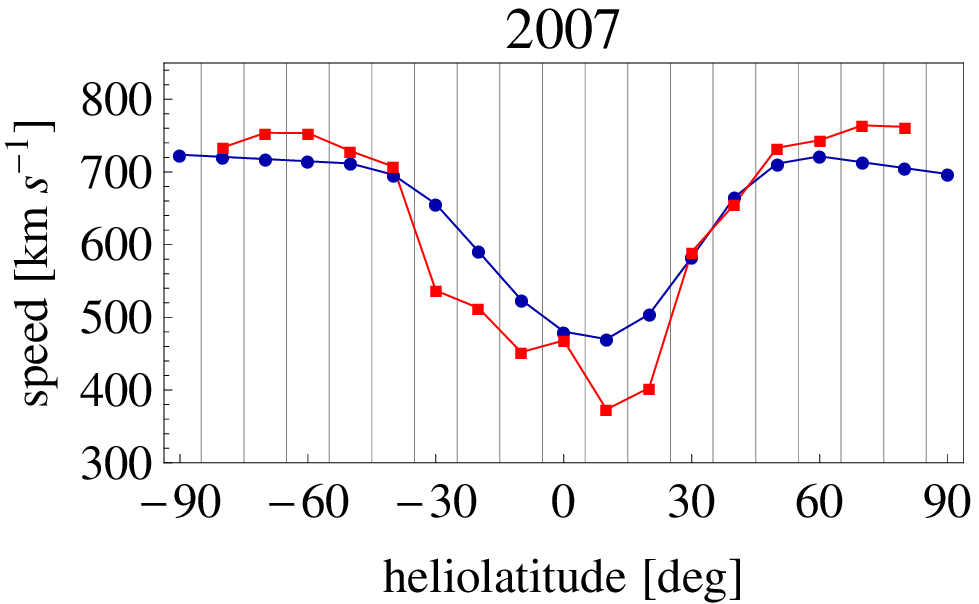}\\	
		\includegraphics[scale=0.4]{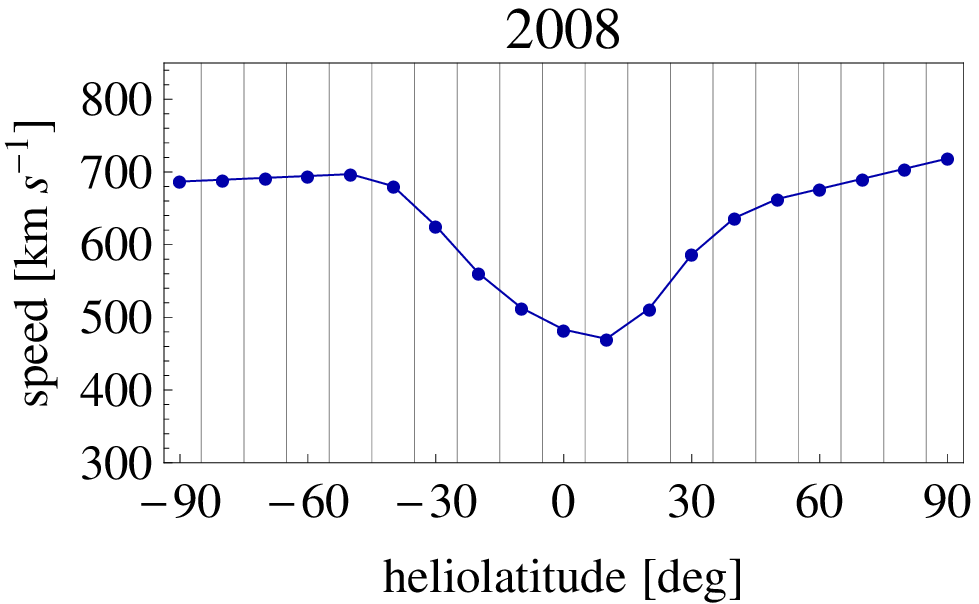}&\includegraphics[scale=0.4]{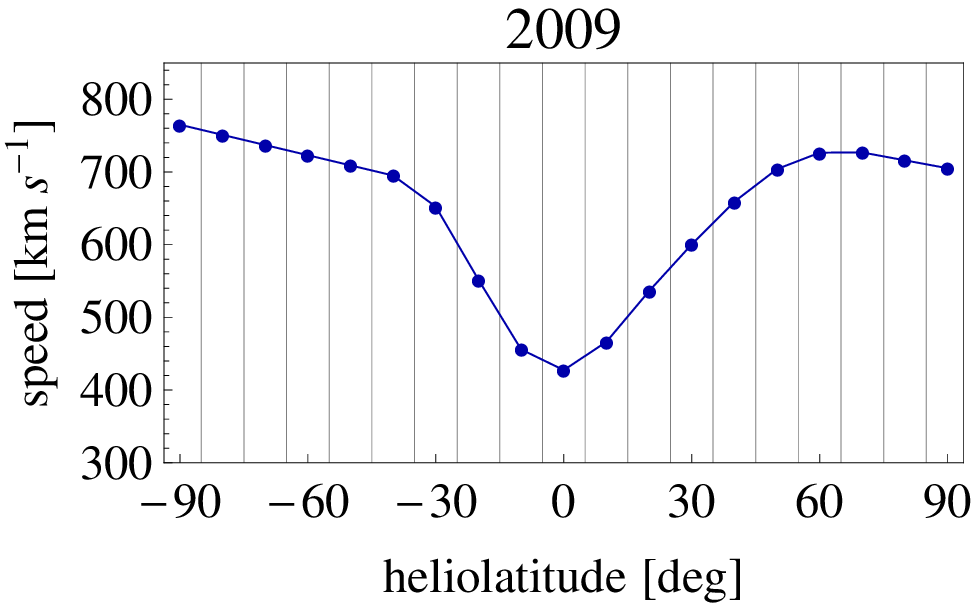}\\
		\end{tabular}
		\caption{Latitude profiles of yearly-averaged solar wind speed obtained from the interim procedure described in the text. The red lines show the speed profiles obtained from the Ulysses fast scans.}
		\label{figResSpeed}
		\end{figure*}
		\clearpage

For further analysis \citet[][]{sokol_etal:12a} smoothed the yearly speed profiles assuming that the variation of solar wind speed at the high latitude close to the poles is linear and the variation outside the polar caps can be approximated by a series of smoothly-transitioning parabolae. The resulting smoothed yearly heliolatitude profiles of solar wind speeds are shown in Fig.~\ref{figResSpeed}. These smoothed profiles will be used in the further part of this chapter to obtain the density and flux profiles of solar wind.
		
\subsection{Latitude structure of solar wind density and flux}  
The time- and latitude-dependence of solar wind flux and density can be obtained using two different methods. The first one, which we regard as interim, relies on an approximate correlation between the solar wind speed and density inferred from the data from fast latitude scans by Ulysses. We discuss it first and show the results. The other one is expected to be the ultimate one and will be based on future results of analysis of the SWAN measurements of the Lyman-alpha helioglow, once it is completed. Here we will show just the method itself.

\subsubsection{Solar wind density and flux from density-speed correlation}

\begin{figure}
\centering
\includegraphics[scale=0.6]{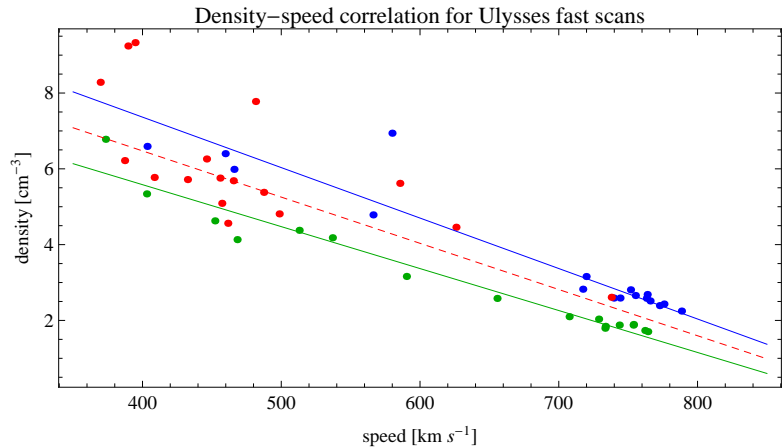}
\caption{Correlation between the solar wind density and speed obtained from Ulysses fast latitude scans. Blue corresponds to the first scan (see Fig. \ref{figUlyssesComposite}), green to the third scan and red to the second scan, performed during solar maximum conditions. The dots correspond to the pairs of speeds and densities averaged over the 10-degree heliolatitude bins, the blue and green lines are the linear correlations specified in Eq.~\ref{eqUlyDensCorr}. The dotted red line is the density -- speed relation proposed for the transition interval close to the solar minimum of 2002, calculated as the mean of the correlation relations obtained from the first and third latitude scans. }
\label{figUlyDensSpeedCorr}
\end{figure}

Solar wind speed and density seem to be correlated in heliolatitude, at least during the solar minimum conditions and for the observations collected during the fast latitude scans. The correlations seem to be slightly different between the first and third latitude scans, as illustrated in Fig. \ref{figUlyDensSpeedCorr}:
\begin{equation}
\label{eqUlyDensCorr}
n_{\mathrm{Ulysses}} \left(v\right) = a_{scan} + b_{scan}\, v
\end{equation}
where $a_{scan}, b_{scan}$ are fitted separately for the speed and density values averaged over 10-degree bins using the ordinary least squares bisector method \citep{isobe_etal:90a} (which allows for uncertainty in both ordinate and abscissa), separately for the first and third latitude scans. For the first scan (blue line and points in Fig. \ref{figUlyDensSpeedCorr}) we obtain $a_{first} = 12.69, b_{first} = -0.01332$, for the third scan (green line and dots in Fig.~\ref{figUlyDensSpeedCorr}) the correlation formula parameters are $a_{third} = 10.01, b_{third} = -0.01107$. Thus, the slopes are almost identical and the main difference between the two formula is in the intercept, which reflects the overall secular decrease in solar wind density.

\begin{figure}
\centering
\includegraphics[scale=0.6]{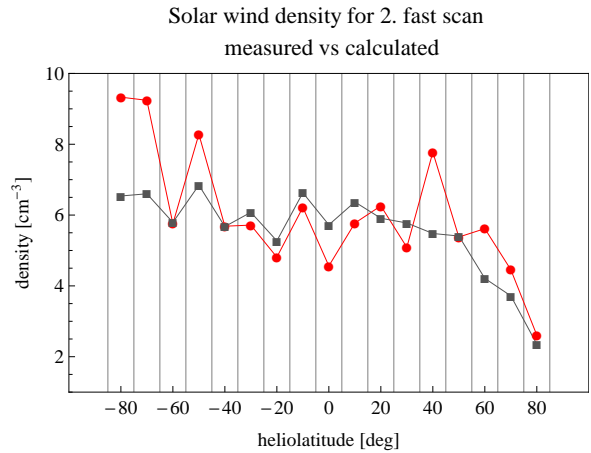}
\caption{Comparison of solar wind density averaged over 10-degree heliolatitude bins actually measured by Ulysses (red) and calculated from the correlation formula in Eq.~\ref{eqUlyDensCorr} (gray).}
\label{figUlyDens2}
\end{figure}

The relation between density and speed for the second scan, which occurred during solar maximum, does not seem to be linear, but in this case the spatial and temporal effects seem to be convolved (as discussed earlier in this section). Therefore, we propose to use a relation that will be an arithmetic mean of the relations for the first and third scans: $a_{second} = \left(a_{first} + a_{third} \right)/2$ and $b_{second} = \left(b_{first} + b_{third} \right)/2$. This relation is shown in Fig.~\ref{figUlyDensSpeedCorr} as the red broken line and a comparison of the density values actually measured during the second latitude scan and calculated from the correlation formula is shown in Fig.~\ref{figUlyDens2}.

The interval of applicability of the latter formula is from ~1998 until 2002. The formula from the first scan is applicable to the interval before 1998 and the formula from the third scan for the interval after 2002. 

\begin{figure*}[t]
		\centering
		\begin{tabular}{ccc}		
\includegraphics[scale=0.4]{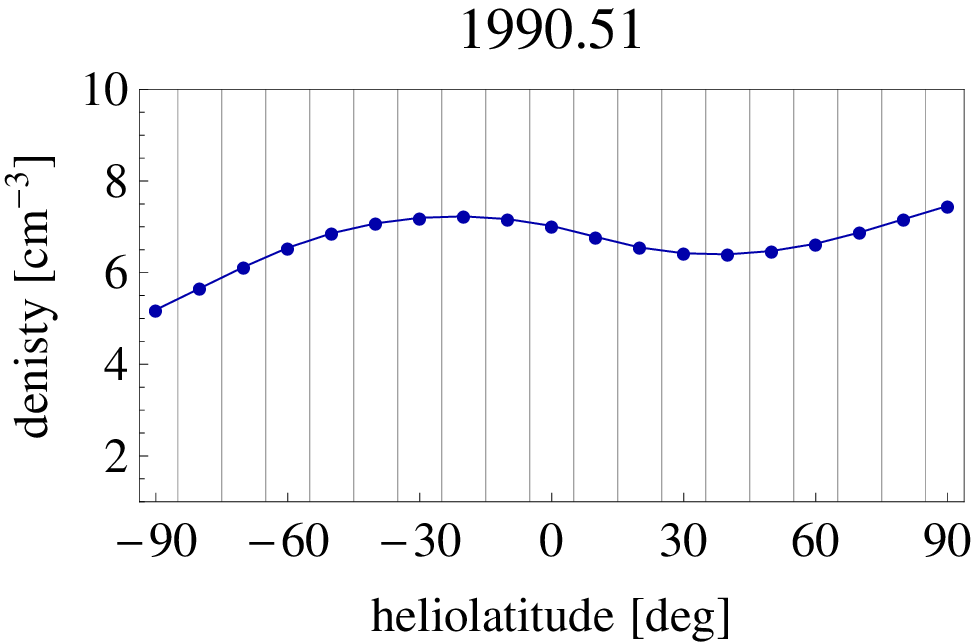}&\includegraphics[scale=0.4]{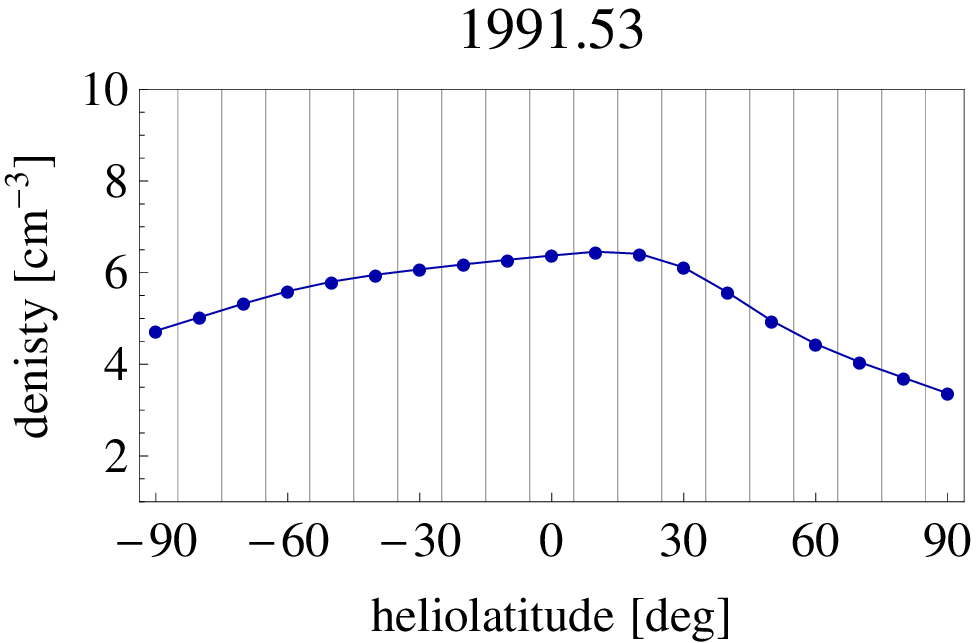}&\includegraphics[scale=0.4]{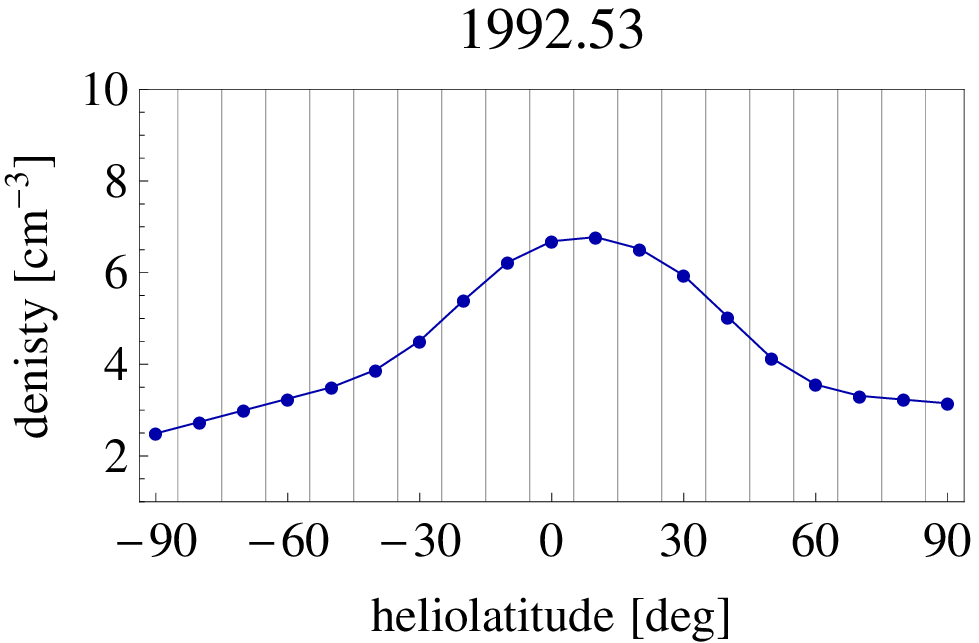}\\	\includegraphics[scale=0.4]{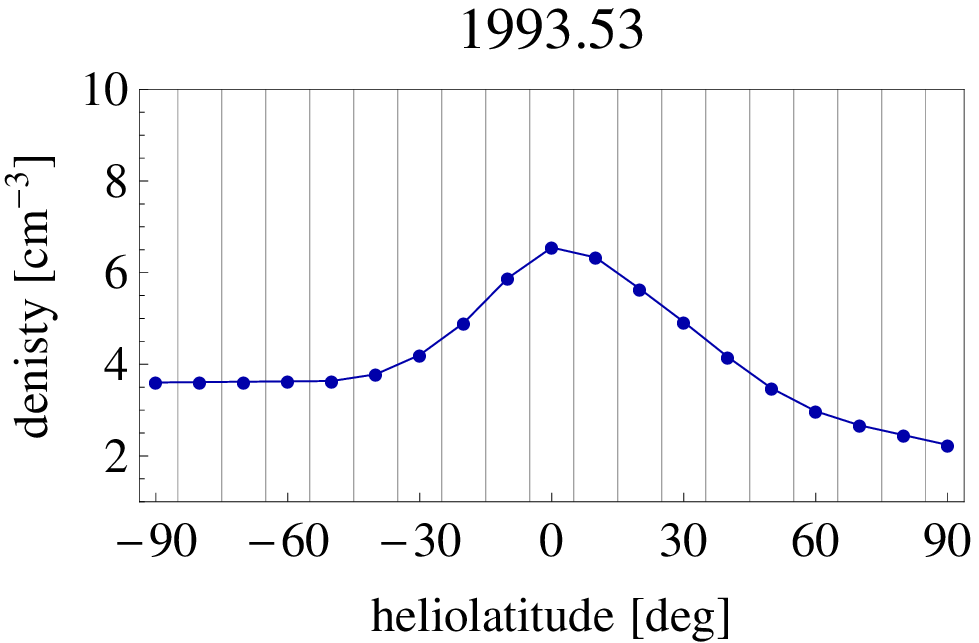}&\includegraphics[scale=0.4]{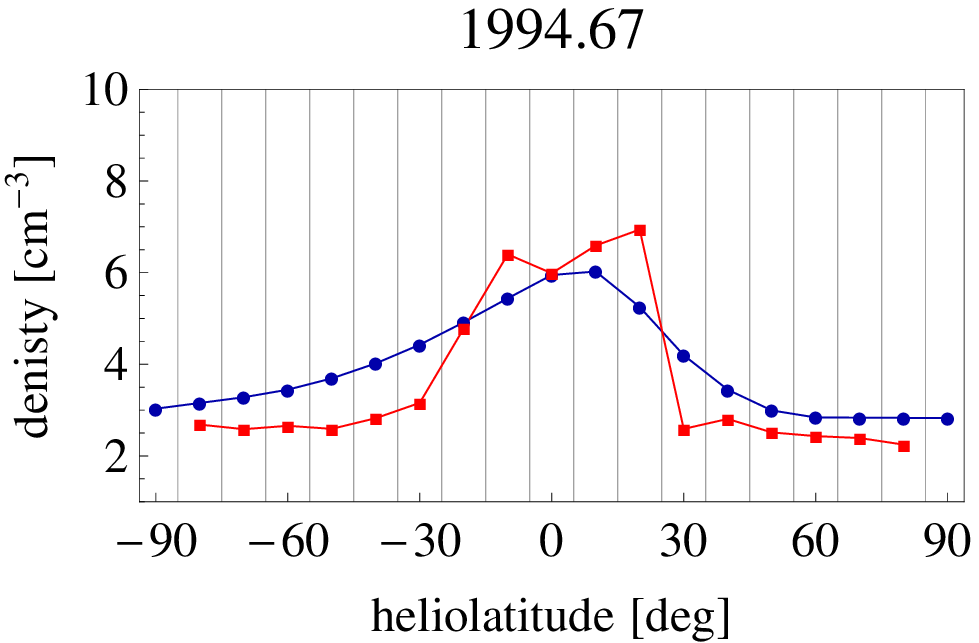}&\includegraphics[scale=0.4]{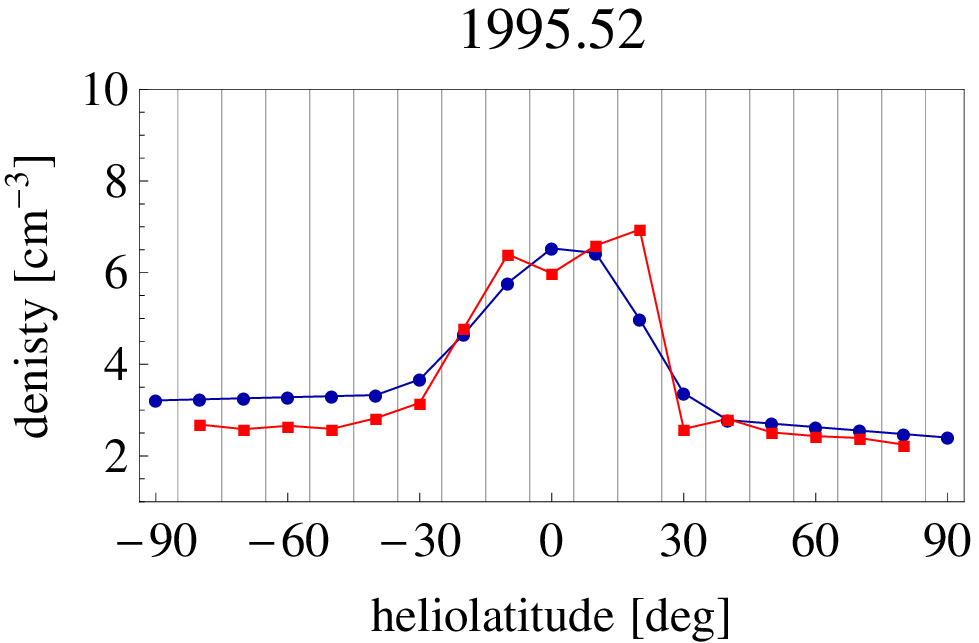}\\	\includegraphics[scale=0.4]{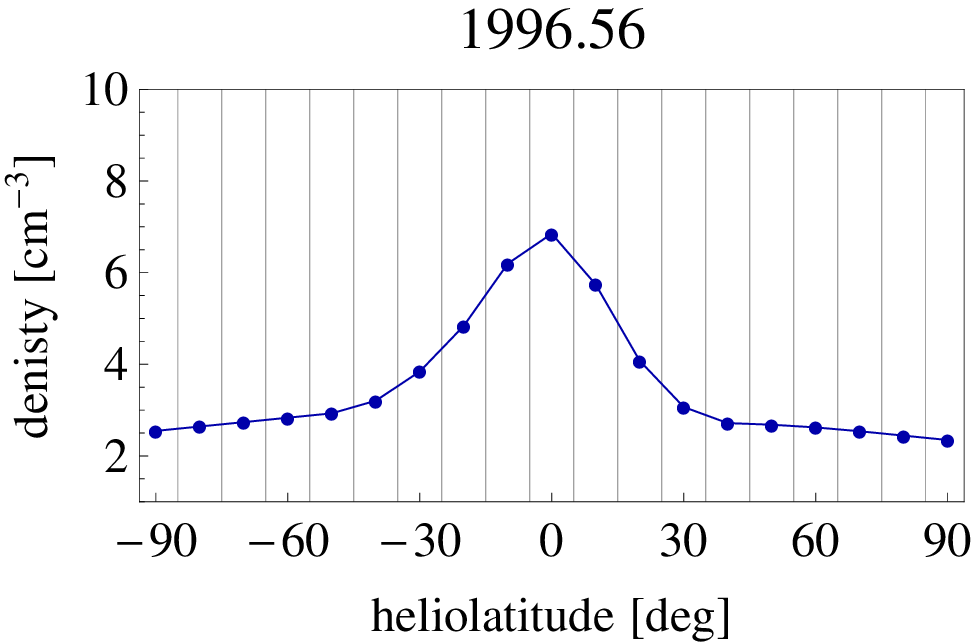}&\includegraphics[scale=0.4]{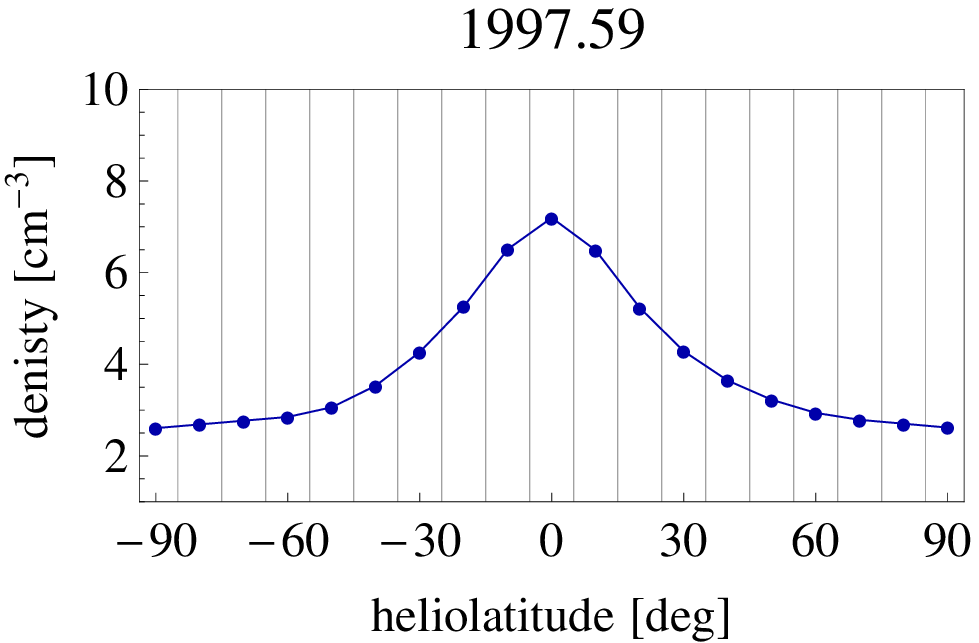}&\includegraphics[scale=0.4]{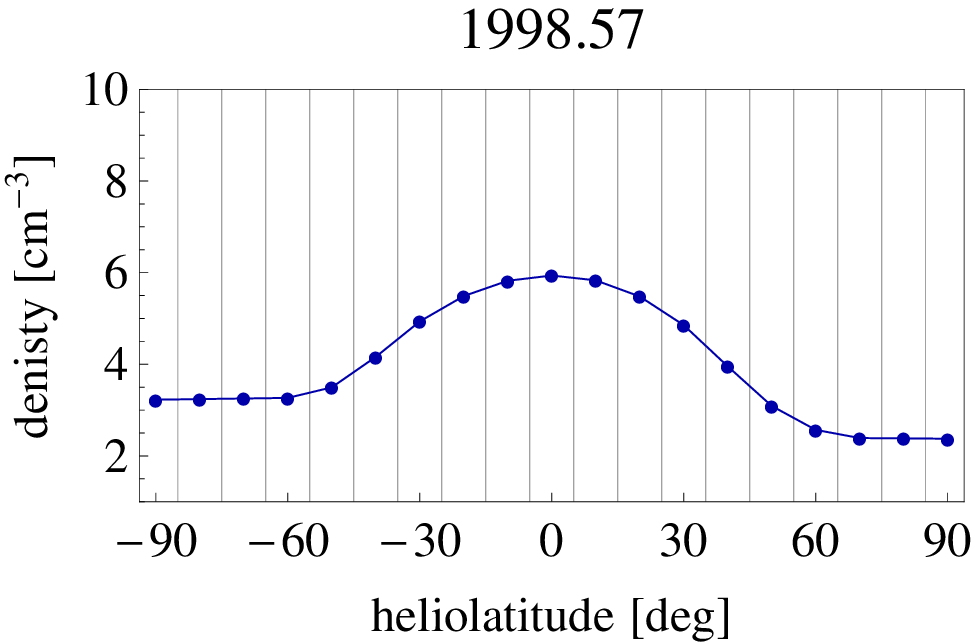}\\	\includegraphics[scale=0.4]{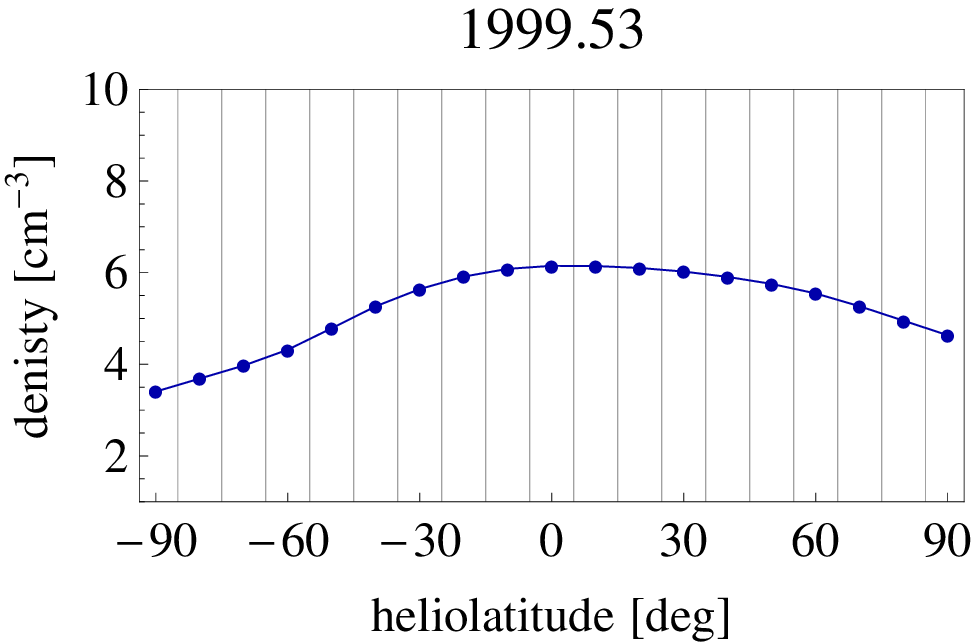}&\includegraphics[scale=0.4]{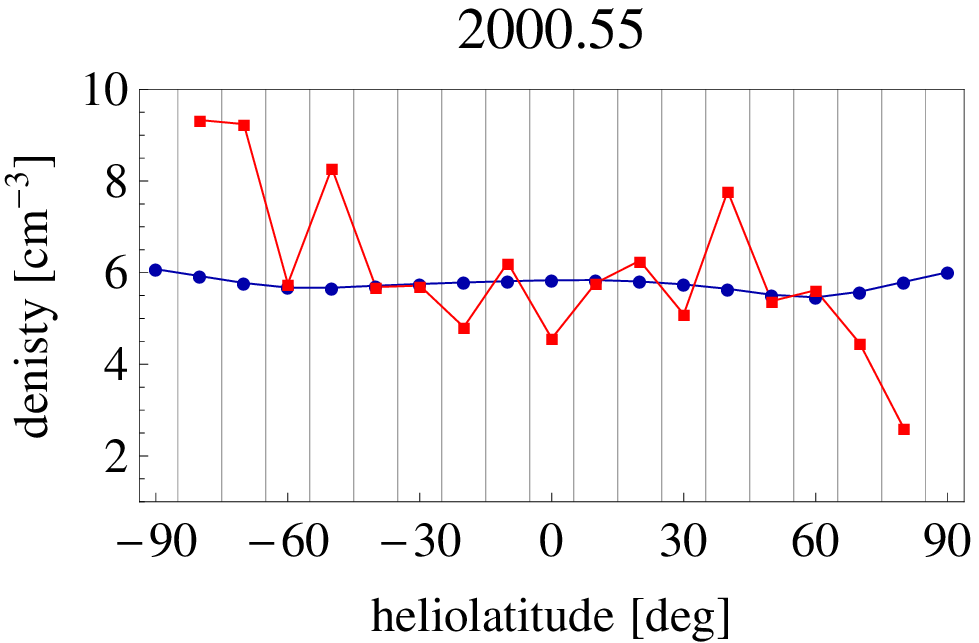}&\includegraphics[scale=0.4]{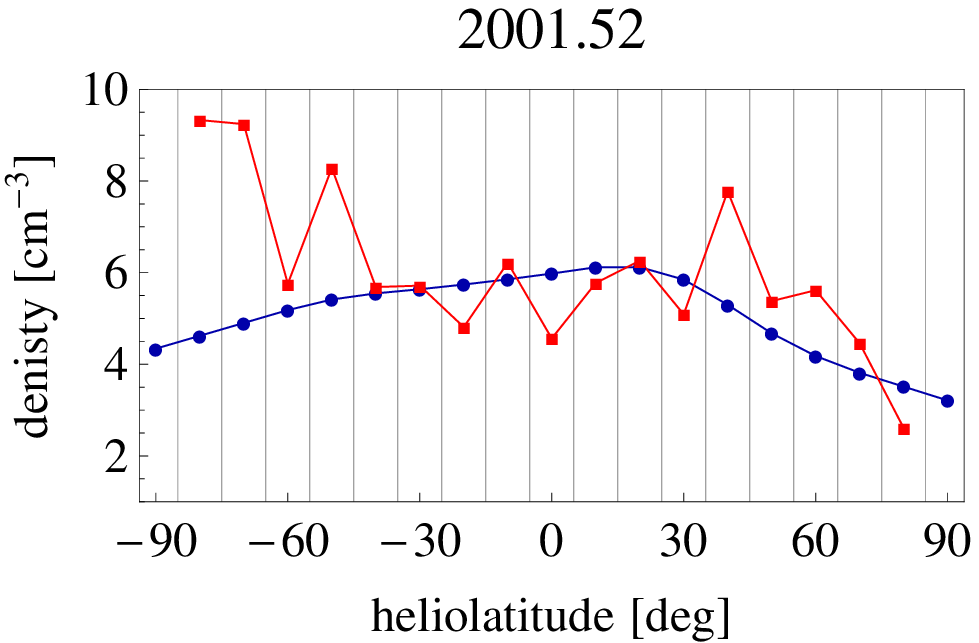}\\	\includegraphics[scale=0.4]{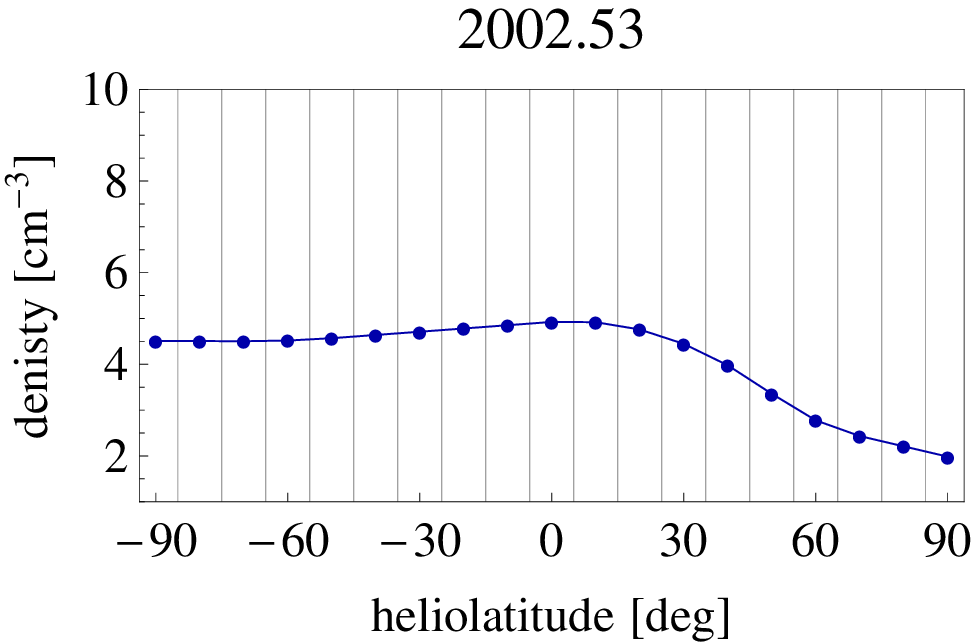}&\includegraphics[scale=0.4]{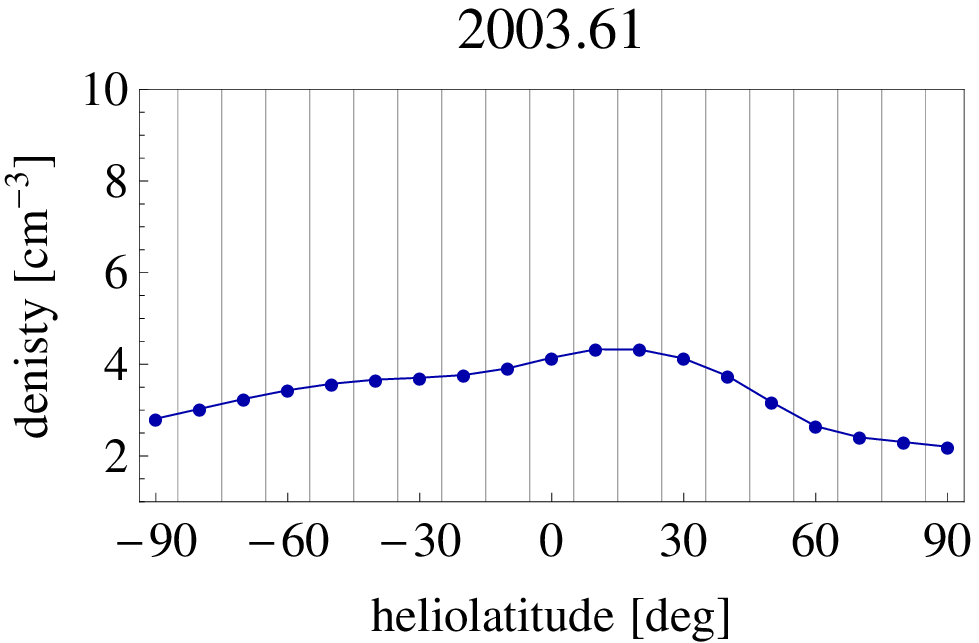}&\includegraphics[scale=0.4]{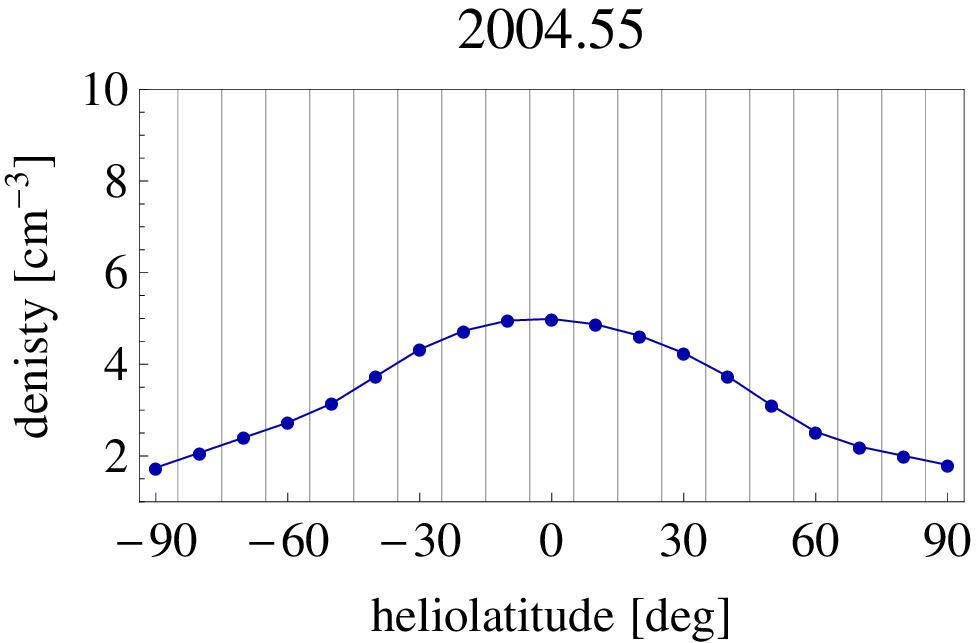}\\	\includegraphics[scale=0.4]{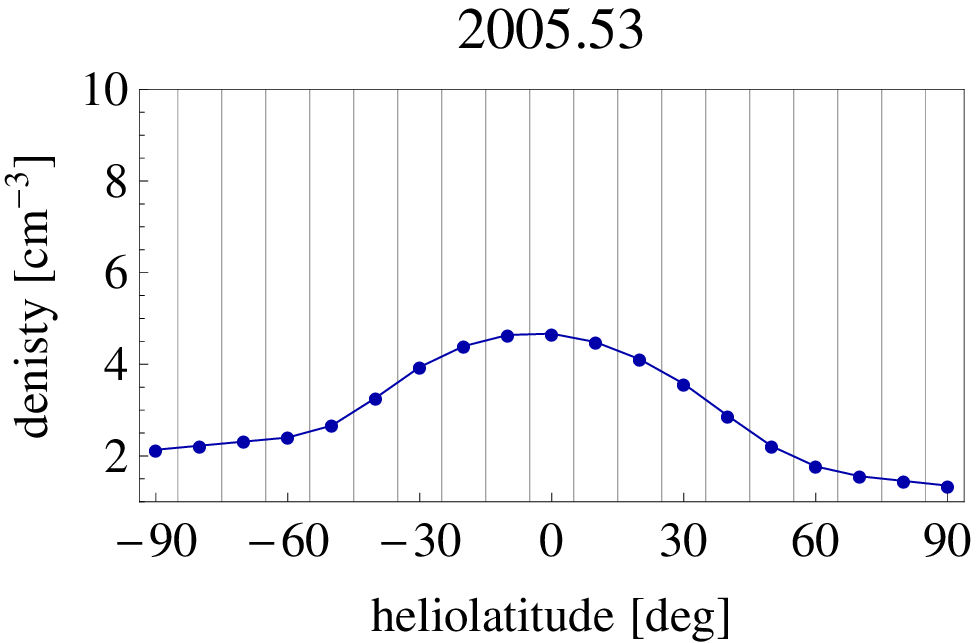}&\includegraphics[scale=0.4]{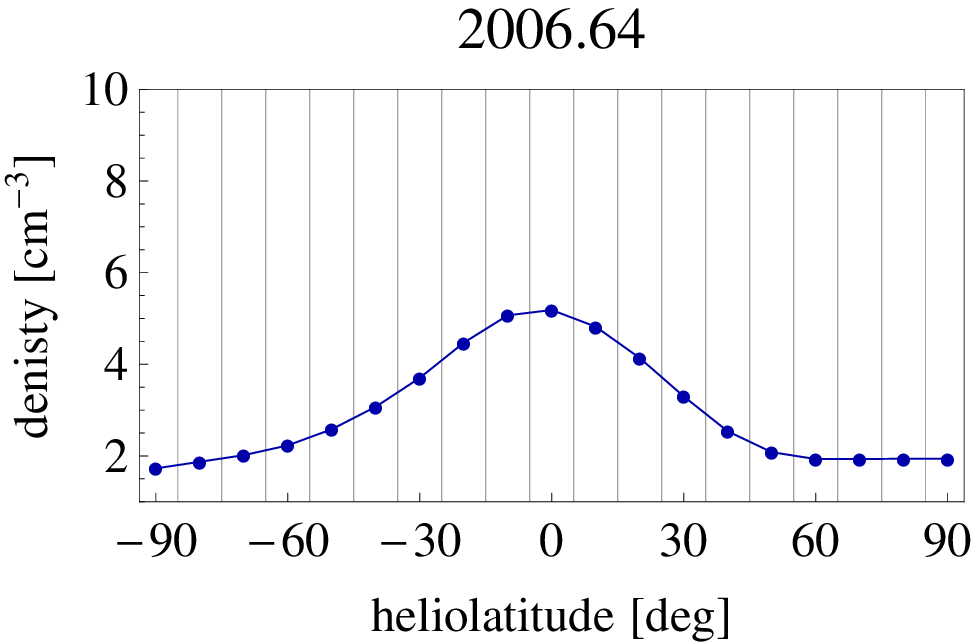}&\includegraphics[scale=0.4]{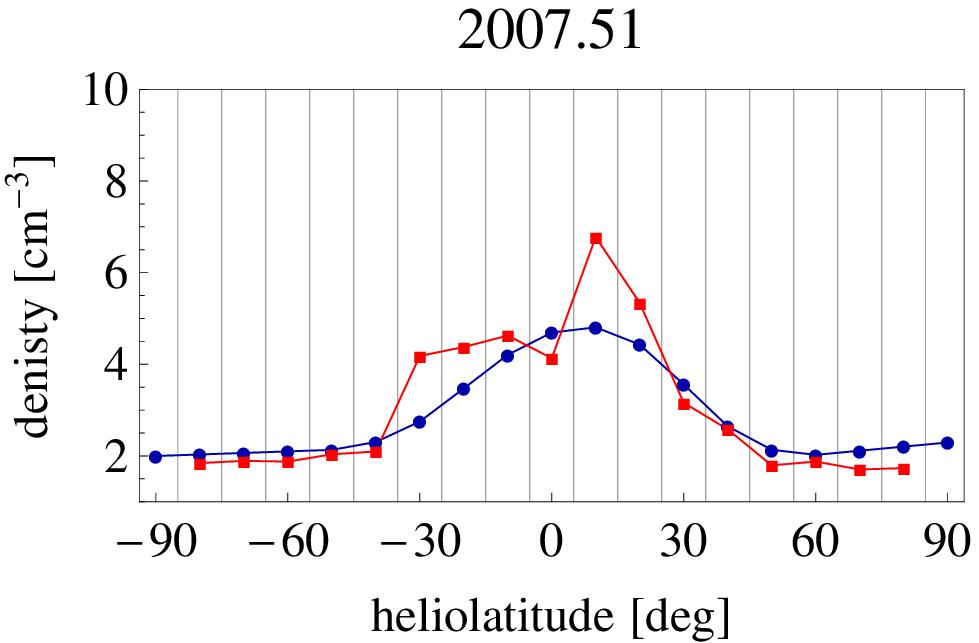}\\	
		\includegraphics[scale=0.4]{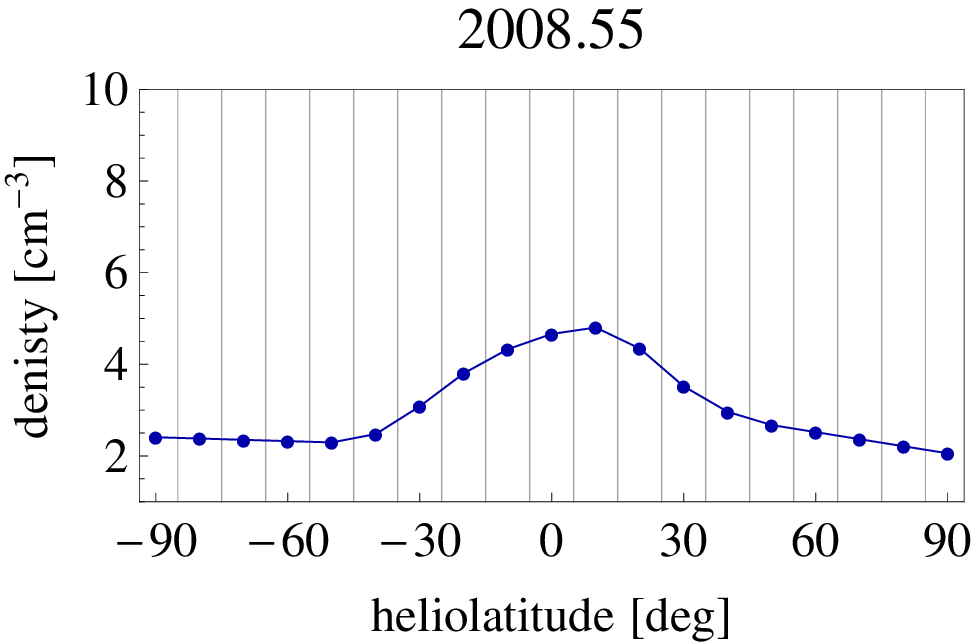}&\includegraphics[scale=0.4]{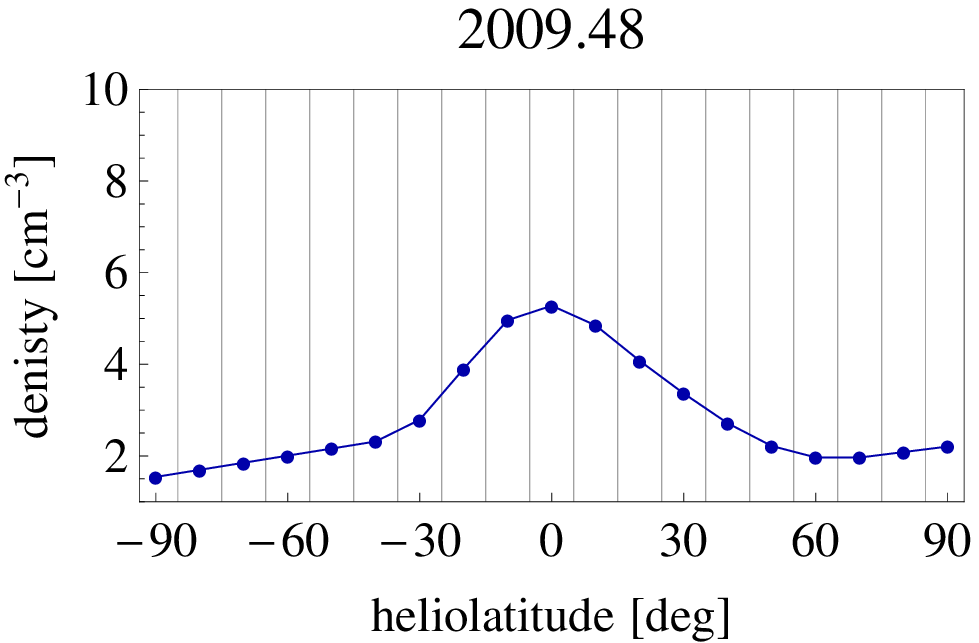}\\
		\end{tabular}
		\caption{Latitude profiles of yearly averaged solar wind density obtained from the interim procedure. The red lines show the Ulysses fast scan profiles.}
		\label{figResDens}
		\end{figure*}

We calculate the interim yearly profiles of solar wind density as a function of heliolatitude by applying Eq.~(\ref{eqUlyDensCorr}) to the speed profiles obtained in the preceding subsection. Since this is an interim and very approximate solution, which by its nature is not very accurate, we do not attempt to further adjust it at the equator to the corresponding OMNI densities. The results are shown in Fig.~\ref{figResDens}. The accuracy of the results in the polar regions is on a level of 20\% to 40\%.

		\clearpage
		
\begin{figure*}[t]
		\centering	
		\includegraphics[scale=0.65]{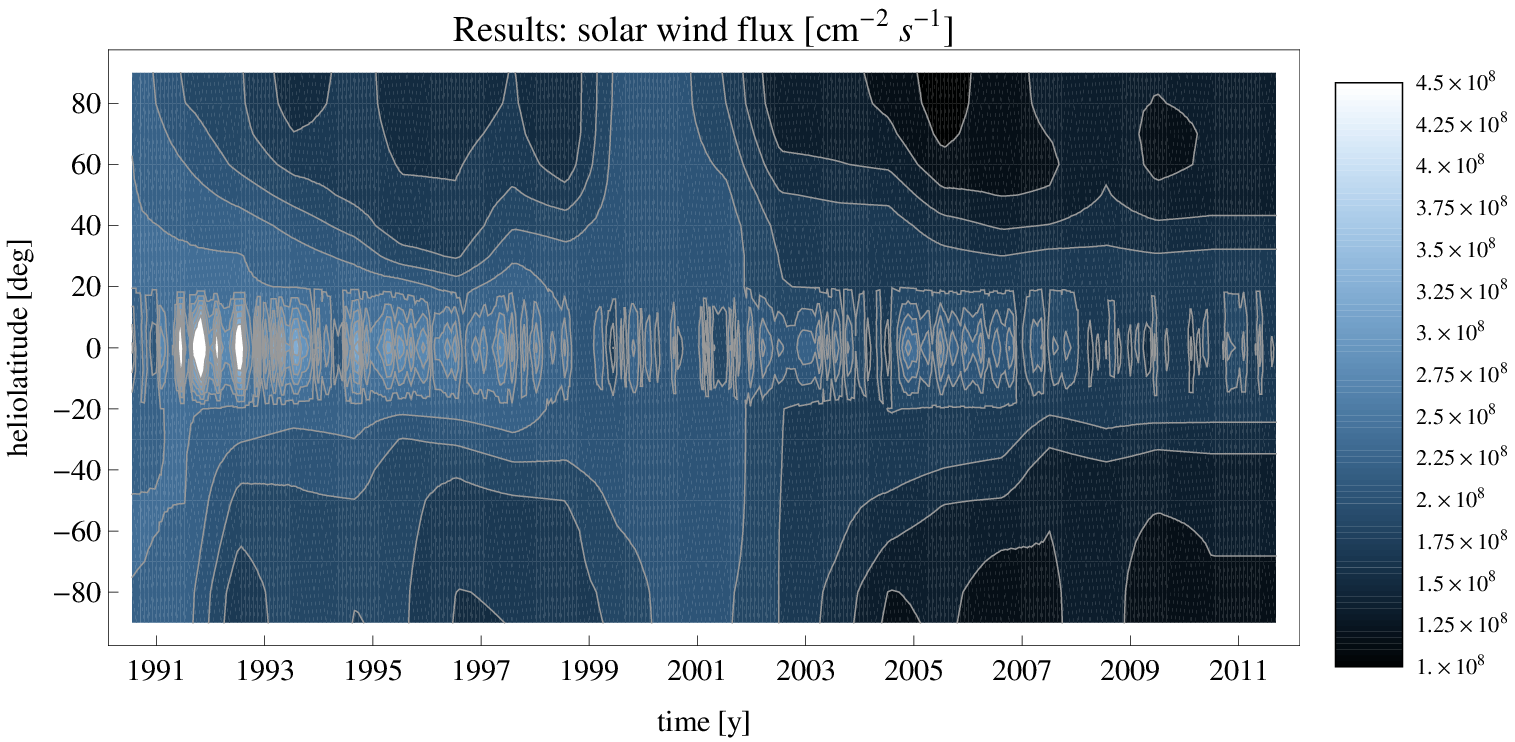}\\
		\includegraphics[scale=0.65]{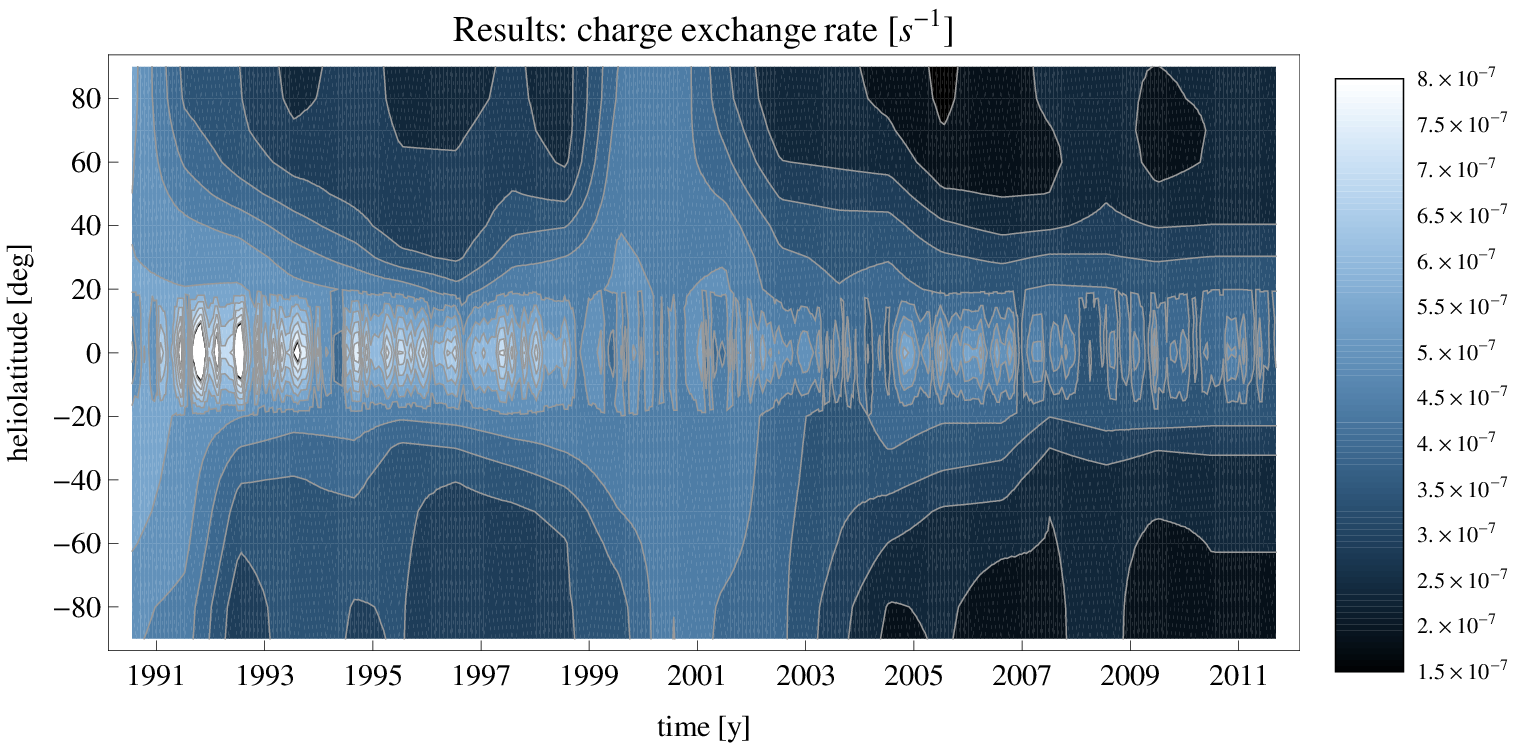}\\
		\includegraphics[scale=0.65]{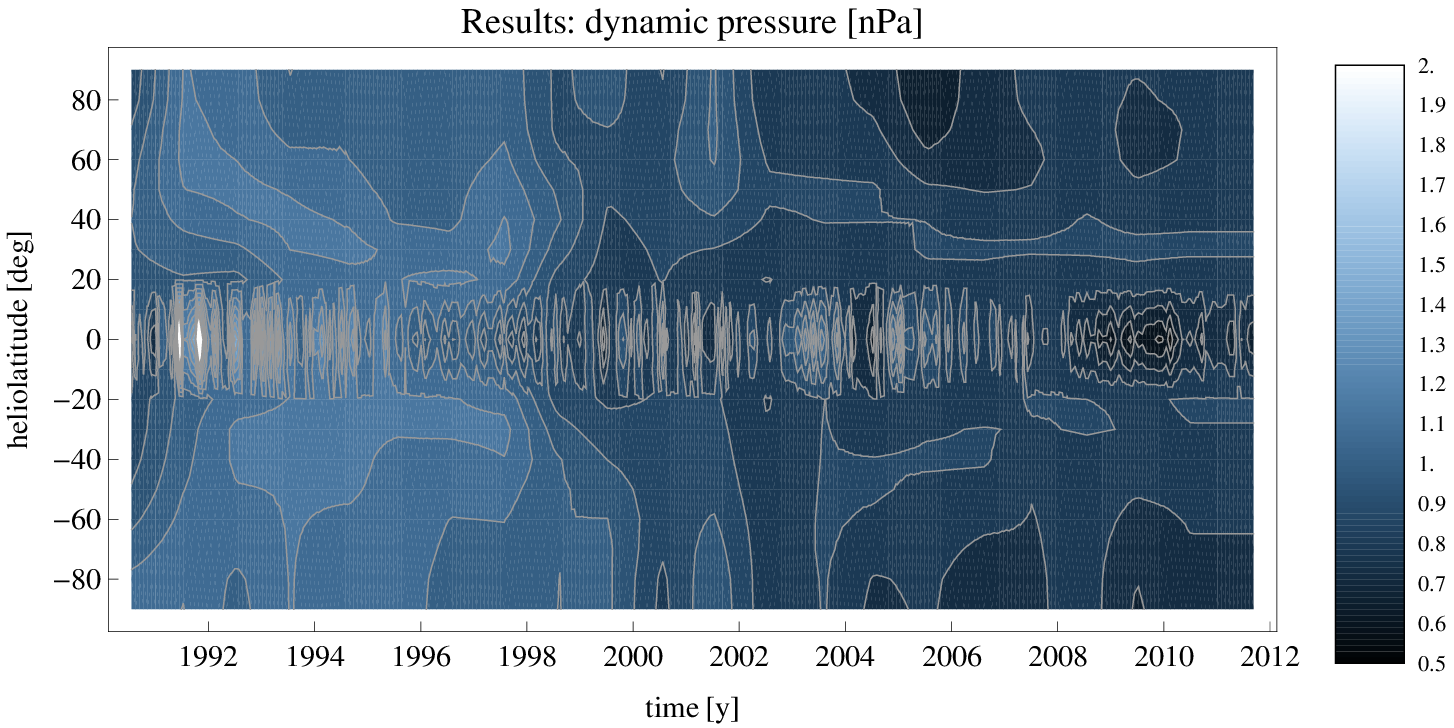}\\
		\caption{Contour maps of solar wind flux (in cm$^{-2}$~$s^{-1}$), charge exchange (in s$^{-1}$) and dynamic pressure (in nanopascals) shown as a function of time and heliolatitude.}
		\label{figResFluxCXDynPress}
		\end{figure*}

With the density and speed profiles on hand, one can easily calculate the flux:
		\begin{equation}
		\label{eqFluxDef}		
		F \left(\phi_j, t_i\right)= v\left( \phi_j, t_i \right) n \left( \phi_j, t_i \right),
		\end{equation}
dynamic pressure:
		\begin{equation}
		\label{eqDynPressDef}		
		p_{dyn} \left(\phi_j, t_i\right)= \frac{1}{2} m_p n \left( \phi_j, t_i \right) \left(v \left( \phi_j, t_i \right)\right)^2 ,
		\end{equation}
and charge exchange rate (in the stationary H atom approximation;  Eq.~\ref{eqBetaCXSWIsGas}). They are collectively shown in  Fig.~\ref{figResFluxCXDynPress}.

To obtain the solar wind parameters (speed, density, flux, dynamic pressure, charge exchange rate and total ionization rate) at a monthly resolution, we replace the equatorial bin directly with the Carrington rotation averaged series from OMNI, linearly interpolated to halves of Carrington rotations. The $\pm 10 $~deg bins are replaced with values linearly interpolated between the $\pm 20$~deg bins and the equatorial bin. The pole values are calculated from the the parabolic interpolation between the $\pm 70$ and $\pm 80$~deg bins, because due to the problems discussed earlier in this chapter direct measurements over the poles are not available. The remaining latitudinal bins are linearly interpolated in time between the yearly profiles. As a result of such a treatment, we utilize all the available information on the equatorial bin of the solar wind. Away from the equatorial bin, where such an information is not available, we have a smooth transition into the region of low time-resolution model. 

Summing up this portion, we have the structure of solar wind speed from the smoothed IPS profiles (Fig.~\ref{figResSpeed}). The density structure is obtained (Fig.~\ref{figResDens}) from the density-speed correlation from Ulysses (Eq.~(\ref{eqUlyDensCorr}), Fig.~\ref{figUlyDensSpeedCorr}). From these, we calculate the solar wind flux, dynamic pressure, and charge exchange rate between solar wind protons and neutral H atoms (Fig.~\ref{figResFluxCXDynPress}).

		\clearpage

\subsubsection{Outlook: latitudinal structure of solar wind flux and density from IPS and Lyman-alpha helioglow observations \emph{(a sketch of the method)}}

We assume that the total ionization rate from SWAN is the sum of the charge exchange and photoionization rates of H, with electron impact ionization neglected (because $\beta_{\mathrm{el}} \ll \beta_{\mathrm{CX}}$ for the MER distance and beyond). Thus the heliolatitude- and time-dependent charge exchange rate can be calculated as
		\begin{equation}
		\label{eqCXRate}		
		\beta_{\mathrm{CX}}\left(\phi_j, t_i\right)=\beta_{\mathrm{tot}}\left( \phi_j, t_i \right) - \beta_{\mathrm{Hph}}\left( 0, t_i \right)
		\end{equation}
where $\beta_{\mathrm{tot}}\left( \phi_j, t_i \right)$ is obtained from inversion of a SWAN map and the photoionization rate $\beta_{\mathrm{Hph}}\left( 0, t_i \right)$, here assumed to be spherically symmetric, is obtained from one of the formulae specified in Eqs.~\ref {eqSEMLyaToTimed}, \ref{eqMgIIToTimed} or \ref{eqF107ToTimed}. 

Since the total ionization rate data from SWAN for the equator do not always agree with the rate derived from direct in situ measurements, we calculate the contrasts $\kappa_{\beta,\mathrm{SWAN}}\left(\phi_j, t_i\right)$ of the SWAN-derived ionization rates as a function of heliolatitude using:
	\begin{equation}
		\label{eqBetaKappa}		
		\kappa_{\mathrm{\beta,SWAN}}\left(\phi_j, t_i\right)=\frac{\beta_{\mathrm{SWAN}}\left( \phi_j, t_i\right)}{\beta_{\mathrm{SWAN}}\left( 0, t_i\right)},
		\end{equation}
where $\beta_{\mathrm{SWAN}}\left( \phi_j, t_i\right)$ is an ionization rate at $\phi_j$-th heliolatitude and $\beta_{\mathrm{SWAN}}\left( 0, t_i\right)$ is an ionization rate at the equator. In this way we can build yearly contrasts for the 19 heliolatitudes bins and for the years from 1996 to 2010. Now, multiplying the contrasts $\kappa_{\beta,\mathrm{SWAN}}\left(\phi_j, t_i\right)$ with the monthly averages of the total equatorial ionization rates (see Fig.~\ref{figOMNIIonProcess}), we obtain latitudinal profiles of the total ionization rate that agree with  the baseline values for the  equator. 

Now we can calculate the absolute charge exchange ionization rate for all heliolatitudes by subtracting the photoionization rate from the total rate Eq.~(\ref{eqCXRate}). In view of the formula for charge exchange rate given in Eq.~(\ref{eqBetaCXSWIsGas}), with the solar wind velocity profile obtained from IPS observations $v\left(\phi_j, t_i\right)$ we can now calculate the total solar wind flux as a function of heliolatitude in the following way: 
		\begin{equation}
		\label{eqFlux}		
		F\left(\phi_j, t_i\right)=\beta_{\mathrm{CX}}\left( \phi_j, t_i \right) / \sigma_{\mathrm{CX}}\left(v\left( \phi_j, t_i\right) \right)
		\end{equation}
With the flux and solar wind speed profile on hand, it is straightforward to calculate the profile of density:
		\begin{equation}
		\label{eqDens}		
		n\left(\phi_j, t_i\right)=F\left( \phi_j, t_i \right) /v \left(\phi_j, t_i \right) 
		\end{equation}
In that way we can obtain a model of evolution of solar wind speed and density as a function of time and heliolatitude that can be used to calculate the evolution of ionization rate for H atoms traveling at arbitrary speeds and to be input to global models of the heliosphere to calculate the flux and dynamic pressure of solar wind. This calculation will be performed once the intercalibration of the heliospheric EUV measurements, which is the goal of this working team, is completed and the final inversion of the SWAN photometric observations for the ionization rate profiles is performed.

\section{Summary}
Intercalibration of heliospheric UV and EUV measurements requires a common basis of heliospheric ionization processes, which affect the distribuition of neutral interstellar gas in the heliosphere and thus influence both the spectrum and intensity distribution of the heliospheric backscatter glow. In this chapter we present a review of the solar factors affecting the distribution of neutral interstellar gas in the heliosphere, namely we discuss the radiation pressure, solar EUV ionizing radiation, and solar wind. We review the history of measurements of these factors and develop a model of time and heliolatitude evolution of solar wind speed and density based on data available from in situ measurements of the solar wind parameters, remote sensing interplanetary scintillation observations of the solar wind speed structure, and from remote sensing observations of the heliospheric Lyman-alpha backscatter glow. The results of this model are used as input in the global heliospheric models, discussed by Izmodenov et al. in another chapter of this book. The results of the global heliospheric modeling can in turn be used to fine tune the absolute calibration of heliospheric EUV measurements and, in a new iteration, to further refine the 
model of evolution of solar wind. The refined model will cover at least the two recent solar cycles and thus will provide a common homogeneous basis for interpretation of the present and past heliospheric experiments. It will also be used to interpret the observations of Energetic Neutral Atoms by the Interstellar Boundary Explorer (IBEX) \citep{mccomas_etal:09a, mccomas_etal:09c}.

\acknowledgments
M.B. and J.S. are obliged to Marty Snow for sharing his insight into the realm of EUV measurements. 
The authors acknowledge the use of NASA/GSFC's Space Physics Data Facility's ftp service for Ulysses/SWOOPS and TIMED/SEE data, SOHO/CELIAS/SEM\newline
(\url{http://www.usc.edu/dept/space_science/semdatafolder/long/daily_avg/}), and OMNI2 data collection 
(\url{ftp://nssdcftp.gsfc.nasa.gov/spacecraft_data/omni/}). \newline
The F$_{10.7}$ solar radio flux was provided by the NOAA and Pentincton Solar Radio Monitaring Programme operated jointly by the National Research Council and the Canadian Space Agency \newline
(\url{ftp://ftp.ngdc.noaa.gov/STP/SOLAR_DATA/SOLAR_RADIO/FLUX/Penticton_Adjusted/} \newline
and \url{ftp://ftp.geolab.nrcan.gc.ca/data/solar_flux/daily_flux_values/}). \newline
The composite Lyman-alpha flux and MgII$_{\mathrm{c/w}}$ were obtained from LASP, accessed through the LISIRD Web page at  (\url{http://lasp.colorado.edu/lisird/lya/}). \newline
The SOLAR2000 Research Grade historical irradiances are provided courtesy of W. Kent Tobiska and SpaceWx.com. These historical irradiances have been developed with funding from the NASA UARS, TIMED, and SOHO missions.



M.B. and J.S. were supported by the Polish Ministry for Science and Higher Education grants NS-1260-11-09 and N-N203-513-038. P.B. acknowledges support from NASA SR\&T Grant NX09AW32G. Contributions from D.M. were supported by NASA's IBEX Explorer mission. Support from the International Space Science Institute in Bern, Switzerland, the organizer of the Fully-Online Database for Ultraviolet Experiments (FONDUE), is gratefully acknowledged. 		

\bibliographystyle{issi2008}
\bibliography{iplbib}

\begin{thebibliography}{216}
\providecommand{\natexlab}[1]{#1}
\providecommand{\url}[1]{\texttt{#1}}
\providecommand{\urlprefix}{URL }
\expandafter\ifx\csname urlstyle\endcsname\relax
  \providecommand{\doi}[1]{doi:\discretionary{}{}{}#1}\else
  \providecommand{\doi}{doi:\discretionary{}{}{}\begingroup
  \urlstyle{rm}\Url}\fi

\bibitem[{\emph{Amblard et~al.}(2008)\emph{Amblard, Moussaoui, {Dudok de Wit},
  Aboudarham, Kretzschmar, Lilensten, and Auchere}}]{amblard_etal:08a}
Amblard, P.~O., Moussaoui, S., {Dudok de Wit}, T., Aboudarham, J., Kretzschmar,
  M., Lilensten, J., and Auchere, F., 2008, The {EUV} {S}un as the
  superposition of elementary suns, \emph{\aap}, \textbf{487}, L13--L16.

\bibitem[{\emph{{Artzner} et~al.}(1978)\emph{{Artzner}, {Vial}, {Lemaire},
  {Gouttebroze}, and {Leibacher}}}]{artzner_etal:78a}
{Artzner}, G., {Vial}, J.~C., {Lemaire}, P., {Gouttebroze}, P., and
  {Leibacher}, J., 1978, {Simultaneous time-resolved observations of the H
  L-alpha, MG K 2795 A, and CA K solar lines}, \emph{\apjl}, \textbf{224},
  L83--L85, \doi{10.1086/182765}.

\bibitem[{\emph{{Asai} et~al.}(1998)\emph{{Asai}, {Kojima}, {Tokumaru},
  {Yokobe}, {Jackson}, {Hick}, and {Manoharan}}}]{asai_etal:98a}
{Asai}, K., {Kojima}, M., {Tokumaru}, M., {Yokobe}, A., {Jackson}, B.~V.,
  {Hick}, P.~L., and {Manoharan}, P.~K., 1998, {Heliospheric tomography using
  interplanetary scintillation observations. III - Correlation between speed
  and electron density fluctuations in the solar wind}, \emph{\jgr},
  \textbf{103}, 1991--2001, \doi{10.1029/97JA02750}.

\bibitem[{\emph{Asbridge et~al.}(1976)\emph{Asbridge, Bame, Feldman, and
  Montgomery}}]{asbridge_etal:76a}
Asbridge, J.~R., Bame, S.~J., Feldman, W.~C., and Montgomery, M.~D., 1976,
  Helium and hydrogen velocity differences in the solar wind, \emph{\jgr},
  \textbf{81}, 2719--2727, \doi{10.1029/JA081i016p02719}.

\bibitem[{\emph{{Auch{\`e}re}}(2005)}]{auchere:05}
{Auch{\`e}re}, F., 2005, {Effect of the H I Ly{$\alpha$} chromospheric flux
  anisotropy on the total intensity of the resonantly scattered coronal
  radiation}, \emph{\apj}, \textbf{622}, 737--743.

\bibitem[{\emph{{Auch{\`e}re} et~al.}(2005{\natexlab{a}})\emph{{Auch{\`e}re},
  {Cook}, {Newmark}, {McMullin}, {von Steiger}, and
  {Witte}}}]{auchere_etal:05b}
{Auch{\`e}re}, F., {Cook}, J.~W., {Newmark}, J.~S., {McMullin}, D.~R., {von
  Steiger}, R., and {Witte}, M., 2005{\natexlab{a}}, {Model of the all-sky He
  II 30.4 nm solar flux}, \emph{\asr}, \textbf{35}, 388--392.

\bibitem[{\emph{{Auch{\`e}re} et~al.}(2005{\natexlab{b}})\emph{{Auch{\`e}re},
  {McMullin}, {Cook}, and {et al.}}}]{auchere_etal:05c}
{Auch{\`e}re}, F., {McMullin}, D.~R., {Cook}, J.~W., and {et al.},
  2005{\natexlab{b}}, A model for solar {EUV} flux helium photoionization
  throughout the 3-dimensional heliosphere, in \emph{ESA SP-592: Solar Wind
  11/SOHO 16, Connecting Sun and Heliosphere}, pp. 327--329.

\bibitem[{\emph{Bame et~al.}(1971)\emph{Bame, E.~J.~Hones, and
  Asbridge}}]{bame_etal:71a}
Bame, S.~J., E.~J.~Hones, M. D.~M., S. I.~Akasofu, and Asbridge, J.~R., 1971,
  Geomagnetic storm particles in the high-latitude magnetotail, \emph{\jgr},
  \textbf{76}, 7566.

\bibitem[{\emph{Bame et~al.}(1978{\natexlab{a}})\emph{Bame, Asbridge,
  Felthauser, Glore, Hawk, and Chavez}}]{bame_etal:78a}
Bame, S.~J., Asbridge, J.~R., Felthauser, H.~E., Glore, J.~P., Hawk, H.~L., and
  Chavez, J., 1978{\natexlab{a}}, {ISEE-C} solar wind plasma experiment,
  \emph{IEEE Transactions on Geoscience Electronics}, \textbf{16}, 160--162.

\bibitem[{\emph{Bame et~al.}(1978{\natexlab{b}})\emph{Bame, Asbridge,
  Felthauser, Glore, Paschmann, Hemmerich, Lehmann, and
  Rosenbauer}}]{bame_etal:78b}
Bame, S.~J., Asbridge, J.~R., Felthauser, H.~E., Glore, J.~P., Paschmann, G.,
  Hemmerich, P., Lehmann, K., and Rosenbauer, H., 1978{\natexlab{b}}, {ISEE-1}
  and {ISEE-2} fast plasma experiment and the {ISEE-1} solar wind experiment,
  \emph{IEEE Transactions on Geoscience Electronics}, \textbf{16}, 216--220.

\bibitem[{\emph{{Bame} et~al.}(1992)\emph{{Bame}, {McComas}, {Barraclough},
  {Phillips}, {Sofaly}, {Chavez}, {Goldstein}, and {Sakurai}}}]{bame_etal:92a}
{Bame}, S.~J., {McComas}, D.~J., {Barraclough}, B.~L., {Phillips}, J.~L.,
  {Sofaly}, K.~J., {Chavez}, J.~C., {Goldstein}, B.~E., and {Sakurai}, R.~K.,
  1992, {The ULYSSES solar wind plasma experiment}, \emph{\aaps}, \textbf{92},
  237--265.

\bibitem[{\emph{Baranov}(2006{\natexlab{a}})}]{baranov:06a}
Baranov, V.~B., 2006{\natexlab{a}}, Kinetic and hydrodynamic approaches in
  space plasma, in \emph{The physics of the heliospheric boundaries}, edited by
  V.~Izmodenov and R.~Kallenbach, vol. SR-005 of \emph{ISSI Sci.Rep}, pp.
  1--26.

\bibitem[{\emph{Baranov}(2006{\natexlab{b}})}]{baranov:06b}
Baranov, V.~B., 2006{\natexlab{b}}, Early concepts of the heliospheric
  interface: plasma, in \emph{The physics of the heliospheric boundaries},
  edited by V.~Izmodenov and R.~Kallenbach, vol. SR-005 of \emph{ISSI Sci.Rep},
  pp. 27--44.

\bibitem[{\emph{Baranov and Malama}(1993)}]{baranov_malama:93}
Baranov, V.~B. and Malama, Y.~G., 1993, Model of the solar wind interaction
  with the {L}ocal {I}nterstellar {M}edium: numerical solution of
  self-consistent problem, \emph{\jgr}, \textbf{98}, 15\,157--15\,163.

\bibitem[{\emph{Baranov et~al.}(1991)\emph{Baranov, Lebedev, and
  Malama}}]{baranov_etal:91}
Baranov, V.~B., Lebedev, M.~G., and Malama, Y.~G., 1991, The influence of the
  interface between the heliosphere and the {L}ocal {I}nterstellar {M}edium on
  the penetration of the {H} atoms to the solar system, \emph{\apj},
  \textbf{375}, 347--351.

\bibitem[{\emph{Barnett et~al.}(1990)\emph{Barnett, Hunter, Kirkpatrick,
  Alvarez, Cisneros, and Phaneuf}}]{barnett_etal:90}
Barnett, C.~F., Hunter, H.~T., Kirkpatrick, M.~I., Alvarez, I., Cisneros, C.,
  and Phaneuf, R.~A., 1990, \emph{Atomic data for fusion. Collisions of {H},
  {H}$_2$, {H}e and {L}i atoms and ions with atoms and molecules}, vol.
  ORNL-6086/V1, Oak Ridge National Laboratories, Oak Ridge, Tenn.

\bibitem[{\emph{Bertaux and Blamont}(1971)}]{bertaux_blamont:71}
Bertaux, J.~L. and Blamont, J.~E., 1971, Evidence for a source of an
  extraterrstrial hydrogen {L}yman {A}lpha emission: {T}he interstellar wind,
  \emph{\aap}, \textbf{11}, 200--217.

\bibitem[{\emph{Bertaux et~al.}(1995)\emph{Bertaux, Kyr{\"o}l{\"a},
  Qu{\'e}merais, Pellinen, Lallement, Schmidt, Berth{\'e}, Dimarellis, Goutail,
  Taulemasse, Bernard, Leppelmeier, Summanen, Hannula, Huomo, Kehl{\"a},
  Korpela, Lepp{\"a}l{\"a}, Str{\"o}mmer, Torsti, Viherkanto, Hochedez,
  Chretiennot, and Holzer}}]{bertaux_etal:95}
Bertaux, J.~L., Kyr{\"o}l{\"a}, E., Qu{\'e}merais, E., Pellinen, R., Lallement,
  R., Schmidt, W., Berth{\'e}, M., Dimarellis, E., Goutail, J.~P., Taulemasse,
  C., Bernard, C., Leppelmeier, G., Summanen, T., Hannula, H., Huomo, H.,
  Kehl{\"a}, V., Korpela, S., Lepp{\"a}l{\"a}, K., Str{\"o}mmer, Torsti, J.,
  Viherkanto, K., Hochedez, J.~F., Chretiennot, G., and Holzer, T., 1995,
  {SWAN:} a study of solar wind anisotropies on {SOHO} with {L}yman {A}lpha sky
  mapping, \emph{\solphys}, \textbf{162}, 403--439.

\bibitem[{\emph{Bertaux et~al.}(1996)\emph{Bertaux, Qu{\'e}merais, and
  Lallement}}]{bertaux_etal:96a}
Bertaux, J.~L., Qu{\'e}merais, E., and Lallement, R., 1996, Observations of a
  sky {L}yman~$\alpha$ groove related to enhanced solar wind mass flux in the
  neutral sheet, \emph{\grl}, \textbf{23}, 3675--3678.

\bibitem[{\emph{Bertaux et~al.}(1997)\emph{Bertaux, Qu{\'e}merais, Lallement,
  Kyr{\"o}l{\"a}, Schmidt, Summanen, M{\"a}kinen, and
  Holzer}}]{bertaux_etal:97a}
Bertaux, J.~L., Qu{\'e}merais, E., Lallement, R., Kyr{\"o}l{\"a}, E., Schmidt,
  W., Summanen, T., M{\"a}kinen, T., and Holzer, T., 1997, The first 1.5 year
  of observation from {SWAN} {L}yman-{A}lpha solar wind mapper on {SOHO}, in
  \emph{Proceedings of the {F}ifth {SOHO} {W}orkshop, {T}he {C}orona and
  {S}olar {W}ind near {M}inimum {A}ctivity}, no. 404 in ESA SP-, pp. 29--36.

\bibitem[{\emph{Bertaux et~al.}(1999)\emph{Bertaux, Kyr{\"o}l{\"a},
  Qu{\'e}merais, Lallement, Schmidt, Costa, and M{\"a}kinen}}]{bertaux_etal:99}
Bertaux, J.-L., Kyr{\"o}l{\"a}, E., Qu{\'e}merais, E., Lallement, R., Schmidt,
  W., Costa, J., and M{\"a}kinen, T., 1999, Swan observations of the solar wind
  latitude distribution and its evolution since launch, \emph{\ssr},
  \textbf{87}, 129--132.

\bibitem[{\emph{Bertaux et~al.}(2000)\emph{Bertaux, Qu{\'e}merais, Lallement,
  Lamassoure, Schmidt, and Kyr{\"o}l{\"a}}}]{bertaux_etal:00}
Bertaux, J.-L., Qu{\'e}merais, E., Lallement, R., Lamassoure, E., Schmidt, W.,
  and Kyr{\"o}l{\"a}, E., 2000, Monitoring solar activity on the far side of
  the {S}un from sky reflected {L}yman~$\alpha$ radiation, \emph{\grl},
  \textbf{27}, 1331--1334.

\bibitem[{\emph{Bochsler et~al.}(2011)\emph{Bochsler, Bzowski, Didkovsky,
  Kucharek, Sok{\'o}{\l}, and Woods}}]{bochsler_etal:11b}
Bochsler, P., Bzowski, M., Didkovsky, L., Kucharek, H., Sok{\'o}{\l}, J.~M.,
  and Woods, T., 2011, Ionization rates (preliminary), \emph{in preparation},
  \textbf{00}, 00.

\bibitem[{\emph{Bonetti et~al.}(1969)\emph{Bonetti, Moreno, Cantarano, Egidi,
  Marconero, Palutan, and Pizella}}]{bonetti_etal:69a}
Bonetti, A., Moreno, G., Cantarano, S., Egidi, A., Marconero, R., Palutan, F.,
  and Pizella, G., 1969, Solar wind observations with satellite esro heos-1 in
  december 1969, \emph{Nuovo Cimento B Series}, \textbf{46}, 307--323,
  \doi{10.1007/BF02711013}.

\bibitem[{\emph{{Bonnet} et~al.}(1978)\emph{{Bonnet}, {Lemaire}, {Vial},
  {Artzner}, {Gouttebroze}, {Jouchoux}, {Vidal-Madjar}, {Leibacher}, and
  {Skumanich}}}]{bonnet_etal:78a}
{Bonnet}, R.~M., {Lemaire}, P., {Vial}, J.~C., {Artzner}, G., {Gouttebroze},
  P., {Jouchoux}, A., {Vidal-Madjar}, A., {Leibacher}, J.~W., and {Skumanich},
  A., 1978, {The LPSP instrument on OSO 8. II - In-flight performance and
  preliminary results}, \emph{\apj}, \textbf{221}, 1032--1053,
  \doi{10.1086/156109}.

\bibitem[{\emph{{Brandt} et~al.}(1972)\emph{{Brandt}, {Roosen}, and
  {Harrington}}}]{brandt_etal:72a}
{Brandt}, J.~C., {Roosen}, R.~G., and {Harrington}, R.~S., 1972,
  {Interplanetary gas. XVII. an astrometric determination of solar wind
  velocities from orientations of ionic comet tails}, \emph{\apj},
  \textbf{177}, 277--284, \doi{10.1086/151706}.

\bibitem[{\emph{{Brandt} et~al.}(1975)\emph{{Brandt}, {Harrington}, and
  {Roosen}}}]{brandt_etal:75a}
{Brandt}, J.~C., {Harrington}, R.~S., and {Roosen}, R.~G., 1975,
  {Interplanetary gas. XX - Does the radial solar wind speed increase with
  latitude}, \emph{\apj}, \textbf{196}, 877--878, \doi{10.1086/153478}.

\bibitem[{\emph{{Brasken} and {Kyrola}}(1998)}]{brasken_kyrola:98}
{Brasken}, M. and {Kyrola}, E., 1998, {Resonance scattering of Lyman alpha from
  interstellar hydrogen}, \emph{\aap}, \textbf{332}, 732--738.

\bibitem[{\emph{Bridge et~al.}(1965)\emph{Bridge, Egidi, Lazarus, and
  Lyon}}]{bridge_etal:65a}
Bridge, J.~S., Egidi, A., Lazarus, A.~J., and Lyon, E., 1965, Preliminary
  results of plasma measurements on imp-a, \emph{Space Research}, \textbf{V}.

\bibitem[{\emph{Bzowski}(2001{\natexlab{a}})}]{bzowski:01a}
Bzowski, M., 2001{\natexlab{a}}, Time dependent radiation pressure and time
  dependent {2D} ionisation rate for heliospheric modelling, in \emph{The
  {O}uter {H}eliosphere: {T}he {N}ext {F}rontiers}, edited by K.~Scherer,
  H.~Fichtner, H.~J. Fahr, and E.~Marsch, COSPAR Colloquia Series Vol. 11, pp.
  69--72, Elsevier, Pergamon.

\bibitem[{\emph{Bzowski}(2001{\natexlab{b}})}]{bzowski:01b}
Bzowski, M., 2001{\natexlab{b}}, A model of charge exchange of interstellar
  hydrogen on a time-dependent, {2D} solar wind, \emph{\ssr}, \textbf{97},
  379--383.

\bibitem[{\emph{Bzowski}(2003)}]{bzowski:03}
Bzowski, M., 2003, Response of the groove in heliospheric {L}yman-$\alpha$ glow
  to latitude-dependent ionization rate, \emph{\aap}, \textbf{408}, 1155--1164.

\bibitem[{\emph{Bzowski}(2008)}]{bzowski:08a}
Bzowski, M., 2008, Survival probability and energy modification of hydrogen
  energetic neutral atoms on their way from the termination shock to earth
  orbit, \emph{\aap}, \textbf{488}, 1057--1068,
  \doi{10.1051/0004-6361:200809393}.

\bibitem[{\emph{Bzowski and
  Ruci{\'n}ski}(1995{\natexlab{a}})}]{bzowski_rucinski:95a}
Bzowski, M. and Ruci{\'n}ski, D., 1995{\natexlab{a}}, Solar cycle modulation of
  the interstellar hydrogen density distribution in the heliosphere,
  \emph{\ssr}, \textbf{72}, 467--470.

\bibitem[{\emph{Bzowski and
  Ruci{\'n}ski}(1995{\natexlab{b}})}]{bzowski_rucinski:95b}
Bzowski, M. and Ruci{\'n}ski, D., 1995{\natexlab{b}}, Variability of the
  neutral hydrogen density distribution due to solar cycle related effects,
  \emph{\asr}, \textbf{16}, 131--134.

\bibitem[{\emph{Bzowski et~al.}(1997)\emph{Bzowski, Fahr, Ruci{\'n}ski, and
  Scherer}}]{bzowski_etal:97}
Bzowski, M., Fahr, H.~J., Ruci{\'n}ski, D., and Scherer, H., 1997, Variation of
  bulk velocity and temperature anisotropy of neutral heliospheric hydrogen
  during the solar cycle, \emph{\aap}, \textbf{326}, 396--411.

\bibitem[{\emph{Bzowski et~al.}(2002)\emph{Bzowski, Summanen, Ruci{\'n}ski, and
  Kyr{\"o}l{\"a}}}]{bzowski_etal:02}
Bzowski, M., Summanen, T., Ruci{\'n}ski, D., and Kyr{\"o}l{\"a}, E., 2002,
  Response of interplanetary glow to global variations of hydrogen ionization
  rate and solar {L}yman-$\alpha$ flux, \emph{\jgr}, \textbf{107},
  10.1029/2001JA00\,141.

\bibitem[{\emph{Bzowski et~al.}(2003)\emph{Bzowski, M{\"a}kinen,
  Kyr{\"o}l{\"a}, Summanen, and Qu{\`e}merais}}]{bzowski_etal:03a}
Bzowski, M., M{\"a}kinen, T., Kyr{\"o}l{\"a}, E., Summanen, T., and
  Qu{\`e}merais, E., 2003, Latitudinal structure and north-south asymmetry of
  the solar wind from {L}yman-$\alpha$ remote sensing by {SWAN}, \emph{\aap},
  \textbf{408}, 1165--1177.

\bibitem[{\emph{{Bzowski} et~al.}(2008)\emph{{Bzowski}, {M{\"o}bius},
  {Tarnopolski}, {Izmodenov}, and {Gloeckler}}}]{bzowski_etal:08a}
{Bzowski}, M., {M{\"o}bius}, E., {Tarnopolski}, S., {Izmodenov}, V., and
  {Gloeckler}, G., 2008, {Density of neutral interstellar hydrogen at the
  termination shock from Ulysses pickup ion observations}, \emph{\aap},
  \textbf{491}, 7--19, \doi{10.1051/0004-6361:20078810}.

\bibitem[{\emph{{Bzowski} et~al.}(2009)\emph{{Bzowski}, {M{\"o}bius},
  {Tarnopolski}, {Izmodenov}, and {Gloeckler}}}]{bzowski_etal:09a}
{Bzowski}, M., {M{\"o}bius}, E., {Tarnopolski}, S., {Izmodenov}, V., and
  {Gloeckler}, G., 2009, {Neutral H density at the termination shock: a
  consolidation of recent results}, \emph{\ssr}, \textbf{143}, 177--190.

\bibitem[{\emph{Chabrillat and Kockarts}(1997)}]{chabrillat_kockarts:97}
Chabrillat, S. and Kockarts, G., 1997, Simple parameterization of the
  absorption of the solar lyman-alpha line, \emph{\grl}, \textbf{24},
  2659--2662.

\bibitem[{\emph{{Coles} and {Maagoe}}(1972)}]{coles_maagoe:72a}
{Coles}, W.~A. and {Maagoe}, S., 1972, {Solar-wind velocity from IPS
  observations}, \emph{\jgr}, \textbf{77}, 5622--5624,
  \doi{10.1029/JA077i028p05622}.

\bibitem[{\emph{Coles and Rickett}(1976)}]{coles_rickett:76a}
Coles, W.~A. and Rickett, B.~J., 1976, Ips observations of the solar wind speed
  out of the ecliptic, \emph{\jgr}, \textbf{81}, 4797--4799.

\bibitem[{\emph{{Cook} et~al.}(1980)\emph{{Cook}, {Brueckner}, and {van
  Hoosier}}}]{cook_etal:80a}
{Cook}, J.~W., {Brueckner}, G.~E., and {van Hoosier}, M.~E., 1980, {Variability
  of the solar flux in the far ultraviolet 1175-2100 A}, \emph{\jgr},
  \textbf{85}, 2257--2268.

\bibitem[{\emph{{Cook} et~al.}(1981)\emph{{Cook}, {Meier}, {Brueckner}, and
  {van Hoosier}}}]{cook_etal:81a}
{Cook}, J.~W., {Meier}, R.~R., {Brueckner}, G.~E., and {van Hoosier}, M.~E.,
  1981, {Latitudinal anisotropy of the solar far ultraviolet flux - Effect on
  the Lyman alpha sky background}, \emph{\aap}, \textbf{97}, 394--397.

\bibitem[{\emph{{Covington}}(1947)}]{covington:47}
{Covington}, A.~E., 1947, {Micro-Wave Solar Noise Observations During the
  Partial Eclipse of November 23, 1946}, \emph{\nat}, \textbf{159}, 405--406,
  \doi{10.1038/159405a0}.

\bibitem[{\emph{Danby and Camm}(1957)}]{danby_camm:57}
Danby, J. M.~A. and Camm, J.~L., 1957, Statistical dynamics and accretion,
  \emph{\mnras}, \textbf{117}, 150.

\bibitem[{\emph{de~Toma}(2011)}]{deToma:11a}
de~Toma, G., 2011, Evolution of coronal holes and implications for high-speed
  solar wind during the minimum between cycles 23 and 24, \emph{\solphys},
  \doi{10.1007/s11207-010-9677-2}.

\bibitem[{\emph{{Dennison} and {Hewish}}(1967)}]{dennison_hewish:67a}
{Dennison}, P.~A. and {Hewish}, A., 1967, {The Solar Wind outside the Plane of
  the Ecliptic}, \emph{\nat}, \textbf{213}, 343--346, \doi{10.1038/213343a0}.

\bibitem[{\emph{{Dudok de Wit} et~al.}(2005)\emph{{Dudok de Wit}, {Lilensten},
  {Aboudarham}, {Amblard}, and {Kretzschmar}}}]{dudokdewit_etal:05a}
{Dudok de Wit}, T., {Lilensten}, J., {Aboudarham}, J., {Amblard}, P.-O., and
  {Kretzschmar}, M., 2005, Retrieving the solar euv spectrum from a reduced set
  of spectral lines, \emph{\ag}, \textbf{23}, 3055--353.

\bibitem[{\emph{{Dudok de Wit} et~al.}(2008)\emph{{Dudok de Wit},
  {Kretzschmar}, {Aboudarham}, {Amblard}, {Auch{\`e}re}, and
  {Lilensten}}}]{dudokdewit_etal:08a}
{Dudok de Wit}, T., {Kretzschmar}, M., {Aboudarham}, J., {Amblard}, P.-O.,
  {Auch{\`e}re}, F., and {Lilensten}, J., 2008, Which solar {EUV} indices are
  best for reconstructing the solar {EUV} irradiance?, \emph{\asr},
  \textbf{42}, 903--911, \doi{10.1016/j.asr.2007.04.019}.

\bibitem[{\emph{{Dudok de Wit} et~al.}(2009)\emph{{Dudok de Wit},
  {Kretzschmar}, {Lilensten}, and {Woods}}}]{dudokdewit_etal:09a}
{Dudok de Wit}, T., {Kretzschmar}, M., {Lilensten}, J., and {Woods}, T., 2009,
  {Finding the best proxies for the solar UV irradiance}, \emph{\grl},
  \textbf{36}, L10\,107, \doi{10.1029/2009GL037825}.

\bibitem[{\emph{{Ebert} et~al.}(2009)\emph{{Ebert}, {McComas}, {Elliott},
  {Forsyth}, and {Gosling}}}]{ebert_etal:09a}
{Ebert}, R.~W., {McComas}, D.~J., {Elliott}, H.~A., {Forsyth}, R.~J., and
  {Gosling}, J.~T., 2009, {Bulk properties of the slow and fast solar wind and
  interplanetary coronal mass ejections measured by Ulysses: Three polar orbits
  of observations}, \emph{\jgr}, \textbf{114}, A1109,
  \doi{10.1029/2008JA013631}.

\bibitem[{\emph{{Emerich} et~al.}(2005)\emph{{Emerich}, {Lemaire}, {Vial},
  {Curdt}, {Sch{\"u}hle}, and {Wilhelm}}}]{emerich_etal:05}
{Emerich}, C., {Lemaire}, P., {Vial}, J.-C., {Curdt}, W., {Sch{\"u}hle}, U.,
  and {Wilhelm}, K., 2005, {A new relation between the central spectral solar H
  I Lyman {$\alpha$} irradiance and the line irradiance measured by SUMER/SOHO
  during the cycle 23}, \emph{Icarus}, \textbf{178}, 429--433.

\bibitem[{\emph{Fahr et~al.}(2007)\emph{Fahr, Fichtner, and
  Scherer}}]{fahr_etal:07a}
Fahr, H., Fichtner, H., and Scherer, K., 2007, Theoretical aspects of energetic
  neutral atoms as messengers from distant plasma sites with emphasis on the
  heliosphere, \emph{Rev. Geophys.}, \textbf{45}, RG4003,
  \doi{10.1029/2006RG000214}.

\bibitem[{\emph{Fahr}(1973)}]{fahr:73}
Fahr, H.~J., 1973, Non-thermal solar wind heating by supra-thermal ions,
  \emph{\solphys}, \textbf{30}, 193--206.

\bibitem[{\emph{Fahr}(1978)}]{fahr:78}
Fahr, H.~J., 1978, Change of interstellar gas parameters in stellar wind
  dominated astrospheres: solar case, \emph{\aap}, \textbf{66}, 103--117.

\bibitem[{\emph{Fahr}(1979)}]{fahr:79}
Fahr, H.~J., 1979, Interstellar hydrogen subject to a net repulsive solar force
  field, \emph{\aap}, \textbf{77}, 101--109.

\bibitem[{\emph{Fahr}(2004)}]{fahr:04a}
Fahr, H.~J., 2004, The {3D} heliosphere: three decades of growing knowledge,
  \emph{\asr}, \textbf{32}, 3--13.

\bibitem[{\emph{Fahr and Ruci{\'n}ski}(1999)}]{fahr_rucinski:99}
Fahr, H.~J. and Ruci{\'n}ski, D., 1999, Neutral interstellar gas atoms reducing
  the solar wind number and fractionally neutralizing the solar wind,
  \emph{\aap}, \textbf{350}, 1071--1078.

\bibitem[{\emph{{Fahr} and {Ruci{\'n}ski}}(2001)}]{fahr_rucinski:01a}
{Fahr}, H.~J. and {Ruci{\'n}ski}, D., 2001, Modification of properties and
  dynamics of distant solar wind due to its interaction with neutral
  interstellar gas, \emph{\ssr}, \textbf{97}, 407--412,
  \doi{10.1023/A:1011874311272}.

\bibitem[{\emph{{Fahr} and {Ruci{\'n}ski}}(2002)}]{fahr_rucinski:02a}
{Fahr}, H.~J. and {Ruci{\'n}ski}, D., 2002, Heliospheric pick-up ions
  influencing thermodynamics and dynamics of the distant solar wind,
  \emph{Nonlinear Processes in Geophysics}, \textbf{9}, 377--386.

\bibitem[{\emph{{Feldman} et~al.}(1973)\emph{{Feldman}, {Asbridge}, {Bame}, and
  {Montgomery}}}]{feldman_etal:73a}
{Feldman}, W.~C., {Asbridge}, J.~R., {Bame}, S.~J., and {Montgomery}, M.~D.,
  1973, {Double ion streams in the solar wind.}, \emph{\jgr}, \textbf{78},
  2017--2027, \doi{10.1029/JA078i013p02017}.

\bibitem[{\emph{Fite et~al.}(1962)\emph{Fite, Smith, and
  Stebbins}}]{fite_etal:62}
Fite, W.~L., Smith, A. C.~S., and Stebbins, R.~F., 1962, Charge transfer in
  collisions involving symmetric and asymmetric resonance, \emph{Proc. R. Soc.
  London Ser. A}, \textbf{268}, 527.

\bibitem[{\emph{{Floyd} et~al.}(2003)\emph{{Floyd}, {Rottman}, {Deland}, and
  {Pap}}}]{floyd_etal:03a}
{Floyd}, L., {Rottman}, G., {Deland}, M., and {Pap}, J., 2003, {11 years of
  solar UV irradiance measurements from UARS}, in \emph{Solar Variability as an
  Input to the Earth's Environment}, edited by A.~{Wilson}, vol. 535 of
  \emph{ESA Special Publication}, pp. 195--203.

\bibitem[{\emph{{Floyd} et~al.}(2005)\emph{{Floyd}, {Newmark}, {Cook},
  {Herring}, and {McMullin}}}]{floyd_etal:05a}
{Floyd}, L., {Newmark}, J., {Cook}, J., {Herring}, L., and {McMullin}, D.,
  2005, {Solar EUV and UV spectral irradiances and solar indices},
  \emph{Journal of Atmospheric and Solar-Terrestrial Physics}, \textbf{67},
  3--15, \doi{10.1016/j.jastp.2004.07.013}.

\bibitem[{\emph{{Floyd} et~al.}(2002)\emph{{Floyd}, {Prinz}, {Crane}, and
  {Herring}}}]{floyd_etal:02a}
{Floyd}, L.~E., {Prinz}, D.~K., {Crane}, P.~C., and {Herring}, L.~C., 2002,
  {Solar UV irradiance variation during cycles 22 and 23}, \emph{\asr},
  \textbf{29}, 1957--1962.

\bibitem[{\emph{{Frisch} et~al.}(2009)\emph{{Frisch}, {Bzowski}, {Gr{\"u}n},
  {Izmodenov}, {Kr{\"u}ger}, {Linsky}, {McComas}, {M{\"o}bius}, {Redfield},
  {Schwadron}, {Shelton}, {Slavin}, and {Wood}}}]{frisch_etal:09a}
{Frisch}, P.~C., {Bzowski}, M., {Gr{\"u}n}, E., {Izmodenov}, V., {Kr{\"u}ger},
  H., {Linsky}, J.~L., {McComas}, D.~J., {M{\"o}bius}, E., {Redfield}, S.,
  {Schwadron}, N., {Shelton}, R., {Slavin}, J.~D., and {Wood}, B.~E., 2009, The
  galactic environment of the {S}un: {I}nterstellar material inside and outside
  of the heliosphere, \emph{\ssr}, \textbf{146}, 235--273,
  \doi{10.1007/s11214-009-9502-0}.

\bibitem[{\emph{{Frisch} et~al.}(2011)\emph{{Frisch}, {Redfield}, and
  {Slavin}}}]{frisch_etal:11a}
{Frisch}, P.~C., {Redfield}, S., and {Slavin}, J.~D., 2011, The interstellar
  medium surrounding the {Sun}, \emph{\araa}, \textbf{49}, 237--279,
  \doi{10.1146/annurev-astro-081710-102613}.

\bibitem[{\emph{{Fujiki} et~al.}(2003{\natexlab{a}})\emph{{Fujiki}, {Kojima},
  {Tokumaru}, {Ohmi}, {Yokobe}, and {Hayashi}}}]{fujiki_etal:03c}
{Fujiki}, K., {Kojima}, M., {Tokumaru}, M., {Ohmi}, T., {Yokobe}, A., and
  {Hayashi}, K., 2003{\natexlab{a}}, {Solar Cycle Dependence of High-Latitude
  Solar Wind}, in \emph{Solar Wind Ten}, edited by {M.~Velli, R.~Bruno,
  F.~Malara, \& B.~Bucci}, vol. 679 of \emph{American Institute of Physics
  Conference Series}, pp. 141--143, \doi{10.1063/1.1618561}.

\bibitem[{\emph{{Fujiki} et~al.}(2003{\natexlab{b}})\emph{{Fujiki}, {Kojima},
  {Tokumaru}, {Ohmi}, {Yokobe}, {Hayashi}, {McComas}, and
  {Elliott}}}]{fujiki_etal:03a}
{Fujiki}, K., {Kojima}, M., {Tokumaru}, M., {Ohmi}, T., {Yokobe}, A.,
  {Hayashi}, K., {McComas}, D.~J., and {Elliott}, H.~A., 2003{\natexlab{b}},
  {Solar wind velocity structure around the solar maximum observed by
  interplanetary scintillation}, in \emph{Solar Wind Ten}, edited by {M.~Velli,
  R.~Bruno, F.~Malara, \& B.~Bucci}, vol. 679 of \emph{American Institute of
  Physics Conference Series}, pp. 226--229, \doi{10.1063/1.1618583}.

\bibitem[{\emph{{Fujiki} et~al.}(2003{\natexlab{c}})\emph{{Fujiki}, {Kojima},
  {Tokumaru}, {Ohmi}, {Yokobe}, {Hayashi}, {McComas}, and
  {Elliott}}}]{fujiki_etal:03b}
{Fujiki}, K., {Kojima}, M., {Tokumaru}, M., {Ohmi}, T., {Yokobe}, A.,
  {Hayashi}, K., {McComas}, D.~J., and {Elliott}, H.~A., 2003{\natexlab{c}},
  {How did the solar wind structure change around the solar maximum? From
  interplanetary scintillation observation}, \emph{\ag}, \textbf{21},
  1257--1261, \doi{10.5194/angeo-21-1257-2003}.

\bibitem[{\emph{{Gloeckler} and {Geiss}}(2001)}]{gloeckler_geiss:01a}
{Gloeckler}, G. and {Geiss}, J., 2001, {Heliospheric and Interstellar Phenomena
  Deduced From Pickup ion Observations}, \emph{\ssr}, \textbf{97}, 169--181.

\bibitem[{\emph{{Gloeckler} et~al.}(1992)\emph{{Gloeckler}, {Geiss},
  {Balsiger}, {Bedini}, {Cain}, {Fisher}, {Fisk}, {Galvin}, {Gliem}, and
  {Hamilton}}}]{gloeckler_etal:92}
{Gloeckler}, G., {Geiss}, J., {Balsiger}, H., {Bedini}, P., {Cain}, J.~C.,
  {Fisher}, J., {Fisk}, L.~A., {Galvin}, A.~B., {Gliem}, F., and {Hamilton},
  D.~C., 1992, {The Solar Wind Ion Composition Spectrometer}, \emph{\aaps},
  \textbf{92}, 267--289.

\bibitem[{\emph{Gloeckler et~al.}(1993)\emph{Gloeckler, Geiss, Balsiger, Fisk,
  Galvin, Ipavich, Ogilvie, von Steiger, and Wilken}}]{gloeckler_etal:93a}
Gloeckler, G., Geiss, J., Balsiger, H., Fisk, L.~A., Galvin, A.~B., Ipavich,
  F.~M., Ogilvie, K.~W., von Steiger, R., and Wilken, B., 1993, Detection of
  interstellar pickup hydrogen in the solar system, \emph{Science},
  \textbf{261}, 70--73.

\bibitem[{\emph{{Gloeckler} et~al.}(2004)\emph{{Gloeckler}, {M{\"o}bius},
  {Geiss}, {Bzowski}, {Chalov}, {Fahr}, {McMullin}, {Noda}, {Oka},
  {Ruci{\'n}ski}, {Skoug}, {Terasawa}, {von Steiger}, {Yamazaki}, and
  {Zurbuchen}}}]{gloeckler_etal:04b}
{Gloeckler}, G., {M{\"o}bius}, E., {Geiss}, J., {Bzowski}, M., {Chalov}, S.,
  {Fahr}, H., {McMullin}, D.~R., {Noda}, H., {Oka}, M., {Ruci{\'n}ski}, D.,
  {Skoug}, R., {Terasawa}, T., {von Steiger}, R., {Yamazaki}, A., and
  {Zurbuchen}, T., 2004, {Observations of the helium focusing cone with pickup
  ions}, \emph{\aap}, \textbf{426}, 845--854.

\bibitem[{\emph{Gringauz et~al.}(1960)\emph{Gringauz, Bezrukih, Ozerov, and
  Ribchinsky}}]{gringauz_etal:60a}
Gringauz, K., Bezrukih, V., Ozerov, V., and Ribchinsky, R., 1960, A study of
  the interplanetary ionized gas, high-energy electrons and corpuscular
  radiation from the {S}un by means of the three electrode trap for charged
  particles on the second soviet cosmic rocket, \emph{Soviet Physics Doklady},
  \textbf{5}, 361.

\bibitem[{\emph{{Harmon}}(1975)}]{harmon:75a}
{Harmon}, J.~K., 1975, \emph{{Scintillation studies of density microstructure
  in the solar wind plasma}}, Ph.D. thesis, California Univ., San Diego.

\bibitem[{\emph{Harvey and Recely}(2002)}]{harvey_recely:02}
Harvey, K.~L. and Recely, F., 2002, Polar coronal holes during cycles 22 and
  23, \emph{\solphys}, \textbf{211}, 31--52.

\bibitem[{\emph{{Hayashi} et~al.}(2003)\emph{{Hayashi}, {Kojima}, {Tokumaru},
  and {Fujiki}}}]{hayashi_etal:03a}
{Hayashi}, K., {Kojima}, M., {Tokumaru}, M., and {Fujiki}, K., 2003, {MHD
  tomography using interplanetary scintillation measurement}, \emph{\jgr},
  \textbf{108}, 1102, \doi{10.1029/2002JA009567}.

\bibitem[{\emph{{Heath} and {Schlesinger}}(1986)}]{heath_schlesinger:86}
{Heath}, D.~F. and {Schlesinger}, B.~M., 1986, {The Mg 280-nm doublet as a
  monitor of changes in solar ultraviolet irradiance}, \emph{\jgr},
  \textbf{91}, 8672--8682, \doi{10.1029/JD091iD08p08672}.

\bibitem[{\emph{{Hewish} and {Symonds}}(1969)}]{hewish_symonds:69a}
{Hewish}, A. and {Symonds}, M.~D., 1969, {Radio investigation of the solar
  plasma}, \emph{\pss}, \textbf{17}, 313, \doi{10.1016/0032-0633(69)90064-6}.

\bibitem[{\emph{{Hewish} et~al.}(1964)\emph{{Hewish}, {Scott}, and
  {Wills}}}]{hewish_etal:64a}
{Hewish}, A., {Scott}, P.~F., and {Wills}, D., 1964, Interplanetary
  scintillation of small diameter radio sources, \emph{\nat}, \textbf{203},
  1214--1217, \doi{10.1038/2031214a0}.

\bibitem[{\emph{{Hinteregger} et~al.}(1981)\emph{{Hinteregger}, {Fukui}, and
  {Gilson}}}]{hinteregger_etal:81a}
{Hinteregger}, H.~E., {Fukui}, K., and {Gilson}, B.~R., 1981, {Observational,
  reference and model data on solar EUV, from measurements on AE-E},
  \emph{\grl}, \textbf{8}, 1147--1150, \doi{10.1029/GL008i011p01147}.

\bibitem[{\emph{{Houminer}}(1971)}]{houminer:71a}
{Houminer}, Z., 1971, {Radio source scintillation-Evidence of plasma streams
  corotating about the Sun}, \emph{\nat}, \textbf{231}, 165.

\bibitem[{\emph{{Hovestadt} et~al.}(1995)\emph{{Hovestadt}, {Hilchenbach},
  {B{\"u}rgi}, {Klecker}, {Laeverenz}, {Scholer}, {Gr{\"u}nwaldt}, {Axford},
  {Livi}, {Marsch}, {Wilken}, {Winterhoff}, {Ipavich}, {Bedini}, {Coplan},
  {Galvin}, {Gloeckler}, {Bochsler}, {Balsiger}, {Fischer}, {Geiss},
  {Kallenbach}, {Wurz}, {Reiche}, {Gliem}, {Judge}, {Ogawa}, {Hsieh},
  {M{\"o}bius}, {Lee}, {Managadze}, {Verigin}, and
  {Neugebauer}}}]{hovestadt_etal:95a}
{Hovestadt}, D., {Hilchenbach}, M., {B{\"u}rgi}, A., {Klecker}, B.,
  {Laeverenz}, P., {Scholer}, M., {Gr{\"u}nwaldt}, H., {Axford}, W.~I., {Livi},
  S., {Marsch}, E., {Wilken}, B., {Winterhoff}, H.~P., {Ipavich}, F.~M.,
  {Bedini}, P., {Coplan}, M.~A., {Galvin}, A.~B., {Gloeckler}, G., {Bochsler},
  P., {Balsiger}, H., {Fischer}, J., {Geiss}, J., {Kallenbach}, R., {Wurz}, P.,
  {Reiche}, K.-U., {Gliem}, F., {Judge}, D.~L., {Ogawa}, H.~S., {Hsieh}, K.~C.,
  {M{\"o}bius}, E., {Lee}, M.~A., {Managadze}, G.~G., {Verigin}, M.~I., and
  {Neugebauer}, M., 1995, {CELIAS - Charge, Element and Isotope Analysis System
  for SOHO}, \emph{\solphys}, \textbf{162}, 441--481, \doi{10.1007/BF00733436}.

\bibitem[{\emph{Hundhausen et~al.}(1967)\emph{Hundhausen, Asbridge, Gilbert,
  and Strong}}]{hundhausen_etal:67a}
Hundhausen, A.~J., Asbridge, J.~R., Gilbert, S. J. B. H.~E., and Strong, I.~B.,
  1967, Vela 3 satellite observations of solar wind ions, \emph{\jgr},
  \textbf{72}, 1979, \doi{10.1029/JZ072i007p01979}.

\bibitem[{\emph{{Isobe} et~al.}(1990)\emph{{Isobe}, {Feigelson}, {Akritas}, and
  {Babu}}}]{isobe_etal:90a}
{Isobe}, T., {Feigelson}, E.~D., {Akritas}, M.~G., and {Babu}, G.~J., 1990,
  {Linear regression in astronomy.}, \emph{\apj}, \textbf{364}, 104--113,
  \doi{10.1086/169390}.

\bibitem[{\emph{{Issautier}}(2009)}]{issautier:09a}
{Issautier}, K., 2009, {Diagnostics of the Solar Wind Plasma}, in
  \emph{Turbulence in Space Plasmas}, edited by {P.~Cargill \& L.~Vlahos}, vol.
  778 of \emph{Lecture Notes in Physics, Berlin Springer Verlag}, pp. 223--246.

\bibitem[{\emph{{Issautier} et~al.}(1998)\emph{{Issautier}, {Meyer-Vernet},
  {Moncuquet}, and {Hoang}}}]{issautier_etal:98}
{Issautier}, K., {Meyer-Vernet}, N., {Moncuquet}, M., and {Hoang}, S., 1998,
  {Solar wind radial and latitudinal structure - Electron density and core
  temperature from ULYSSES thermal noise spectroscopy}, \emph{\jgr},
  \textbf{103}, 1969--1979.

\bibitem[{\emph{{Issautier} et~al.}(2001)\emph{{Issautier}, {Skoug}, {Gosling},
  {Gary}, and {McComas}}}]{issautier_etal:01a}
{Issautier}, K., {Skoug}, R.~M., {Gosling}, J.~T., {Gary}, S.~P., and
  {McComas}, D.~J., 2001, {Solar wind plasma parameters on Ulysses: Detailed
  comparison between the URAP and SWOOPS experiments}, \emph{\jgr},
  \textbf{106}, 15\,665--15\,676, \doi{10.1029/2000JA000412}.

\bibitem[{\emph{Izmodenov and Baranov}(2006)}]{izmodenov_baranov:06a}
Izmodenov, V.~V. and Baranov, V.~B., 2006, Modern multi-component models of the
  heliospheric interface, in \emph{The physics of the heliospheric boundaries},
  edited by V.~Izmodenov and R.~Kallenbach, vol. SR-005 of \emph{ISSI Sci.Rep},
  pp. 67--136.

\bibitem[{\emph{{Izmodenov} et~al.}(2000)\emph{{Izmodenov}, {Malama},
  {Kalinin}, {Gruntman}, {Lallement}, and {Rodionova}}}]{izmodenov_etal:00a}
{Izmodenov}, V.~V., {Malama}, Y.~G., {Kalinin}, A.~P., {Gruntman}, M.,
  {Lallement}, R., and {Rodionova}, I.~P., 2000, Hot neutral {H} in the
  heliosphere: elastic {H-H, H-p} collisions, \emph{\apss}, \textbf{274},
  71--76, \doi{10.1023/A:1026531519864}.

\bibitem[{\emph{Izmodenov et~al.}(2009)\emph{Izmodenov, Alexashov, Chalov,
  Katushkina, Malama, and Provornikova}}]{izmodenov_etal:09a}
Izmodenov, V.~V., Alexashov, D.~B., Chalov, S.~V., Katushkina, O.~A., Malama,
  Y.~G., and Provornikova, E.~A., 2009, Kinetic-gasdynamic modeling of the
  heliospheric interface: global structure, interstellar atoms and heliospheric
  enas, \emph{\ssr}, \textbf{146}, 329--351, \doi{10.1007/s11214-009-9528-3}.

\bibitem[{\emph{{Jackson} et~al.}(1997)\emph{{Jackson}, {Hick}, {Kojima}, and
  {Yokobe}}}]{jackson_etal:97a}
{Jackson}, B.~V., {Hick}, P.~L., {Kojima}, M., and {Yokobe}, A., 1997,
  {Heliospheric tomography using interplanetary scintillation observations},
  \emph{\asr}, \textbf{20}, 23--26, \doi{10.1016/S0273-1177(97)00474-2}.

\bibitem[{\emph{Jackson et~al.}(1998)\emph{Jackson, Hick, Kojima, and
  Yokobe}}]{jackson_etal:98a}
Jackson, B.~V., Hick, P.~L., Kojima, M., and Yokobe, A., 1998, Heliospheric
  tomography using interplanetary scintillation observations. i. combined
  nagoya and cambridge data, \emph{\jgr}, \textbf{103}, 12\,049--12\,067.

\bibitem[{\emph{{Judge} et~al.}(1998)\emph{{Judge}, {McMullin}, {Ogawa},
  {Hovestadt}, {Klecker}, {Hilchenbach}, {M{\"o}bius}, {Canfield}, {Vest},
  {Watts}, {Tarrio}, {Kuehne}, and {Wurz}}}]{judge_etal:98}
{Judge}, D.~L., {McMullin}, D.~R., {Ogawa}, H.~S., {Hovestadt}, D., {Klecker},
  B., {Hilchenbach}, M., {M{\"o}bius}, E., {Canfield}, L.~R., {Vest}, R.~E.,
  {Watts}, R., {Tarrio}, C., {Kuehne}, M., and {Wurz}, P., 1998, {First Solar
  EUV Irradiances Obtained from SOHO by the CELIAS/SEM}, \emph{\solphys},
  \textbf{177}, 161--173.

\bibitem[{\emph{{Kasper}}(2002)}]{kasper:02a}
{Kasper}, J.~C., 2002, \emph{{Solar wind plasma: Kinetic properties and micro-
  instabilities}}, Ph.D. thesis, MASSACHUSETTS INSTITUTE OF TECHNOLOGY.

\bibitem[{\emph{{Kasper} et~al.}(2006)\emph{{Kasper}, {Lazarus}, {Steinberg},
  {Ogilvie}, and {Szabo}}}]{kasper_etal:06a}
{Kasper}, J.~C., {Lazarus}, A.~J., {Steinberg}, J.~T., {Ogilvie}, K.~W., and
  {Szabo}, A., 2006, {Physics-based tests to identify the accuracy of solar
  wind ion measurements: A case study with the Wind Faraday Cups}, \emph{\jgr},
  \textbf{111}, A03\,105, \doi{10.1029/2005JA011442}.

\bibitem[{\emph{{Katushkina} and {Izmodenov}}(2010)}]{katushkina_izmodenov:10}
{Katushkina}, O.~A. and {Izmodenov}, V.~V., 2010, {Effect of the heliospheric
  interface on the distribution of interstellar hydrogen atom inside the
  heliosphere}, \emph{Astronomy Letters}, \textbf{36}, 297--306,
  \doi{10.1134/S1063773710040080}.

\bibitem[{\emph{{King} and {Papitashvili}}(2005)}]{king_papitashvili:05}
{King}, J.~H. and {Papitashvili}, N.~E., 2005, {Solar wind spatial scales in
  and comparisons of hourly Wind and ACE plasma and magnetic field data},
  \emph{\jgr}, \textbf{110}, 2104--2111, \doi{10.1029/2004JA010649}.

\bibitem[{\emph{{Kiselman} et~al.}(2011)\emph{{Kiselman}, {Pereira},
  {Gustafsson}, {Asplund}, {Mel{\'e}ndez}, and {Langhans}}}]{kiselman_etal:11a}
{Kiselman}, D., {Pereira}, T., {Gustafsson}, B., {Asplund}, M., {Mel{\'e}ndez},
  J., and {Langhans}, K., 2011, {Is the solar spectrum latitude dependent? An
  investigation with SST/TRIPPEL}, \emph{\aanda}, \textbf{535}, A18.

\bibitem[{\emph{{Kojima} and {Kakinuma}}(1987)}]{kojima_kakinuma:87a}
{Kojima}, M. and {Kakinuma}, T., 1987, {Solar cycle evolution of solar wind
  speed structure between 1973 and 1985 observed with the interplanetary
  scintillation method}, \emph{\jgr}, \textbf{92}, 7269--7279,
  \doi{10.1029/JA092iA07p07269}.

\bibitem[{\emph{{Kojima} and {Kakinuma}}(1990)}]{kojima_kakinuma:90a}
{Kojima}, M. and {Kakinuma}, T., 1990, {Solar cycle dependence of global
  distribution of solar wind speed}, \emph{\ssr}, \textbf{53}, 173--222,
  \doi{10.1007/BF00212754}.

\bibitem[{\emph{Kojima et~al.}(1998)\emph{Kojima, Tokumaru, Watanabe, Yokobe,
  Asai, Jackson, and Hick}}]{kojima_etal:98}
Kojima, M., Tokumaru, M., Watanabe, H., Yokobe, A., Asai, K., Jackson, B.~V.,
  and Hick, P.~L., 1998, Heliospheric tomography using interplanetary
  scintillation observations. 2. {L}atitude and heliocentric distance
  dependence of solar wind structure at 0.1--1~{AU}, \emph{\jgr}, \textbf{103},
  1981--1989.

\bibitem[{\emph{{Kojima} et~al.}(1999)\emph{{Kojima}, {Fujiki}, {Ohmi},
  {Tokumaru}, {Yokobe}, and {Hakamada}}}]{kojima_etal:99a}
{Kojima}, M., {Fujiki}, K., {Ohmi}, T., {Tokumaru}, M., {Yokobe}, A., and
  {Hakamada}, K., 1999, {The Highest Solar Wind Velocity in a Polar Region
  Estimated from IPS Tomography Analysis}, \emph{\ssr}, \textbf{87}, 237--239,
  \doi{10.1023/A:1005108820106}.

\bibitem[{\emph{Kojima et~al.}(2001)\emph{Kojima, Fujiki, Ohmi, Tokumaru,
  Yokobe, and Hakamada}}]{kojima_etal:01}
Kojima, M., Fujiki, K., Ohmi, T., Tokumaru, M., Yokobe, A., and Hakamada, K.,
  2001, Latitudinal velocity structures up to the solar poles estimated from
  interplanetary scintillation tomography analysis, \emph{\jgr}, \textbf{106},
  15\,677--15\,686.

\bibitem[{\emph{{Kojima} et~al.}(2007)\emph{{Kojima}, {Tokumaru}, {Fujiki},
  {Hayashi}, and {Jackson}}}]{kojima_etal:07a}
{Kojima}, M., {Tokumaru}, M., {Fujiki}, K., {Hayashi}, K., and {Jackson},
  B.~V., 2007, {IPS tomographic observations of 3D solar wind structure},
  \emph{Astronomical and Astrophysical Transactions}, \textbf{26}, 467--476.

\bibitem[{\emph{{Kretzschmar} et~al.}(2006)\emph{{Kretzschmar}, {Lilensten},
  and {Aboudarham}}}]{kretzschmar_etal:06a}
{Kretzschmar}, M., {Lilensten}, J., and {Aboudarham}, J., 2006, {Retrieving the
  solar EUV spectral irradiance from the observation of 6 lines}, \emph{\asr},
  \textbf{37}, 341--346, \doi{10.1016/j.asr.2005.02.029}.

\bibitem[{\emph{Kumar and Broadfoot}(1978)}]{kumar_broadfoot:78}
Kumar, S. and Broadfoot, A.~L., 1978, Evidence from {M}ariner~10 of solar wind
  flux depletion at high ecliptic latitudes, \emph{\aap}, \textbf{69}, L5--L8.

\bibitem[{\emph{Kumar and Broadfoot}(1979)}]{kumar_broadfoot:79}
Kumar, S. and Broadfoot, A.~L., 1979, Signatures of solar wind latitudinal
  structure in interplanetary {L}yman-$\alpha$ emissions: {M}ariner~10
  observations, \emph{\apj}, \textbf{228}, 302--311.

\bibitem[{\emph{Kyr{\"o}l{\"a} et~al.}(1994)\emph{Kyr{\"o}l{\"a}, Summanen, and
  R{\aa}back}}]{kyrola_etal:94}
Kyr{\"o}l{\"a}, E., Summanen, T., and R{\aa}back, P., 1994, Solar cycle and
  interplanetary hydrogen, \emph{\aap}, \textbf{288}, 299--314.

\bibitem[{\emph{Kyr{\"o}l{\"a} et~al.}(1998)\emph{Kyr{\"o}l{\"a}, Summanen,
  M{\"a}kinen, Qu{\'e}merais, Bertaux, Lallement, and Costa}}]{kyrola_etal:98}
Kyr{\"o}l{\"a}, E., Summanen, T., M{\"a}kinen, T., Qu{\'e}merais, E., Bertaux,
  J.-L., Lallement, R., and Costa, J., 1998, Preliminary retrieval of solar
  wind anisotropies / {SOHO} observations, \emph{\jgr}, \textbf{103},
  14\,523--14\,538.

\bibitem[{\emph{Lallement and Stewart}(1990)}]{lallement_stewart:90}
Lallement, R. and Stewart, A.~I., 1990, Out-of-ecliptic {L}yman-alpha
  observations with {P}ioneer-{V}enus: solar wind anisotropy degree in 1986,
  \emph{\aap}, \textbf{227}, 600--608.

\bibitem[{\emph{Lallement et~al.}(1985{\natexlab{a}})\emph{Lallement, Bertaux,
  and Dalaudier}}]{lallement_etal:85b}
Lallement, R., Bertaux, J.~L., and Dalaudier, F., 1985{\natexlab{a}},
  Interplanetary {L}yman $\alpha$ spectral profiles and intensities for both
  repulsive and attractive solar force fields: {P}redicted absorption pattern
  by a hydrogen cell, \emph{\aap}, \textbf{150}, 21--32.

\bibitem[{\emph{Lallement et~al.}(1985{\natexlab{b}})\emph{Lallement, Bertaux,
  and Kurt}}]{lallement_etal:85a}
Lallement, R., Bertaux, J.~L., and Kurt, V.~G., 1985{\natexlab{b}}, Solar wind
  decrease at high heliographic latitudes detected from {P}rognoz
  interplanetary {L}yman {A}lpha mapping, \emph{\jgr}, \textbf{90}, 1413--1420.

\bibitem[{\emph{Lallement et~al.}(1986)\emph{Lallement, Holzer, and
  Munro}}]{lallement_etal:86}
Lallement, R., Holzer, T.~E., and Munro, R.~H., 1986, Solar wind expansion in a
  polar coronal hole: inferences from coronal white light and interplanetary
  {L}yman alpha observations, \emph{\jgr}, \textbf{91}, 6751--6759.

\bibitem[{\emph{{Lallement} et~al.}(2010)\emph{{Lallement}, {Qu{\'e}merais},
  {Lamy}, {Bertaux}, {Ferron}, and {Schmidt}}}]{lallement_etal:10b}
{Lallement}, R., {Qu{\'e}merais}, E., {Lamy}, P., {Bertaux}, J.-L., {Ferron},
  S., and {Schmidt}, W., 2010, The solar wind as seen by {SOHO/SWAN} since
  1996: {C}omparison with {SOHO/LASCO C2} coronal densities, in \emph{SOHO-23:
  Understanding a Peculiar Solar Minimum}, edited by {S.~R.~Cranmer,
  J.~T.~Hoeksema, \& J.~L.~Kohl}, vol. 428 of \emph{Astronomical Society of the
  Pacific Conference Series}, pp. 253--258.

\bibitem[{\emph{{Lazarus} and {Paularena}}(1998)}]{lazarus_paularena:98a}
{Lazarus}, A.~J. and {Paularena}, K.~I., 1998, {A Comparison of Solar Wind
  Parameters from Experiments on the IMP 8 and Wind Spacecraft}, in
  \emph{Measurement Techniques in Space Plasmas -- Particles}, edited by
  {R.~F.~Pfaff, J.~E.~Borovsky, \& D.~T.~Young}, pp. 85--90.

\bibitem[{\emph{{Le Chat} et~al.}(2010)\emph{{Le Chat}, {Issautier},
  {Meyer-Vernet}, {Zouganelis}, {Moncuquet}, and {Hoang}}}]{leChat_etal:10a}
{Le Chat}, G., {Issautier}, K., {Meyer-Vernet}, N., {Zouganelis}, I.,
  {Moncuquet}, M., and {Hoang}, S., 2010, {Quasi-thermal noise spectroscopy:
  preliminary comparison between kappa and sum of two Maxwellian
  distributions}, \emph{Twelfth International Solar Wind Conference},
  \textbf{1216}, 316--319, \doi{10.1063/1.3395864}.

\bibitem[{\emph{{Le Chat} et~al.}(2011)\emph{{Le Chat}, {Issautier},
  {Meyer-Vernet}, and {Hoang}}}]{leChat_etal:11a}
{Le Chat}, G., {Issautier}, K., {Meyer-Vernet}, N., and {Hoang}, S., 2011,
  Large-scale variation of solar wind electron properties from quasi-thermal
  noise spectroscopy: {Ulysses} measurements, \emph{\solphys}, \textbf{271},
  141--148, \doi{10.1007/s11207-011-9797-3}.

\bibitem[{\emph{{Lean} et~al.}(2003)\emph{{Lean}, {Warren}, {Mariska}, and
  {Bishop}}}]{lean_etal:03a}
{Lean}, J.~L., {Warren}, H.~P., {Mariska}, J.~T., and {Bishop}, J., 2003, {A
  new model of solar EUV irradiance variability 2. Comparisons with empirical
  models and observations and implications for space weather}, \emph{\jgr},
  \textbf{108}, 1059, \doi{10.1029/2001JA009238}.

\bibitem[{\emph{{Lean} et~al.}(2011)\emph{{Lean}, {Woods}, {Eparvier}, {Meier},
  {Strickland}, {Correira}, and {Evans}}}]{lean_etal:11a}
{Lean}, J.~L., {Woods}, T.~N., {Eparvier}, F.~G., {Meier}, R.~R., {Strickland},
  D.~J., {Correira}, J.~T., and {Evans}, J.~S., 2011, {Solar extreme
  ultraviolet irradiance: Present, past, and future}, \emph{\jgr},
  \textbf{116}, A01\,102, \doi{10.1029/2010JA015901}.

\bibitem[{\emph{{Lemaire} et~al.}(1978)\emph{{Lemaire}, {Charra}, {Jouchoux},
  {Vidal-Madjar}, {Artzner}, {Vial}, {Bonnet}, and
  {Skumanich}}}]{lemaire_etal:78a}
{Lemaire}, P., {Charra}, J., {Jouchoux}, A., {Vidal-Madjar}, A., {Artzner},
  G.~E., {Vial}, J.~C., {Bonnet}, R.~M., and {Skumanich}, A., 1978, {Calibrated
  full disk solar H I Lyman-alpha and Lyman-beta profiles}, \emph{\apjl},
  \textbf{223}, L55--L58, \doi{10.1086/182727}.

\bibitem[{\emph{Lemaire et~al.}(1998)\emph{Lemaire, Emerich, Curdt,
  Sch{\"u}hle, and Wilhelm}}]{lemaire_etal:98}
Lemaire, P., Emerich, C., Curdt, W., Sch{\"u}hle, U., and Wilhelm, K., 1998,
  Solar {HI} {L}yman~$\alpha$ full disk profile obtained with the {SUMER/SOHO}
  spectrometer, \emph{\aap}, \textbf{334}, 1095--1098.

\bibitem[{\emph{{Lemaire} et~al.}(2005)\emph{{Lemaire}, {Emerich}, {Vial},
  {Curdt}, {Sch{\"u}hle}, and {Wilhelm}}}]{lemaire_etal:05}
{Lemaire}, P., {Emerich}, C., {Vial}, J.-C., {Curdt}, W., {Sch{\"u}hle}, U.,
  and {Wilhelm}, K., 2005, {Variation of the full Sun hydrogen Lyman profiles
  through solar cycle 23}, \emph{Advances in Space Research}, \textbf{35},
  384--387.

\bibitem[{\emph{Lemaire et~al.}(2002)\emph{Lemaire, Emerich, Vial, Curdt,
  Sch{\"u}le, and Wilhelm}}]{lemaire_etal:02}
Lemaire, P.~L., Emerich, C., Vial, J.~C., Curdt, W., Sch{\"u}le, U., and
  Wilhelm, K., 2002, Variation of the full sun hydrogen lyman $\alpha$ and
  $\beta$ profiles with the activity cycle, in \emph{ESA SP-508: From Solar Min
  to Max: Half a Solar Cycle with SOHO}, pp. 219--222.

\bibitem[{\emph{{Liewer} et~al.}(1993)\emph{{Liewer}, {Goldstein}, and
  {Omidi}}}]{liewer_etal:93a}
{Liewer}, P.~C., {Goldstein}, B.~E., and {Omidi}, N., 1993, {Hybrid simulations
  of the effects of interstellar pickup hydrogen on the solar wind termination
  shock}, \emph{\jgr}, \textbf{981}, 15\,211--15\,220, \doi{10.1029/93JA01172}.

\bibitem[{\emph{{Lindsay} and {Stebbings}}(2005)}]{lindsay_stebbings:05a}
{Lindsay}, B.~G. and {Stebbings}, R.~F., 2005, {Charge transfer cross sections
  for energetic neutral atom data analysis}, \emph{\jgr}, \textbf{110},
  A12\,213, \doi{10.1029/2005JA011298}.

\bibitem[{\emph{Lotz}(1967{\natexlab{a}})}]{lotz:67}
Lotz, W., 1967{\natexlab{a}}, An empirical formula for the electron-impact
  ionization cross-section, \emph{Z. Phys.}, \textbf{206}, 205--211.

\bibitem[{\emph{Lotz}(1967{\natexlab{b}})}]{lotz:67a}
Lotz, W., 1967{\natexlab{b}}, Electron-impact ionization cross-sections and
  ionization rate coefficients for atoms and ions, \emph{\apjs}, \textbf{14},
  207--238.

\bibitem[{\emph{{Lyon} et~al.}(1967)\emph{{Lyon}, {Bridge}, and
  {Binsack}}}]{lyon_etal:67a}
{Lyon}, E.~F., {Bridge}, H.~S., and {Binsack}, J.~H., 1967, Explorer 35 plasma
  measurements in the vicinity of the {M}oon, \emph{\jgr}, \textbf{72},
  6113--6117, \doi{10.1029/JZ072i023p06113}.

\bibitem[{\emph{Lyon et~al.}(1968)\emph{Lyon, Egidi, Pizella, Bridge, Binsack,
  Baker, and Butler}}]{lyon_etal:68a}
Lyon, E.~F., Egidi, A., Pizella, G., Bridge, H.~S., Binsack, J.~S., Baker, R.,
  and Butler, R., 1968, Plasma measurements on {Explorer 33 (I)} interplanetary
  region, \emph{Space Research}, \textbf{VIII}, 99.

\bibitem[{\emph{Maher and Tinsley}(1977)}]{maher_tinsley:77}
Maher, L.~J. and Tinsley, B.~A., 1977, Atomic hydrogen escape rate due to
  charge exchange with hot plasmaspheric ions, \emph{\jgr}, \textbf{82},
  689--695.

\bibitem[{\emph{{Maksimovic} et~al.}(2000)\emph{{Maksimovic}, {Gary}, and
  {Skoug}}}]{maksimovic_etal:00a}
{Maksimovic}, M., {Gary}, S.~P., and {Skoug}, R.~M., 2000, {Solar wind electron
  suprathermal strength and temperature gradients: Ulysses observations},
  \emph{\jgr}, \textbf{105}, 18\,337--18\,350.

\bibitem[{\emph{{Maksimovic} et~al.}(2005)\emph{{Maksimovic}, {Zouganelis},
  {Chaufray}, {Issautier}, {Scime}, {Littleton}, {Marsch}, {McComas}, {Salem},
  {Lin}, and {Elliott}}}]{maksimovic_etal:05a}
{Maksimovic}, M., {Zouganelis}, I., {Chaufray}, J.-Y., {Issautier}, K.,
  {Scime}, E.~E., {Littleton}, J.~E., {Marsch}, E., {McComas}, D.~J., {Salem},
  C., {Lin}, R.~P., and {Elliott}, H., 2005, {Radial evolution of the electron
  distribution functions in the fast solar wind between 0.3 and 1.5 AU},
  \emph{\jgr}, \textbf{110}, A9104, \doi{10.1029/2005JA011119}.

\bibitem[{\emph{Malama et~al.}(2006)\emph{Malama, Izmodenov, and
  Chalov}}]{malama_etal:06}
Malama, Y., Izmodenov, V.~V., and Chalov, S.~V., 2006, Modeling of the
  heliospheric interface: multi-component nature of the heliospheric plasma,
  \emph{\aap}, \textbf{445}, 693--701.

\bibitem[{\emph{{Manoharan}}(1993)}]{manoharan:93b}
{Manoharan}, P.~K., 1993, {Three-dimensional structure of the solar wind:
  Variation of density with the solar cycle}, \emph{\solphys}, \textbf{148},
  153--167, \doi{10.1007/BF00675541}.

\bibitem[{\emph{Marsden and Smith}(1997)}]{marsden_smith:97}
Marsden, R.~G. and Smith, E.~J., 1997, Ulysses: a summary of the first
  high-latitude survey, \emph{\asr}, \textbf{19}, (6)825--(6)834.

\bibitem[{\emph{McComas et~al.}(1998)\emph{McComas, Bame, Barker, Feldman, and
  Phillips}}]{mccomas_etal:98a}
McComas, D.~J., Bame, S., Barker, P., Feldman, W., and Phillips, J., 1998,
  Solar wind electron proton alpha monitor {(SWEPAM)} for the {Advanced
  Composition Explorer}, \emph{\ssr}, \textbf{86}, 563--612.

\bibitem[{\emph{{McComas} et~al.}(1998)\emph{{McComas}, {Bame}, {Barraclough},
  {Feldman}, {Funsten}, {Gosling}, {Riley}, {Skoug}, {Balogh}, {Forsyth},
  {Goldstein}, and {Neugebauer}}}]{mccomas_etal:98b}
{McComas}, D.~J., {Bame}, S.~J., {Barraclough}, B.~L., {Feldman}, W.~C.,
  {Funsten}, H.~O., {Gosling}, J.~T., {Riley}, P., {Skoug}, R., {Balogh}, A.,
  {Forsyth}, R., {Goldstein}, B.~E., and {Neugebauer}, M., 1998, {Ulysses'
  return to the slow solar wind}, \emph{\grl}, \textbf{25}, 1--4,
  \doi{10.1029/97GL03444}.

\bibitem[{\emph{{McComas} et~al.}(1999)\emph{{McComas}, Funsten, Gosling, and
  Pryor}}]{mccomas_etal:99}
{McComas}, D.~J., Funsten, H.~O., Gosling, J.~T., and Pryor, W.~R., 1999,
  Ulysses measurements of variations in the solar wind --- interstellar
  hydrogen charge exchange rate, \emph{\grl}, \textbf{26}, 2701--2704.

\bibitem[{\emph{{McComas} et~al.}(2000{\natexlab{a}})\emph{{McComas},
  Barraclough, Funsten, Gosling, Santiago-Munoz, Goldstein, Neugebauer, Riley,
  and Balogh}}]{mccomas_etal:00b}
{McComas}, D.~J., Barraclough, B.~L., Funsten, H.~O., Gosling, J.~T.,
  Santiago-Munoz, Goldstein, B.~E., Neugebauer, M., Riley, P., and Balogh, A.,
  2000{\natexlab{a}}, Solar wind observations over {U}lysses first full polar
  orbit, \emph{\jgr}, \textbf{105}, 10\,419--10\,433.

\bibitem[{\emph{{McComas} et~al.}(2000{\natexlab{b}})\emph{{McComas}, Gosling,
  and Skoug}}]{mccomas_etal:00a}
{McComas}, D.~J., Gosling, J.~T., and Skoug, R.~M., 2000{\natexlab{b}}, Ulysses
  observations of the irregularly structured mid-latitude solar wind during the
  approach to solar maximum, \emph{\grl}, \textbf{27}, 2437--2440.

\bibitem[{\emph{{McComas} et~al.}(2002{\natexlab{a}})\emph{{McComas}, Elliot,
  and von Steiger}}]{mccomas_etal:02a}
{McComas}, D.~J., Elliot, H.~A., and von Steiger, R., 2002{\natexlab{a}}, Solar
  wind from high-latitude coronal holes at solar maximum, \emph{\grl},
  \textbf{29}, 10.1029/2001GL013\,940.

\bibitem[{\emph{{McComas} et~al.}(2002{\natexlab{b}})\emph{{McComas},
  {Elliott}, {Gosling}, {Reisenfeld}, {Skoug}, {Goldstein}, {Neugebauer}, and
  {Balogh}}}]{mccomas_etal:02b}
{McComas}, D.~J., {Elliott}, H.~A., {Gosling}, J.~T., {Reisenfeld}, D.~B.,
  {Skoug}, R.~M., {Goldstein}, B.~E., {Neugebauer}, M., and {Balogh}, A.,
  2002{\natexlab{b}}, {Ulysses' second fast-latitude scan: Complexity near
  solar maximum and the reformation of polar coronal holes}, \emph{\grl},
  \textbf{29}, 1290, \doi{10.1029/2001GL014164}.

\bibitem[{\emph{{McComas} et~al.}(2003)\emph{{McComas}, Elliot, Schwadron,
  Gosling, Skoug, and Goldstein}}]{mccomas_etal:03a}
{McComas}, D.~J., Elliot, H.~A., Schwadron, N.~A., Gosling, J.~T., Skoug,
  R.~M., and Goldstein, B.~E., 2003, The three-dimensional solar wind around
  solar maximum, \emph{\grl}, \textbf{30}, 10.1029/2003GL017\,136.

\bibitem[{\emph{{McComas} et~al.}(2006)\emph{{McComas}, {Allegrini},
  {Bartolone}, {Bochsler}, {Bzowski}, {Collier}, {Fahr}, {Fichtner}, {Frisch},
  {Funsten}, {Fuselier}, {Gloeckler}, {Gruntman}, {Izmodenov}, {Knappenberger},
  {Lee}, {Livi}, {Mitchell}, {M{\"o}bius}, {Moore}, {Pope}, {Reisenfeld},
  {Roelof}, {Runge}, {Scherrer}, {Schwadron}, {Tyler}, {Wieser}, {Witte},
  {Wurz}, and {Zank}}}]{mccomas_etal:06b}
{McComas}, D.~J., {Allegrini}, F., {Bartolone}, L., {Bochsler}, P., {Bzowski},
  M., {Collier}, M., {Fahr}, H., {Fichtner}, H., {Frisch}, P., {Funsten}, H.,
  {Fuselier}, S., {Gloeckler}, G., {Gruntman}, M., {Izmodenov}, V.,
  {Knappenberger}, P., {Lee}, M., {Livi}, S., {Mitchell}, D., {M{\"o}bius}, E.,
  {Moore}, T., {Pope}, S., {Reisenfeld}, D., {Roelof}, E., {Runge}, H.,
  {Scherrer}, J., {Schwadron}, N., {Tyler}, R., {Wieser}, M., {Witte}, M.,
  {Wurz}, P., and {Zank}, G., 2006, {The Interstellar Boundary Explorer (IBEX):
  Update at the end of phase B}, in \emph{Physics of the Inner Heliosheath},
  edited by J.~{Heerikhuisen}, V.~{Florinski}, G.~P. {Zank}, and N.~V.
  {Pogorelov}, vol. 858 of \emph{American Institute of Physics Conference
  Series}, pp. 241--250.

\bibitem[{\emph{{McComas} et~al.}(2008)\emph{{McComas}, Ebert, Elliot,
  Goldstein, Gosling, Schwadron, and Skoug}}]{mccomas_etal:08a}
{McComas}, D.~J., Ebert, R.~W., Elliot, H.~A., Goldstein, B.~E., Gosling,
  J.~T., Schwadron, N.~A., and Skoug, R.~M., 2008, Weaker solar wind from the
  polar coronal holes and the whole {S}un, \emph{\grl}, \textbf{35}, L18\,103,
  \doi{10.1029/2008GL034896}.

\bibitem[{\emph{{McComas} et~al.}(2009{\natexlab{a}})\emph{{McComas},
  {Allegrini}, {Bochsler}, {Bzowski}, {Christian}, {Crew}, {DeMajistre},
  {Fahr}, {Fichtner}, {Frisch}, {Funsten}, {Fuselier}, {Gloeckler}, {Gruntman},
  {Heerikhuisen}, {Izmodenov}, {Janzen}, {Knappenberger}, {Krimigis},
  {Kucharek}, {Lee}, {Livadiotis}, {Livi}, {MacDowall}, {Mitchell},
  {M{\"o}bius}, {Moore}, {Pogorelov}, {Reisenfeld}, {Roelof}, {Saul},
  {Schwadron}, {Valek}, {Vanderspek}, {Wurz}, and {Zank}}}]{mccomas_etal:09c}
{McComas}, D.~J., {Allegrini}, F., {Bochsler}, P., {Bzowski}, M., {Christian},
  E.~R., {Crew}, G.~B., {DeMajistre}, R., {Fahr}, H., {Fichtner}, H., {Frisch},
  P.~C., {Funsten}, H.~O., {Fuselier}, S.~A., {Gloeckler}, G., {Gruntman}, M.,
  {Heerikhuisen}, J., {Izmodenov}, V., {Janzen}, P., {Knappenberger}, P.,
  {Krimigis}, S., {Kucharek}, H., {Lee}, M., {Livadiotis}, G., {Livi}, S.,
  {MacDowall}, R.~J., {Mitchell}, D., {M{\"o}bius}, E., {Moore}, T.,
  {Pogorelov}, N.~V., {Reisenfeld}, D., {Roelof}, E., {Saul}, L., {Schwadron},
  N.~A., {Valek}, P.~W., {Vanderspek}, R., {Wurz}, P., and {Zank}, G.~P.,
  2009{\natexlab{a}}, {Global Observations of the Interstellar Interaction from
  the Interstellar Boundary Explorer (IBEX)}, \emph{Science}, \textbf{326},
  959--962, \doi{10.1126/science.1180906}.

\bibitem[{\emph{{McComas} et~al.}(2009{\natexlab{b}})\emph{{McComas},
  {Allegrini}, {Bochsler}, {Bzowski}, {Collier}, {Fahr}, {Fichtner}, {Frisch},
  {Funsten}, {Fuselier}, {Gloeckler}, {Gruntman}, {Izmodenov}, {Knappenberger},
  {Lee}, {Livi}, {Mitchell}, {M{\"o}bius}, {Moore}, {Pope}, {Reisenfeld},
  {Roelof}, {Scherrer}, {Schwadron}, {Tyler}, {Wieser}, {Witte}, {Wurz}, and
  {Zank}}}]{mccomas_etal:09a}
{McComas}, D.~J., {Allegrini}, F., {Bochsler}, P., {Bzowski}, M., {Collier},
  M., {Fahr}, H., {Fichtner}, H., {Frisch}, P., {Funsten}, H.~O., {Fuselier},
  S.~A., {Gloeckler}, G., {Gruntman}, M., {Izmodenov}, V., {Knappenberger}, P.,
  {Lee}, M., {Livi}, S., {Mitchell}, D., {M{\"o}bius}, E., {Moore}, T., {Pope},
  S., {Reisenfeld}, D., {Roelof}, E., {Scherrer}, J., {Schwadron}, N., {Tyler},
  R., {Wieser}, M., {Witte}, M., {Wurz}, P., and {Zank}, G.,
  2009{\natexlab{b}}, {IBEX -- Interstellar Boundary Explorer}, \emph{\ssr},
  \textbf{146}, 11--33, \doi{10.1007/s11214-009-9499-4}.

\bibitem[{\emph{{M{\"o}bius} et~al.}(1985)\emph{{M{\"o}bius}, {Hovestadt},
  {Klecker}, {Scholer}, and {Gloeckler}}}]{mobius_etal:85a}
{M{\"o}bius}, E., {Hovestadt}, D., {Klecker}, B., {Scholer}, M., and
  {Gloeckler}, G., 1985, {Direct observation of {H}e$^+$ pick-up ions of
  interstellar origin in the solar wind}, \emph{\nat}, \textbf{318}, 426--429.

\bibitem[{\emph{Neugebauer}(1970)}]{neugebauer:70a}
Neugebauer, M., 1970, Initial deceleration of solar wind positive ions in the
  earth's bow shock, \emph{\jgr}, \textbf{75}, 717--733.

\bibitem[{\emph{{Neugebauer} and {Snyder}}(1962)}]{neugebauer_snyder:62a}
{Neugebauer}, M. and {Snyder}, C.~W., 1962, {Solar Plasma Experiment},
  \emph{Science}, \textbf{138}, 1095--1097, \doi{10.1029/JA075i004p00717}.

\bibitem[{\emph{{Ogawa} et~al.}(1995)\emph{{Ogawa}, {Wu}, {Gangopadhyay}, and
  {Judge}}}]{ogawa_etal:95}
{Ogawa}, H.~S., {Wu}, C.~Y.~R., {Gangopadhyay}, P., and {Judge}, D.~L., 1995,
  {Solar photoionization as a loss mechanism of neutral interstellar hydrogen
  in interplanetary space}, \emph{\jgr}, \textbf{100}, 3455--3462.

\bibitem[{\emph{{Ogilvie} et~al.}(1968)\emph{{Ogilvie}, {Burlaga}, and
  {Wilkerson}}}]{ogilvie_etal:68a}
{Ogilvie}, K.~W., {Burlaga}, L.~F., and {Wilkerson}, T.~D., 1968, {Plasma
  observations on Explorer 34}, \emph{\jgr}, \textbf{73}, 6809--6824,
  \doi{10.1029/JA073i021p06809}.

\bibitem[{\emph{{Ohmi} et~al.}(2001)\emph{{Ohmi}, {Kojima}, {Yokobe},
  {Tokumaru}, {Fujiki}, and {Hakamada}}}]{ohmi_etal:01a}
{Ohmi}, T., {Kojima}, M., {Yokobe}, A., {Tokumaru}, M., {Fujiki}, K., and
  {Hakamada}, K., 2001, {Polar low-speed solar wind at the solar activity
  maximum}, \emph{\jgr}, \textbf{106}, 24\,923--24\,936,
  \doi{10.1029/2001JA900094}.

\bibitem[{\emph{{Ohmi} et~al.}(2003)\emph{{Ohmi}, {Kojima}, {Fujiki},
  {Tokumaru}, {Hayashi}, and {Hakamada}}}]{ohmi_etal:03a}
{Ohmi}, T., {Kojima}, M., {Fujiki}, K., {Tokumaru}, M., {Hayashi}, K., and
  {Hakamada}, K., 2003, {Polar low-speed solar wind reappeared at the solar
  activity maximum of cycle 23}, \emph{\grl}, \textbf{30}, 1409,
  \doi{10.1029/2002GL016347}.

\bibitem[{\emph{Osterbart and Fahr}(1992)}]{osterbart_fahr:92}
Osterbart, R. and Fahr, H.~J., 1992, A {B}oltzmann--kinetic approach to
  describe entrance of neutral interstellar hydrogen into the heliosphere,
  \emph{\aap}, \textbf{264}, 260--269.

\bibitem[{\emph{{Owocki} et~al.}(1983)\emph{{Owocki}, {Holzer}, and
  {Hundhausen}}}]{owocki_scudder:83a}
{Owocki}, S.~P., {Holzer}, T.~E., and {Hundhausen}, A.~J., 1983, {The solar
  wind ionization state as a coronal temperature diagnostic}, \emph{\apj},
  \textbf{275}, 354--366.

\bibitem[{\emph{Parker}(1958)}]{parker:57a}
Parker, E.~N., 1958, Dynamics of the interplanetary gas and magnetic fields,
  \emph{\apj}, \textbf{128}, 664--676.

\bibitem[{\emph{Phillips et~al.}(1995{\natexlab{a}})\emph{Phillips, Bame,
  Barnes, Barrcalough, Feldman, Goldstein, Gosling, Hoogveen, McComas,
  Neugebauer, and Suess}}]{phillips_etal:95c}
Phillips, J.~L., Bame, S.~J., Barnes, A., Barrcalough, B.~L., Feldman, W.~C.,
  Goldstein, B.~E., Gosling, J.~T., Hoogveen, G.~W., McComas, D.~J.,
  Neugebauer, M., and Suess, S.~T., 1995{\natexlab{a}}, Ulysses solar wind
  plasma observations from pole to pole, \emph{\grl}, \textbf{22}, 3301--3304.

\bibitem[{\emph{Phillips et~al.}(1995{\natexlab{b}})\emph{Phillips, Bame,
  Feldman, Gosling, Hammond, McComas, Goldstein, and
  Neugebauer}}]{phillips_etal:95a}
Phillips, J.~L., Bame, S.~J., Feldman, W.~C., Gosling, J.~T., Hammond, C.~M.,
  McComas, D.~J., Goldstein, B.~E., and Neugebauer, M., 1995{\natexlab{b}},
  Ulysses solar wind plasma observations during the declining phase of solar
  cycle 22, \emph{\asr}, \textbf{16}, (9)85--(9)94.

\bibitem[{\emph{{Pilipp} et~al.}(1987{\natexlab{a}})\emph{{Pilipp},
  {Muehlhaeuser}, {Miggenrieder}, {Montgomery}, and
  {Rosenbauer}}}]{pilipp_etal:87b}
{Pilipp}, W.~G., {Muehlhaeuser}, K.-H., {Miggenrieder}, H., {Montgomery},
  M.~D., and {Rosenbauer}, H., 1987{\natexlab{a}}, {Unusual electron
  distribution functions in the solar wind derived from the HELIOS plasma
  experiment - Double-strahl distributions and distributions with an extremely
  anisotropic core}, \emph{\jgr}, \textbf{92}, 1093--1101.

\bibitem[{\emph{{Pilipp} et~al.}(1987{\natexlab{b}})\emph{{Pilipp},
  {Muehlhaeuser}, {Miggenrieder}, {Montgomery}, and
  {Rosenbauer}}}]{pilipp_etal:87c}
{Pilipp}, W.~G., {Muehlhaeuser}, K.-H., {Miggenrieder}, H., {Montgomery},
  M.~D., and {Rosenbauer}, H., 1987{\natexlab{b}}, {Characteristics of electron
  velocity distribution functions in the solar wind derived from the HELIOS
  plasma experiment}, \emph{\jgr}, \textbf{92}, 1075--1092.

\bibitem[{\emph{Pryor et~al.}(1992)\emph{Pryor, Ajello, Barth, Hord, Stewart,
  Simmons, McClintock, Sandel, and Shemansky}}]{pryor_etal:92}
Pryor, W.~R., Ajello, J.~M., Barth, C.~A., Hord, C.~W., Stewart, A. I.~F.,
  Simmons, K.~E., McClintock, W.~E., Sandel, B.~R., and Shemansky, D.~E., 1992,
  The {GALILEO} and {PIONEER VENUS} ultraviolet spectrometer experiments: solar
  {L}yman-$\alpha$ latitude variation at solar maximum from interplanetary
  {L}yman-$\alpha$ observations, \emph{\apj}, \textbf{394}, 363--377.

\bibitem[{\emph{Pryor et~al.}(1998)\emph{Pryor, Witte, and
  Ajello}}]{pryor_etal:98b}
Pryor, W.~R., Witte, M., and Ajello, J.~M., 1998, Interplanetary
  {L}yman~$\alpha$ remote sensing with the {U}lysses interstellar neutral gas
  experiment, \emph{\jgr}, \textbf{103}, 26\,813--26\,831.

\bibitem[{\emph{{Pryor} et~al.}(2003)\emph{{Pryor}, {Ajello}, {McComas},
  {Witte}, and {Tobiska}}}]{pryor_etal:03a}
{Pryor}, W.~R., {Ajello}, J.~M., {McComas}, D.~J., {Witte}, M., and {Tobiska},
  W.~K., 2003, {Hydrogen atom lifetimes in the three-dimensional heliosphere
  over the solar cycle}, \emph{\jgr}, \textbf{108}, 8034,
  \doi{10.1029/2003JA009878}.

\bibitem[{\emph{{Qu{\'e}merais}}(2006)}]{quemerais:06a}
{Qu{\'e}merais}, E., 2006, The interplanetary {L}yman-{$\alpha$} background, in
  \emph{The Physics of the Heliospheric Boundaries}, edited by {V.~V.~Izmodenov
  \& R.~Kallenbach}, pp. 283--310.

\bibitem[{\emph{{Qu{\'e}merais} et~al.}(2006)\emph{{Qu{\'e}merais},
  {Lallement}, {Ferron}, {Koutroumpa}, {Bertaux}, {Kyr{\"o}l{\"a}}, and
  {Schmidt}}}]{quemerais_etal:06b}
{Qu{\'e}merais}, E., {Lallement}, R., {Ferron}, S., {Koutroumpa}, D.,
  {Bertaux}, J.-L., {Kyr{\"o}l{\"a}}, E., and {Schmidt}, W., 2006,
  {Interplanetary hydrogen absolute ionization rates: Retrieving the solar wind
  mass flux latitude and cycle dependence with SWAN/SOHO maps}, \emph{\jgr},
  \textbf{111}, 9114--9131, \doi{10.1029/2006JA011711}.

\bibitem[{\emph{{Richards} et~al.}(1994)\emph{{Richards}, {Fennelly}, and
  {Torr}}}]{richards_etal:94a}
{Richards}, P.~G., {Fennelly}, J.~A., and {Torr}, D.~G., 1994, {EUVAC: A solar
  EUV flux model for aeronomic calculations}, \emph{\jgr}, \textbf{99},
  8981--8992, \doi{10.1029/94JA00518}.

\bibitem[{\emph{Richardson et~al.}(1995)\emph{Richardson, Paularena, Lazarus,
  and Belcher}}]{richardson_etal:95a}
Richardson, J.~D., Paularena, K.~I., Lazarus, A.~J., and Belcher, J.~W., 1995,
  Radial evolution of the solar wind from {IMP}~8 to {V}oyager~2, \emph{\grl},
  \textbf{22}, 325--328.

\bibitem[{\emph{{Richardson} et~al.}(2008)\emph{{Richardson}, {Kasper}, {Wang},
  {Belcher}, and {Lazarus}}}]{richardson_etal:08b}
{Richardson}, J.~D., {Kasper}, J.~C., {Wang}, C., {Belcher}, J.~W., and
  {Lazarus}, A.~J., 2008, {Cool heliosheath plasma and deceleration of the
  upstream solar wind at the termination shock}, \emph{\nat}, \textbf{454},
  63--66, \doi{10.1038/nature07024}.

\bibitem[{\emph{Richardson et~al.}(2008)\emph{Richardson, Liu, Wang, and
  McComas}}]{richardson_etal:08a}
Richardson, J.~D., Liu, Y., Wang, C., and McComas, D., 2008, Determining the
  {LIC} {H} density from the solar wind slowdown, \emph{\aap}, \textbf{491},
  1--5, \doi{10.1051/0004-6361:20078565}.

\bibitem[{\emph{Ruci{\'n}ski and
  Bzowski}(1995{\natexlab{a}})}]{rucinski_bzowski:95a}
Ruci{\'n}ski, D. and Bzowski, M., 1995{\natexlab{a}}, Solar cycle dependence of
  the production of {H}$^+$ pick-up ions in the inner heliosphere, \emph{\asr},
  \textbf{16}, 121--124.

\bibitem[{\emph{Ruci{\'n}ski and
  Bzowski}(1995{\natexlab{b}})}]{rucinski_bzowski:95b}
Ruci{\'n}ski, D. and Bzowski, M., 1995{\natexlab{b}}, Modulation of
  interplanetary hydrogen distribution during the solar cycle, \emph{\aap},
  \textbf{296}, 248--263.

\bibitem[{\emph{Ruci{\'n}ski and Fahr}(1989)}]{rucinski_fahr:89}
Ruci{\'n}ski, D. and Fahr, H.~J., 1989, The influence of electron impact
  ionization on the distribution of interstellar helium in the inner
  heliosphere: {P}ossible consequences for determination of interstellar helium
  parameters, \emph{\aap}, \textbf{224}, 290--298.

\bibitem[{\emph{Ruci{\'n}ski and Fahr}(1991)}]{rucinski_fahr:91}
Ruci{\'n}ski, D. and Fahr, H.~J., 1991, Nonthermal ions of interstellar origin
  at different solar wind conditions, \emph{\ag}, \textbf{9}, 102--110.

\bibitem[{\emph{{Salem} et~al.}(2001)\emph{{Salem}, {Bosqued}, {Larson},
  {Mangeney}, {Maksimovic}, {Perche}, {Lin}, and {Bougeret}}}]{salem_etal:01a}
{Salem}, C., {Bosqued}, J.-M., {Larson}, D.~E., {Mangeney}, A., {Maksimovic},
  M., {Perche}, C., {Lin}, R.~P., and {Bougeret}, J.-L., 2001, {Determination
  of accurate solar wind electron parameters using particle detectors and radio
  wave receivers}, \emph{\jgr}, \textbf{106}, 21\,701--21\,717,
  \doi{10.1029/2001JA900031}.

\bibitem[{\emph{{Salem} et~al.}(2003)\emph{{Salem}, {Hoang}, {Issautier},
  {Maksimovic}, and {Perche}}}]{salem_etal:03a}
{Salem}, C., {Hoang}, S., {Issautier}, K., {Maksimovic}, M., and {Perche}, C.,
  2003, {Wind-Ulysses in-situ thermal noise measurements of solar wind electron
  density and core temperature at solar maximum and minimum}, \emph{\asr},
  \textbf{32}, 491--496.

\bibitem[{\emph{Scherer et~al.}(1999)\emph{Scherer, Bzowski, Fahr, and
  Ruci{\'n}ski}}]{scherer_etal:99}
Scherer, H., Bzowski, M., Fahr, H.~J., and Ruci{\'n}ski, D., 1999, Improved
  analysis of interplanetary hst-h-ly$\alpha$ spectra using time-dependent
  modelings, \emph{\aap}, \textbf{342}, 601--609.

\bibitem[{\emph{{Scherer} et~al.}(2000)\emph{{Scherer}, {Fahr}, {Bzowski}, and
  {Ruci{\'n}ski}}}]{scherer_etal:00a}
{Scherer}, H., {Fahr}, H.~J., {Bzowski}, M., and {Ruci{\'n}ski}, D., 2000, {The
  influence of fluctuations of the solar emission line profile on the {D}oppler
  shift of interplanetary {HL}y{$\alpha$} lines observed by the
  {H}ubble-{S}pace-{T}elescope}, \emph{\apss}, \textbf{274}, 133--141.

\bibitem[{\emph{Scime et~al.}(1994)\emph{Scime, Bame, Feldman, Gary, and
  Phillips}}]{scime_etal:94}
Scime, E.~E., Bame, S.~J., Feldman, W.~C., Gary, S.~P., and Phillips, J.~L.,
  1994, Regulation of the solar wind electron heat flux from 1 to 5~{AU},
  \emph{\jgr}, \textbf{99}, 23\,401--23\,410.

\bibitem[{\emph{Smith et~al.}(1991)\emph{Smith, Johnson, Gao, Smith, and
  Stebbings}}]{smith_etal:91a}
Smith, G.~J., Johnson, L.~K., Gao, R.~S., Smith, K.~A., and Stebbings, R.~F.,
  1991, Absolute differential cross sections for electron capture and loss by
  kilo-electron-volt hydrogen atoms, \emph{Phys. Rev. A}, \textbf{44},
  5647--5652.

\bibitem[{\emph{{Sok\'{o}{\l}} et~al.}(2012)\emph{{Sok\'{o}{\l}}, Bzowski,
  Tokumaru, Fujiki, and McComas}}]{sokol_etal:12a}
{Sok\'{o}{\l}}, J., Bzowski, M., Tokumaru, M., Fujiki, K., and McComas, D.,
  2012, {Heliolatitude and time variations of solar wind structure from in-situ
  measurements and interplanetary scintillation observations}, \emph{\solphys},
  \textbf{in preparation}, 00.

\bibitem[{\emph{Summanen}(1996)}]{summanen:96}
Summanen, T., 1996, The effect of the time and latitude-dependent solar
  ionisation rate on the measured {L}yman-$\alpha$-intensity, \emph{\aap},
  \textbf{314}, 663--671.

\bibitem[{\emph{Summanen et~al.}(1993)\emph{Summanen, Lallement, Bertaux, and
  Kyr{\"o}l{\"a}}}]{summanen_etal:93}
Summanen, T., Lallement, R., Bertaux, J.~L., and Kyr{\"o}l{\"a}, E., 1993,
  Latitudinal distribution of solar wind as deduced from {L}yman {A}lpha
  measurements: an improved method, \emph{\jgr}, \textbf{98}, 13\,215--13\,224.

\bibitem[{\emph{{Tapping}}(1987)}]{tapping:87}
{Tapping}, K.~F., 1987, {Recent solar radio astronomy at centimeter wavelengths
  - The temporal variability of the 10.7-cm flux}, \emph{\jgr}, \textbf{92},
  829--838, \doi{10.1029/JD092iD01p00829}.

\bibitem[{\emph{{Tarnopolski} and
  {Bzowski}}(2008{\natexlab{a}})}]{tarnopolski_bzowski:08a}
{Tarnopolski}, S. and {Bzowski}, M., 2008{\natexlab{a}}, {Detectability of
  neutral interstellar deuterium by a forthcoming SMEX mission IBEX},
  \emph{\aap}, \textbf{483}, L35--L38, \doi{10.1051/0004-6361:200809593}.

\bibitem[{\emph{{Tarnopolski} and
  {Bzowski}}(2008{\natexlab{b}})}]{tarnopolski_bzowski:09}
{Tarnopolski}, S. and {Bzowski}, M., 2008{\natexlab{b}}, {Neutral interstellar
  hydrogen in the inner heliosphere under the influence of wavelength-dependent
  solar radiation pressure}, \emph{\aap}, \textbf{493}, 207--216,
  \doi{10.1051/0004-6361:20077058}.

\bibitem[{\emph{{Tarnopolski}}(2007)}]{tarnopolski:07}
{Tarnopolski}, S.~T., 2007, \emph{{Expected distribution of interstellar
  deuterium in the heliosphere}}, Ph.D. thesis, {Space Research Centre PAS}.

\bibitem[{\emph{Thomas}(1978)}]{thomas:78}
Thomas, G.~E., 1978, The interstellar wind and its influence on the
  interplanetary environment, \emph{Ann. Rev. Earth Planet. Sci.}, \textbf{6},
  173--204.

\bibitem[{\emph{{Tian} et~al.}(2009{\natexlab{a}})\emph{{Tian}, {Curdt},
  {Marsch}, and {Sch{\"u}hle}}}]{tian_etal:09e}
{Tian}, H., {Curdt}, W., {Marsch}, E., and {Sch{\"u}hle}, U.,
  2009{\natexlab{a}}, {Hydrogen Lyman-{$\alpha$} and Lyman-{$\beta$} spectral
  radiance profiles in the quiet Sun}, \emph{\aap}, \textbf{504}, 239--248,
  \doi{10.1051/0004-6361/200811445}.

\bibitem[{\emph{{Tian} et~al.}(2009{\natexlab{b}})\emph{{Tian}, {Curdt},
  {Teriaca}, {Landi}, and {Marsch}}}]{tian_etal:09c}
{Tian}, H., {Curdt}, W., {Teriaca}, L., {Landi}, E., and {Marsch}, E.,
  2009{\natexlab{b}}, {Solar transition region above sunspots}, \emph{\aap},
  \textbf{505}, 307--318, \doi{10.1051/0004-6361/200912114}.

\bibitem[{\emph{{Tian} et~al.}(2009{\natexlab{c}})\emph{{Tian}, {Teriaca},
  {Curdt}, and {Vial}}}]{tian_etal:09d}
{Tian}, H., {Teriaca}, L., {Curdt}, W., and {Vial}, J.-C., 2009{\natexlab{c}},
  {Hydrogen Ly{$\alpha$} and Ly{$\beta$} Radiances and Profiles in Polar
  Coronal Holes}, \emph{\apjl}, \textbf{703}, L152--L156,
  \doi{10.1088/0004-637X/703/2/L152}.

\bibitem[{\emph{{Tobiska} et~al.}(2000)\emph{{Tobiska}, {Woods}, {Eparvier},
  {Viereck}, {Floyd}, {Bouwer}, {Rottman}, and {White}}}]{tobiska_etal:00c}
{Tobiska}, W.~K., {Woods}, T., {Eparvier}, F., {Viereck}, R., {Floyd}, L.,
  {Bouwer}, D., {Rottman}, G., and {White}, O.~R., 2000, {The SOLAR2000
  empirical solar irradiance model and forecast tool}, \emph{J. Atm. Sol. Terr.
  Phys.}, \textbf{62}, 1233--1250.

\bibitem[{\emph{{Tokumaru} et~al.}(2009)\emph{{Tokumaru}, {Kojima}, {Fujiki},
  and {Hayashi}}}]{tokumaru_etal:09a}
{Tokumaru}, M., {Kojima}, M., {Fujiki}, K., and {Hayashi}, K., 2009,
  {Non-dipolar solar wind structure observed in the cycle 23/24 minimum},
  \emph{\grl}, \textbf{360}, L09\,101, \doi{10.1029/2009GL037461}.

\bibitem[{\emph{{Tokumaru} et~al.}(2010)\emph{{Tokumaru}, {Kojima}, and
  {Fujiki}}}]{tokumaru_etal:10a}
{Tokumaru}, M., {Kojima}, M., and {Fujiki}, K., 2010, {Solar cycle evolution of
  the solar wind speed distribution from 1985 to 2008}, \emph{\jgr},
  \textbf{115}, A04\,102, \doi{10.1029/2009JA014628}.

\bibitem[{\emph{{{\v S}tver{\'a}k} et~al.}(2008)\emph{{{\v S}tver{\'a}k},
  {Tr{\'a}vn{\'{\i}}{\v c}ek}, {Maksimovic}, {Marsch}, {Fazakerley}, and
  {Scime}}}]{stverak_etal:08a}
{{\v S}tver{\'a}k}, {\v S}., {Tr{\'a}vn{\'{\i}}{\v c}ek}, P., {Maksimovic}, M.,
  {Marsch}, E., {Fazakerley}, A.~N., and {Scime}, E.~E., 2008, {Electron
  temperature anisotropy constraints in the solar wind}, \emph{\jgr},
  \textbf{113}, A03\,103, \doi{10.1029/2007JA012733}.

\bibitem[{\emph{{{\v S}tver{\'a}k} et~al.}(2009)\emph{{{\v S}tver{\'a}k},
  {Maksimovic}, {Tr{\'a}vn{\'{\i}}{\v c}ek}, {Marsch}, {Fazakerley}, and
  {Scime}}}]{stverak_etal:09a}
{{\v S}tver{\'a}k}, {\v S}., {Maksimovic}, M., {Tr{\'a}vn{\'{\i}}{\v c}ek},
  P.~M., {Marsch}, E., {Fazakerley}, A.~N., and {Scime}, E.~E., 2009, {Radial
  evolution of nonthermal electron populations in the low-latitude solar wind:
  Helios, Cluster, and Ulysses Observations}, \emph{\jgr}, \textbf{114},
  A05\,104, \doi{10.1029/2008JA013883}.

\bibitem[{\emph{Vasyliunas and Siscoe}(1976)}]{vasyliunas_siscoe:76}
Vasyliunas, V. and Siscoe, G., 1976, On the flux and the energy spectrum of
  interstellar ions in the solar wind, \emph{\jgr}, \textbf{81}, 1247--1252.

\bibitem[{\emph{Verner et~al.}(1996)\emph{Verner, Ferland, Korista, and
  Yakovlev}}]{verner_etal:96}
Verner, D.~A., Ferland, G.~J., Korista, T.~K., and Yakovlev, D.~G., 1996,
  Atomic data for astrophysics. {II}. {N}ew fits for photoionization
  cross-sections of atoms and ions, \emph{\apj}, \textbf{465}, 487--498.

\bibitem[{\emph{Vidal-Madjar}(1975)}]{vidal-madjar:75}
Vidal-Madjar, A., 1975, Evolution of the solar {L}yman alpha flux during four
  consecutive years, \emph{\solphys}, \textbf{40}, 69--86.

\bibitem[{\emph{Vidal-Madjar and Phissamay}(1980)}]{vidal-madjar_phissamay:80}
Vidal-Madjar, A. and Phissamay, B., 1980, The solar {L}$\alpha$ flux near solar
  minimum, \emph{\solphys}, \textbf{66}, 259--271.

\bibitem[{\emph{{Viereck} and {Puga}}(1999)}]{viereck_puga:99}
{Viereck}, R.~A. and {Puga}, L.~C., 1999, {The NOAA Mg II core-to-wing solar
  index: Construction of a 20-year time series of chromospheric variability
  from multiple satellites}, \emph{\jgr}, \textbf{104}, 9995--10\,006,
  \doi{10.1029/1998JA900163}.

\bibitem[{\emph{{Wachowicz}}(2006)}]{wachowicz:06}
{Wachowicz}, M.~E., 2006, \emph{{Global model of distribution of ionization
  states of heavy ions from solar plsama in the heliosphere (in {P}olish)}},
  Ph.D. thesis, {Space Research Centre PAS}.

\bibitem[{\emph{{Warren}}(2006)}]{warren:06a}
{Warren}, H.~P., 2006, {NRLEUV 2: A new model of solar EUV irradiance
  variability}, \emph{Advances in Space Research}, \textbf{37}, 359--365,
  \doi{10.1016/j.asr.2005.10.028}.

\bibitem[{\emph{{Warren} et~al.}(1998{\natexlab{a}})\emph{{Warren}, {Mariska},
  and {Lean}}}]{warren_etal:98a}
{Warren}, H.~P., {Mariska}, J.~T., and {Lean}, J., 1998{\natexlab{a}}, {A new
  reference spectrum for the EUV irradiance of the quiet Sun 1. Emission
  measure formulation}, \emph{\jgr}, \textbf{103}, 12\,077--12\,090,
  \doi{10.1029/98JA00810}.

\bibitem[{\emph{{Warren} et~al.}(1998{\natexlab{b}})\emph{{Warren}, {Mariska},
  and {Lean}}}]{warren_etal:98b}
{Warren}, H.~P., {Mariska}, J.~T., and {Lean}, J., 1998{\natexlab{b}}, {A new
  reference spectrum for the EUV irradiance of the quiet Sun 2. Comparisons
  with observations and previous models}, \emph{\jgr}, \textbf{103},
  12\,091--12\,102, \doi{10.1029/98JA00811}.

\bibitem[{\emph{{Warren} et~al.}(1998{\natexlab{c}})\emph{{Warren}, {Mariska},
  and {Wilhelm}}}]{warren_mariska:98a}
{Warren}, H.~P., {Mariska}, J.~T., and {Wilhelm}, K., 1998{\natexlab{c}},
  {High-Resolution Observations of the Solar Hydrogen Lyman Lines in the Quiet
  Sun with the SUMER Instrument on SOHO}, \emph{\apjs}, \textbf{119}, 105--120,
  \doi{10.1086/313151}.

\bibitem[{\emph{{Wenzel} et~al.}(1989)\emph{{Wenzel}, {Marsden}, {Page}, and
  {Smith}}}]{wenzel_etal:89a}
{Wenzel}, K.-P., {Marsden}, R.~G., {Page}, D.~E., and {Smith}, E.~J., 1989,
  {Ulysses: The first high-latitude heliospheric mission}, \emph{\asr},
  \textbf{9}, 25--29, \doi{10.1016/0273-1177(89)90089-6}.

\bibitem[{\emph{{Woods} et~al.}(1995)\emph{{Woods}, {Rottman}, {White},
  {Fontenla}, and {Avrett}}}]{woods_etal:95a}
{Woods}, T.~N., {Rottman}, G.~J., {White}, O.~R., {Fontenla}, J., and {Avrett},
  E.~H., 1995, {The solar Ly-alpha line profile}, \emph{\apj}, \textbf{442},
  898--906, \doi{10.1086/175492}.

\bibitem[{\emph{{Woods} et~al.}(1996)\emph{{Woods}, {Prinz}, {Rottman},
  {London}, {Crane}, {Cebula}, {Hilsenrath}, {Brueckner}, {Andrews}, {White},
  {VanHoosier}, {Floyd}, {Herring}, {Knapp}, {Pankratz}, and
  {Reiser}}}]{woods_etal:96a}
{Woods}, T.~N., {Prinz}, D.~K., {Rottman}, G.~J., {London}, J., {Crane}, P.~C.,
  {Cebula}, R.~P., {Hilsenrath}, E., {Brueckner}, G.~E., {Andrews}, M.~D.,
  {White}, O.~R., {VanHoosier}, M.~E., {Floyd}, L.~E., {Herring}, L.~C.,
  {Knapp}, B.~G., {Pankratz}, C.~K., and {Reiser}, P.~A., 1996, {Validation of
  the UARS solar ultraviolet irradiances: Comparison with the ATLAS 1 and 2
  measurements}, \emph{\jgr}, \textbf{101}, 9541--9570,
  \doi{10.1029/96JD00225}.

\bibitem[{\emph{Woods et~al.}(2000)\emph{Woods, Tobiska, Rottman, and
  Worden}}]{woods_etal:00}
Woods, T.~N., Tobiska, W.~K., Rottman, G.~J., and Worden, J.~R., 2000, Improved
  solar {L}yman irradiance modeling from 1979 through 1999 based on {UARS}
  observations, \emph{\jgr}, \textbf{105}, 27\,195--27\,215.

\bibitem[{\emph{{Woods} et~al.}(2005)\emph{{Woods}, {Eparvier}, {Bailey},
  {Chamberlin}, {Lean}, {Rottman}, {Solomon}, {Tobiska}, and
  {Woodraska}}}]{woods_etal:05a}
{Woods}, T.~N., {Eparvier}, F.~G., {Bailey}, S.~M., {Chamberlin}, P.~C.,
  {Lean}, J., {Rottman}, G.~J., {Solomon}, S.~C., {Tobiska}, W.~K., and
  {Woodraska}, D.~L., 2005, Solar {EUV} {E}xperiment {(SEE)}: Mission overview
  and first results, \emph{\jgr}, \textbf{110}, A01\,312,
  \doi{10.1029/2004JA010765}.

\bibitem[{\emph{Wu and Judge}(1979)}]{wu_judge:79a}
Wu, F.~M. and Judge, D.~L., 1979, Temperature and velocity of the
  interplanetary gases along solar radii, \emph{\apj}, \textbf{231}, 594--605.

\end{thebibliography}

\end{document}